\documentclass[11pt]{article}
\usepackage[hmargin={2.7cm,2.7cm},vmargin={2.7cm,2.7cm}]{geometry}

\usepackage{amsmath,amssymb}
\usepackage{amsfonts}
\usepackage{mathrsfs}
\usepackage{dsfont}

\usepackage{epic}
\usepackage{eepic}
\usepackage{graphicx}

\usepackage{pgf}
\usepackage{tikz}
\usetikzlibrary{arrows.meta}
\usetikzlibrary{decorations.markings}

\usepackage{amsmath}
\usepackage{mathtools}

\usepackage{bm}

\usepackage{empheq}

\usepackage{multirow}

\usepackage[skip=2pt,font=small]{caption}
\usepackage{hyperref}
\hypersetup{
colorlinks,
linkcolor={blue!60!black},
citecolor={blue!90!black},
urlcolor={blue!90!black}}

\usepackage{footnote}

\usepackage{cite}
\usepackage{enumerate}
\usepackage[curve]{xy}

\usepackage[titletoc,title]{appendix}

\usepackage{amsmath,amsthm}
\newtheorem{theorem}{Theorem}[section]

\newtheorem{prop}[theorem]{Proposition}

\theoremstyle{remark}
\newtheorem*{remark}{Remark}

\newcommand{\tocless}[2]{#1{#2}}

\tikzset{->-/.style={decoration={
  markings,
  mark=at position .60 
  with {\arrow{Latex[scale=1.1]}}},postaction={decorate}}}
\tikzset{->>-/.style={decoration={
  markings,
  mark=at position .56 
  with {\arrow{Latex[scale=1.1]}},
  mark=at position .64 
  with {\arrow{Latex[scale=1.1]}}},postaction={decorate}}}
\tikzset{->>>-/.style={decoration={
  markings,
  mark=at position .54 
  with {\arrow{Latex[scale=1.6,color=black]}}},postaction={decorate}}}
\tikzset{->>>>-/.style={decoration={
  markings,
  mark=at position .52 
  with {\arrow{Latex[scale=1.6,color=black]}},
  mark=at position .62 
  with {\arrow{Latex[scale=1.6,color=black]}}},postaction={decorate}}}
\tikzset{-<-/.style={decoration={
  markings,
  mark=at position .50 
  with {\arrow{Latex[reversed]}}},postaction={decorate}}}
\tikzset{-<<-/.style={decoration={
  markings,
  mark=at position .40 
  with {\arrow{Latex}}
  mark=at position .60 
  with {\arrow{Latex}}},postaction={decorate}}}
  
\tikzset{-|-/.style={decoration={
  markings,
  mark=at position .51 with {\arrow{Bar}}},postaction={decorate}}}
\tikzset{-||-/.style={decoration={
  markings,
  mark=at position .49 with {\arrow{Bar[sep=-5pt] Bar}}},postaction={decorate}}}
\tikzset{-!-/.style={decoration={
  markings,
  mark=at position .51 with {\arrow{Tee Barb[length=4pt]}}},postaction={decorate}}}
\tikzset{-!!-/.style={decoration={
  markings,
  mark=at position .51 with {\arrow{Tee Barb[sep=1pt,length=4pt] Tee Barb[length=4pt]}}},postaction={decorate}}}

\newcommand{\ds}{\displaystyle}

\renewcommand{\author}[1]{\large\rm #1\\ \bigskip}
\newcommand{\address}[1]{{\normalsize\it #1\\}\bigskip}
\renewcommand{\title}[1]{\bigskip\bigskip\Large\bf #1\bigskip\bigskip\\}

\newcommand{\Bigpsi}[3]{\phantom{\Psi}_2 \kern -.05em
\Psi_2\left(\genfrac{}{}{0pt}{}{#1}{#2}\biggl|#3\right)}

\newcommand{\bea}{\begin{eqnarray}}
\newcommand{\eea}{\end{eqnarray}}

\newcommand{\ii}{\mathsf{i}}
\newcommand{\Wh}{\hat{W}}
\newcommand{\oW}{\overline{W}}
\newcommand{\oWh}{\hat{\oW}}

\newcommand{\lag}{{\mathcal L}}
\newcommand{\ol}{\overline{\lag}}
\newcommand{\lam}{{\Lambda}}
\newcommand{\olam}{\overline{\lam}}
\newcommand{\lagh}{\hat{\lag}}
\newcommand{\olh}{\hat{\ol}}

\newcommand{\bb}{\mathsf{b}}

\newcommand{\bu}{{\mathbf u}}
\newcommand{\bv}{{\mathbf v}}
\newcommand{\bw}{{\mathbf w}}

\def\EXP{\textrm{{\large e}}}

\def\re{\mathop{\hbox{\rm Re}}\nolimits}
\def\im{\mathop{\hbox{\rm Im}}\nolimits}

\newcommand{\rua}{{u_1}}
\newcommand{\rub}{{u_2}}
\newcommand{\rva}{{v_1}}
\newcommand{\rvb}{{v_2}}
\newcommand{\rwa}{{w_1}}
\newcommand{\rwb}{{w_2}}

\newcommand{\al}{{\bm{\alpha}}}
\newcommand{\bt}{{\bm{\beta}}}
\newcommand{\gm}{{\bm{\gamma}}}
\newcommand{\bxi}{{\bm{\xi}}}

\newcommand{\xv}{\mathbf{x}}

\newcommand{\xli}{{x}^{(l)}_i}
\newcommand{\xlj}{{x}^{(l)}_j}
\newcommand{\xlk}{{x}^{(l)}_k}
\newcommand{\xri}{{x}^{(r)}_i}
\newcommand{\xrj}{{x}^{(r)}_j}
\newcommand{\xrk}{{x}^{(r)}_k}

\newcommand{\mxi}{y_i}
\newcommand{\mxj}{y_j}

\newcommand{\omxi}{\overline{y}_i}
\newcommand{\omxj}{\overline{y}_j}
\newcommand{\myi}{y'_j}
\newcommand{\myj}{y'_i}

\newcommand{\mui}{z_i}
\newcommand{\muj}{z_j}

\newcommand{\mupi}{z'_i}
\newcommand{\mupj}{z'_j}

\newcommand{\mpc}{\alpha}
\newcommand{\mqc}{\beta}
\newcommand{\mpa}{\alpha_1}
\newcommand{\mpb}{\alpha_2}
\newcommand{\mqa}{\beta_1}
\newcommand{\mqb}{\beta_2}

\newcommand{\tmp}{\dot{\alpha}}
\newcommand{\tmq}{\dot{\beta}}

\newcommand{\me}{\wp(\omega_2)}

\newcommand{\oV}{\overline{V}}

\newcommand{\Log}{{\textrm{Log}\hspace{1pt}}}

\newcommand{\lie}{{\textrm{Li}_2}}

\newcommand{\cp}{\mathbb{CP}^1}

\newcommand{\spn}{{\sigma}}

\def\EXP{\textrm{{\large e}}}

\def\re{\mathop{\hbox{\rm Re}}\nolimits}
\def\im{\mathop{\hbox{\rm Im}}\nolimits}

\begin{document}

\vglue 2cm

\begin{center}

\title{Two-component Yang-Baxter maps and star-triangle relations}
\author{Andrew P.~Kels}
\address{Scuola Internatzionale Superiore di Studi Avanzati\\ Via Bonomea 265, 34136 Trieste, Italy}

\end{center}

\begin{abstract}

It is shown how Yang-Baxter maps may be directly obtained from classical counterparts of the star-triangle relations and quantum Yang-Baxter equations.  This is based on reinterpreting the latter equation and its solutions which are given in terms of special functions, as a set-theoretical form of the Yang-Baxter equation whose solutions are given by quadrirational Yang-Baxter maps.   The Yang-Baxter maps obtained through this approach are found to satisfy two different types of Yang-Baxter equations, one that is the usual equation involving a single map, and another equation that involves a pair of maps, which is a case of what is also known as an entwining Yang-Baxter equation.  Apart from the elliptic case, each of these Yang-Baxter maps are quadrirational, but only maps that solve the former type of Yang-Baxter equation are reversible.  The Yang-Baxter maps are expressed in terms of two-component variables, and two-component parameters, and have a natural QRT-like composition of separate maps for each component.  Through this approach, sixteen different Yang-Baxter maps are derived from known solutions of the classical star-triangle relations.

\end{abstract}





\section{Introduction}

The Yang-Baxter equation is an important equation for integrability of models of statistical mechanics where it can be used to solve a model through the use of the commuting transfer matrix method of Baxter \cite{Baxter:1972hz}.  There are known to be several different forms of such Yang-Baxter equations \cite{Baxter:1982zz}, which provide relations satisfied the by Boltzmann weights of different types of integrable models of statistical mechanics.  One of the important forms of the Yang-Baxter equation that is central to the ideas of this paper is known as {the star-triangle relation} \cite{Bax02rip,Bazhanov:2016ajm}, which implies integrability of an associated 2-dimensional lattice ``spin'' model of statistical mechanics.  The latter star-triangle relation first appeared in statistical mechanics for the two-dimensional Ising model \cite{Onsager}, and other important solutions have since been obtained \cite{Zamolodchikov:1980mb,Fateev:1982wi,Kashiwara:1986tu,AuYang:1987zc,Baxter:1987eq,Bazhanov:2007mh,Bazhanov:2010kz,Spiridonov:2010em,Kels:2013ola,Kels:2015bda,Gahramanov:2015cva,GahramanovKels} for various lattice models that generalise the Ising model.

Another type of Yang-Baxter equation was introduced as the set-theoretical Yang-Baxter equation \cite{DrinfeldYBMs} (or functional Yang-Baxter equation), which is an equation for maps that act on the Cartesian product of a set with itself.  This form of the Yang-Baxter equation is not directly related to a model of statistical mechanics, but it has appeared and been studied in a wide range of different contexts \cite{SklyaninYBM,WeinsteinxuYBM,AdlerCuttingYBM,HietarintaYBM,NoumiYamadaYBM,ESSYBMs,LuYanZhuYBM,EtingofYBM,OdesskiiYBM,KNYYBM,KOTYYBM,VeselovYBMs,SurVes2003,ABSYBM,GatevaIvanovaYBM,KunibaYBM,Bazhanov:2015gra,TsuboiYBMs,DMYBM,GatevaIvanovaYBM2,SmoktunowiczYBM,DSYBM,PozsgayYBMs,CGPYBM22}.  
In the context of dynamical systems the set-theoretical solutions of the Yang-Baxter equation are commonly known as Yang-Baxter maps (YBMs), following the work of Veselov \cite{VeselovYBMs}, and this approach has naturally led to considerable interest in understanding connections to discrete integrable systems \cite{GoncharenkoVeselovYBM,SurVes2003,ABSYBM,PapageorgiouYBMQuad,PapageorgiouYBMQuad2,KouloukasEntwining,KMYBM,CCZYBM,DoliwaYBM,Bazhanov:2015gra,FXYBM,AYYBM,KNPT2020}.

Relevant to this paper, Bazhanov and Sergeev have made an important connection of Yang-Baxter maps with quantum groups and Hamiltonian structures \cite{Bazhanov:2015gra}, in which they made use of a particular solution of the star-triangle relation and its implied vertex form of the Yang-Baxter equation.  Motivated by this, this paper aims to develop and give a better understanding of the general connection between the star-triangle relation and Yang-Baxter maps, with a particular focus on the star-triangle relation as a starting point for systematically obtaining Yang-Baxter maps, independently of associated quantum group and Hamiltonian structures.  In particular, this connection will be used to derive several new Yang-Baxter maps from known solutions of classical forms of the star-triangle relations.

Specifically, from two different forms of the star-triangle relation, two different types of Yang-Baxter maps will be obtained, denoted
\begin{equation}\label{YBMintro}
R(\al,\bt)\colon X\times X\to X\times X,\qquad U(\al,\bt)\colon X\times X\to X\times X,
\end{equation}
which solve the following two different types of functional Yang-Baxter equations
\begin{align}
\label{FYBE1}
R_{jk}(\bt,\gm)\circ R_{ik}(\al,\gm)\circ R_{ij}(\al,\bt)&=R_{ij}(\al,\bt)\circ R_{ik}(\al,\gm)\circ R_{jk}(\bt,\gm), \\
\label{FYBE2}
U_{jk}(\bt,\gm)\circ U_{ik}(\al,\gm)\circ R_{ij}(\al,\bt)&= R_{ij}(\al,\bt)\circ U_{ik}(\al,\gm)\circ U_{jk}(\bt,\gm).
\end{align}
In the latter equation, the maps $R_{ij}$ and $U_{ij}$ 
act non-trivially as $R$ and $U$ respectively on the $i$-th and $j$-th components of the Cartesian product $X\times X\times X$ of a set $X$, and $\al,\bt,\gm$ are parameters. The Yang-Baxter equation \eqref{FYBE1} is the usual one for Yang-Baxter maps,  while equation \eqref{FYBE2} is less common but has been studied before as a case of the so-called ``entwining'' Yang-Baxter equation \cite{KouloukasEntwining,KassotakisEntwining}.  In the approach of this paper, all Yang-Baxter maps $R(\al,\bt)$ which satisfy \eqref{FYBE2}, also satisfy \eqref{FYBE1}. 
Also, for a given map $R(\al,\bt)$, there can be found several different maps $U(\al,\bt)$ to satisfy \eqref{FYBE2}.  

This paper will be concerned with the case of $X=\mathbb{CP}^1\times\mathbb{CP}^1$, as opposed to the more common $X=\mathbb{CP}^1$, and for this reason the Yang-Baxter maps \eqref{YBMintro} will be referred to as two-component maps.  These Yang-Baxter maps will also depend on two-component complex parameters $\al,\bt\in\mathbb{C}^2$.  Although they are two-component maps, they are not extensions of some simpler one-component Yang-Baxter maps (as far as the author is aware), and the two-component cases are the simplest forms of Yang-Baxter maps that can be expected to arise from the approach of this paper.

In terms of two-component variables $\bxi_i=(y_i,z_i)\in\mathbb{CP}^1\times\mathbb{CP}^1$, the Yang-Baxter maps $R(\al,\bt),U(\al,\bt)\colon(\bxi_i,\bxi_j)\mapsto(\bxi_i',\bxi_j')$ found in this paper can typically be written in the following form
\begin{equation}\label{introYBMform}
\begin{gathered}
\myj=\Upsilon_{1,\al\bt}(\muj,\mxi;\mxj), \qquad \mupi=Z_{1,\al\bt}(\mxj,\myi;\myj), \\
\myi=\Upsilon_{2,\al\bt}(\mui,\mxj;\mxi), \qquad \mupj=Z_{2,\al\bt}(\mxi,\myj;\myi),
\end{gathered}
\end{equation}
where each of the $\Upsilon_{1,\al\bt}(x,y;z)$, $\Upsilon_{2,\al\bt}(x,y;z)$, $Z_{1,\al\bt}(x,y;z)$, $Z_{2,\al\bt}(x,y;z)$, are rational bilinear functions of $x$ and $y$, but with non-polynomial dependence on $z$ and the components of $\al$ and $\bt$.  From the rational bilinearity of the two-component maps of the form \eqref{introYBMform}, it follows straightforwardly that they satisfy the so-called quadrirational property, such that the inverse maps for the first, second, and both components, are also rational bilinear.

It also follows from the form of the two-component maps \eqref{introYBMform} that they can be naturally split up into the following sequence for the components
\begin{equation}
\begin{split}
(\bxi_i,\bxi_j)\mapsto(\bxi_i',\bxi_j')\colon\bigl((\mxi,\mui),(\mxj,\muj)\bigr)\mapsto\bigl((\myj,\mui),(\myi,\muj)\bigr)\mapsto\bigl((\myj,\mupi),(\myi,\mupj)\bigr).
\end{split}
\end{equation}
This resembles the mapping sequence of the QRT maps \cite{QUISPEL1988419,QUISPEL1989183}, which may be split up into so-called horizontal and vertical ``switches'' \cite{DuistermaatBook}.  For the Yang-Baxter maps, the analogue of the first switch would be $(\mxi,\mxj)\mapsto(\myj,\myi)$ and the analogue of the second switch would be $(\mui,\muj)\mapsto(\mupi,\mupj)$.  However, besides this observation it is not presently known if there is a deeper connection to QRT maps, and unlike the QRT maps the switches for the Yang-Baxter maps are not involutions.  Instead, due to particular symmetry that is satisfied for certain solutions of the star-triangle relations, the Yang-Baxter maps $R(\al,\bt)$ (but not $U(\al,\bt)$) will be found to satisfy the reversibility property, such that $R_{ij}(\al,\bt)\circ R_{ji}(\bt,\al)=\textrm{Id}$, where $\textrm{Id}$ represents the identity map.

Throughout this paper the acronyms CSTR and CYBE respectively refer to the {\it classical star-triangle relations} and {\it classical Yang-Baxter equations}, which are introduced in Section \ref{sec:YBE}, while the acronym FYBE refers to the {\it functional Yang-Baxter equations} whose solutions are Yang-Baxter maps, which are introduced in Section \ref{sec:YBM}.

The layout of this paper is as follows. In Section \ref{sec:YBE}, the CSTRs are introduced which arise from the quasi-classical limit of different forms of the (quantum) star-triangle relation, and it is shown how the solutions of the CSTRs may be used to define solutions of different vertex forms of the CYBE.  In Section \ref{sec:YBM}, it is shown how to convert the latter solutions of the CYBEs into Yang-Baxter maps which solve two different types of FYBEs.  In Section \ref{sec:examples}, explicit solutions of the CSTRs are given, and using the method of Section \ref{sec:YBM} these are used to derive sixteen different Yang-Baxter maps which satisfy the two different types of FYBE.  Finally, the Yang-Baxter maps obtained in this paper is listed separately in Appendix \ref{app:YBMlist}.

\section{Star-triangle relations and Yang-Baxter equations}\label{sec:YBE}

In this section the star-triangle relations and related Yang-Baxter equations will be introduced that will be used for obtaining Yang-Baxter maps.  These equations are mainly based on previous works involving the quasi-classical limit of the star-triangle relations \cite{Bazhanov:2007mh,Bazhanov:2010kz,Bazhanov:2016ajm,Kels:2018xge}.  
Before going into details of the general forms of the equations, a basic example will be presented here to provide some motivation and illustrate some of the main ideas of this paper.

Throughout this paper, the function $\Log(z)$ denotes the principal branch of the complex logarithm.  Define a complex-valued function
\begin{equation}\label{lagex}
\lag_\alpha(y_i,y_j)=\alpha\Log(y_i-y_j)-\frac{\alpha}{2}\Log(\alpha).
\end{equation}
where $y_i,y_j\in\mathbb{C}$ and $\alpha\in\mathbb{C}$ are complex valued variables and parameters, respectively.  This function satisfies the equation
\begin{equation}\label{CSTRex}
\begin{split}
  \lag_{-\alpha_1}(y_0,y_1)+\lag_{\alpha_1+\alpha_3}(y_0,y_2)+\lag_{-\alpha_3}(y_0,y_3)& \\
 -\lag_{\alpha_1}(y_2,y_3)-\lag_{-\alpha_1-\alpha_3}(y_1,y_3)-\lag_{\alpha_3}(y_1,y_2)&=2\pi\ii(n_1\alpha_1+n_3\alpha_3),
 \end{split}
\end{equation}
for some integers $n_1,n_3\in\mathbb{Z}$, on solutions of the following equation
\begin{equation}\label{3legex}
 \frac{\partial}{\partial y_0}\Bigl(\lag_{-\alpha_1}(y_0,y_1)+\lag_{\alpha_1+\alpha_3}(y_0,y_2)+\lag_{-\alpha_3}(y_0,y_3)\Bigr)=0.
\end{equation}
This can be straightforwardly checked by using the latter equation to solve for one of the $y_i$, $i=0,1,2,3$, substituting the result into \eqref{CSTRex}, and using the fact that $\Log(z_1z_2)=\Log(z_1)+\Log(z_2)+2\pi\ii n$, where $n$ takes one of the values $0$, $\pm1$.  The equation \eqref{CSTRex} is an example of a classical star-triangle relation which arises from the quasi-classical expansion of the star-triangle relation of the scalar case of Zamolodchikov's fishnet model \cite{Bazhanov:2016ajm}. 

Next, in terms of \eqref{lagex}, define the following function
\begin{equation}\label{rmatex}
R(y_i,y_j;y'_i,y'_j)=\lag_{\alpha_1-\beta_2}(y_i,y_j)-\lag_{\alpha_2-\beta_2}(y_i,y_j')-\lag_{\alpha_1-\beta_1}(y'_i,y_j)+\lag_{\alpha_2-\beta_1}(y'_i,y'_j),
\end{equation}
where $y_i,y_j,y'_i,y'_j\in\mathbb{C}$ and $\alpha_1,\alpha_2,\beta_1,\beta_2\in\mathbb{C}$, are complex valued variables and parameters, respectively.  Denote the four partial derivatives of $R(y_i,y_j;y'_i,y'_j)$ with respect to its variables as
\begin{equation}\label{implicitYBM}
\begin{gathered}
z_i=\frac{\partial R}{\partial y_i}= \frac{\alpha_1-\beta_2}{y_i - y_j} - \frac{\alpha_2-\beta_2}{y_i - y'_j},\qquad 
z_i'=-\frac{\partial R}{\partial y_i'}=\frac{\alpha_2-\beta_1}{y'_j-y'_i} - \frac{\alpha_1-\beta_1}{y_j-y'_i},\\ 
z_j=\frac{\partial R}{\partial y_j}= \frac{\alpha_1-\beta_2}{y_j-y_i}-\frac{\alpha_1-\beta_1}{y_j-y'_i},\qquad 
z_j'=-\frac{\partial R}{\partial y_j'}=\frac{\alpha_2-\beta_1}{y'_i-y'_j}-\frac{\alpha_2-\beta_2}{y_i-y'_j}.
\end{gathered}
\end{equation}
Each of $z_i,z_j,z'_i,z'_j$, can be written as a rational function with numerator and denominator having linear dependence on two different variables.  For example, $z_i$ can be written as 
\begin{equation}
z_i= \frac{(\alpha_1-\alpha_2)y_i + (\alpha_2-\beta_2)y_j - (\alpha_1-\beta_2)y'_j}{(y_i - y_j)(y_i - y'_j)},
\end{equation}
where the numerator and denominator is linear in both $y_j$ and $y'_j$, respectively.  This bilinearity implies that \eqref{implicitYBM} implicitly defines a quadrirational map $\bigl((y_i,y_j),(z_i,z_j)\bigr)\mapsto\bigl((y'_i,y'_j),(z'_i,z'_j)\bigr)$, such that each of the maps
\begin{equation}
\begin{gathered}
\bigl((y_i,y_j),(z_i,z_j)\bigr)\mapsto\bigl((y_i',y_j'),(z_i',z_j')\bigr),\\
\bigl((y'_i,y'_j),(z'_i,z'_j)\bigr)\mapsto\bigl((y_i,y_j),(z_i,z_j)\bigr),\\
\bigl((y'_i,y'_j),(z_i,z_j)\bigr)\mapsto\bigl((y_i,y_j),(z'_i,z'_j)\bigr),\\
\bigl((y_i,y_j),(z'_i,z'_j)\bigr)\mapsto\bigl((y'_i,y'_j),(z_i,z_j)\bigr),
\end{gathered}
\end{equation}
exist and are bilinear and rational.  This can be straightforwardly verified using the expression given in \eqref{implicitYBM}.  Denoting parameters $\al,\bt,\gm\in\mathbb{C}^2$ as $\al=(\alpha_1,\alpha_2)$, $\bt=(\beta_1,\beta_2)$, $\gm=(\gamma_1,\gamma_2)$, and variables as $(y_i,y_j),(z_i,z_j)\in\mathbb{CP}^1\times\mathbb{CP}^1$, it turns out that as a map $R(\al,\bt)\colon((y_i,y_j),(z_i,z_j)\bigr)\mapsto\bigl((y'_i,y'_j),(z'_i,z'_j)\bigr)$, \eqref{implicitYBM} satisfies the {\it functional Yang-Baxter equation}
\begin{equation}
\label{FYBEex}
R_{jk}(\bt,\gm)\circ R_{ik}(\al,\gm)\circ R_{ij}(\al,\bt)=R_{ij}(\al,\bt)\circ R_{ik}(\al,\gm)\circ R_{jk}(\bt,\gm),
\end{equation}
where the maps $R_{ij}$ act non-trivially as the map \eqref{implicitYBM} on the $i$-th and $j$-th components of $X\times X\times X$, where $X\in\mathbb{CP}^1\times\mathbb{CP}^1$.  This fact can also be straightforwardly verified.  In this way, starting from a particular complex-valued function \eqref{lagex} that satisfies the CSTR \eqref{CSTRex}, a two-component quadrirational Yang-Baxter map \eqref{implicitYBM} has been derived which satisfies the Yang-Baxter equation \eqref{FYBEex}.  The Yang-Baxter map \eqref{implicitYBM} can also be readily verified to be reversible, meaning that it satisfies $R_{ij}(\al,\bt)\circ R_{ji}(\bt,\al)=\mbox{Id}$, where $\mbox{Id}$ is the identity map.

It turns out that the above method for deriving Yang-Baxter maps can be used for any functions $\lag_\alpha(y_i,y_j)$ which satisfy a {CSTR} of the form \eqref{CSTRex}, as will be established in Section \ref{sec:YBM}.  Such functions can be derived from the quasi-classical asymptotics of (quantum) star-triangle relations which are related to hypergeometric integrals \cite{Bazhanov:2007mh,Bazhanov:2010kz,Bazhanov:2016ajm,Kels:2018xge}, and they have also been independently derived for variational principles of discrete integrable systems \cite{LobbNijhoff,BS09}, where they play an important role (in this context the CSTR \eqref{CSTRex} is known as a closure relation \cite{LobbNijhoff}).  The establishing of a direct connection between Yang-Baxter maps and the star-triangle relation has been motivated by the work of Bazhanov and Sergeev \cite{Bazhanov:2015gra}, where they utilised a certain pair of functions which solve a form of the CSTR.  The goal of this paper is to more thoroughly develop and provide a better understanding of such connections, placing an emphasis on the star-triangle relation as a starting point for obtaining Yang-Baxter maps, independently of quantum group and Hamiltonian structures.  In particular, these connections will be used in Section \ref{sec:examples} to explicitly derive several new Yang-Baxter maps from different solutions to the CSTRs.

\subsection{Star-triangle relations and quasi-classical limit}\label{sec:CSTR}

Probably the most well-known form of star-triangle relation arises as an equation for integrability of lattice models of statistical mechanics. Such lattice models are typically defined on a square lattice, where spin variables $\spn_i$ at vertices take values in some set $X$.  Two nearest-neighbour spins $\sigma_i$ and $\sigma_j$ interact through a Boltzmann weight function $W_{pq}(\spn_i,\spn_j)$ or $\oW_{pq}(\spn_i,\spn_j)$ depending on whether they are connected by a horizontal or vertical edge, where $p,q$ are known as rapidity variables.

The most common case is $X=\mathbb{Z}\textrm{ mod}N$, where the star-triangle relations take the form \cite{Bax02rip}
\begin{equation}\label{STRusual}
\begin{split}
\sum_{\spn_d=1}^N\oW_{qr}(\sigma_a,\sigma_d)W_{pr}(\sigma_b,\sigma_d)\oW_{pq}(\sigma_d,\sigma_c)=W_{qr}(\sigma_b,\sigma_c)\oW_{pr}(\sigma_a,\sigma_c)W_{pq}(\sigma_b,\sigma_a), \\
\sum_{\spn_d=1}^N\oW_{qr}(\sigma_d,\sigma_a)W_{pr}(\sigma_d,\sigma_b)\oW_{pq}(\sigma_c,\sigma_d)=W_{qr}(\sigma_c,\sigma_b)\oW_{pr}(\sigma_c,\sigma_a)W_{pq}(\sigma_a,\sigma_b).
\end{split}
\end{equation}
Examples of models that satisfy such star-triangle relations include the Ising model ($N=2$) and the chiral Potts model ($N\geq 2$) \cite{AuYang:1987zc,Baxter:1987eq}.  The former model satisfies $W_{pq}(\sigma_a,\sigma_b)=W_{qp}(\sigma_b,\sigma_a)$ and $\oW_{pq}(\sigma_a,\sigma_b)=
\oW_{qp}(\sigma_b,\sigma_a)$, in which case the two equations \eqref{STRusual} are equivalent.

Another example is when $X$ is an interval of the real line, in which case the star-triangle relations take the form
\begin{equation}\label{STRcontexamples}
\begin{split}
\int_X d\sigma_0\oW_{qr}(\sigma_a,\sigma_d)W_{pr}(\sigma_b,\sigma_d)\oW_{pq}(\sigma_d,\sigma_c)=W_{qr}(\sigma_b,\sigma_c)\oW_{pr}(\sigma_a,\sigma_c)W_{pq}(\sigma_b,\sigma_a), \\
\int_X d\sigma_0\oW_{qr}(\sigma_d,\sigma_a)W_{pr}(\sigma_d,\sigma_b)\oW_{pq}(\sigma_c,\sigma_d)=W_{qr}(\sigma_c,\sigma_b)\oW_{pr}(\sigma_c,\sigma_a)W_{pq}(\sigma_a,\sigma_b).
\end{split}
\end{equation}
Examples of models that satisfy such star-triangle relations include the Faddeev-Volkov model ($X=\mathbb{R}$) \cite{Bazhanov:2007mh} and Bazhanov and Sergeev's master solution of the star-triangle relation ($X=[0,2\pi]$) \cite{Bazhanov:2010kz}.

The star-triangle relations of the form \eqref{STRcontexamples} have various limits which may be analytically continued to identities involving complex spin variables $\sigma_i\in\mathbb{C}$, and complex parameters $p,q,r\in\mathbb{C}$.  Such identities have a general form
\begin{equation}
\label{STRdef}
\begin{split}
\int_C d\sigma_0 \oV_{q-r}(\sigma_a,\sigma_d)V_{p-r}(\sigma_b,\sigma_d)\oWh_{p-q}(\sigma_d,\sigma_c)=V_{q-r}(\sigma_b,\sigma_c)\oV_{p-r}(\sigma_a,\sigma_c)W_{p-q}(\sigma_b,\sigma_a), \\
\int_C d\sigma_0 \oV_{q-r}(\sigma_d,\sigma_a)V_{p-r}(\sigma_d,\sigma_b)\oW_{p-q}(\sigma_c,\sigma_d)=V_{q-r}(\sigma_c,\sigma_b)\oV_{p-r}(\sigma_c,\sigma_a)\Wh_{p-q}(\sigma_a,\sigma_b),
\end{split}
\end{equation}
where $C$ is a contour in the complex plane, and
\begin{equation}\label{BWs}
W_{\theta}(\sigma_1,\sigma_2),\quad\oW_{\theta}(\sigma_1,\sigma_2),\quad\Wh_{\theta}(\sigma_1,\sigma_2),\quad\oWh_{\theta}(\sigma_1,\sigma_2),\quad V_{\theta}(\sigma_1,\sigma_2),\quad\oV_{\theta}(\sigma_1,\sigma_2),
\end{equation}
are complex-valued functions of the spin variables $\sigma_1,\sigma_2$, and (the difference of) parameters $\theta$.  The identities \eqref{STRdef} depend on three independent variables $\spn_1,\spn_2,\spn_3$, and differences of the three parameters $p,q,r$. Several solutions of this form of the star-triangle relations have been given by the author \cite{Kels:2018xge}, which arise as limits of the master solution of the star-triangle relations \cite{Bazhanov:2010kz} of the form \eqref{STRcontexamples}, where the Boltzmann weights \eqref{BWs} are usually parameterised in terms of the gamma function and its generalisations \cite{Ruijsenaars:1997:FOA}.  Depending on the definitions of \eqref{BWs}, the contour $C$ is typically chosen to separate infinite sequences of poles of the integrand that diverge to infinity.  The solutions of the star-triangle relations of the form \eqref{STRdef} given in \cite{Kels:2018xge} are equivalent to different mathematical identities of hypergeometric integrals \cite{GR,BultThesis,Spiridonov-essays}.

\subsubsection{Examples}

The hyperbolic gamma function is defined by \cite{Ruijsenaars:1997:FOA} (it is equivalent up to a simple change of variables to the multiple sine function \cite{Barnes:1901,Barnes1904,Shintani1976,kurokawa1991} and non-compact quantum dilogarithm \cite{Faddeev:1994fw})
\begin{equation}\label{hgfdef}
    \Gamma_h(z;\bb)=\exp\left(\int_0^\infty\frac{dx}{x}\Bigl(\frac{\ii z}{x}-\frac{\sinh(2\ii zx)}{2\sinh(x\bb)\sinh(x\bb^{-1})}\Bigr)\right),\quad |\im(z)|<\re(\eta),
\end{equation}
where the crossing parameter is
\begin{equation}\label{hypcrossingparameter}
\eta=\frac{\bb+\bb^{-1}}{2}.
\end{equation}
For the purposes here, the modular parameter $\bb$ will take positive real values
\begin{equation}
\bb>0,
\end{equation}
such that the crossing parameter \eqref{hypcrossingparameter} is also real-valued and positive.  The hyperbolic gamma function \eqref{hgfdef} satisfies the inversion relation
\begin{equation}
\Gamma_h(z;\bb)=\frac{1}{\Gamma(-z;\bb)},
\end{equation}
and difference equations
\begin{equation}
\begin{split}
\frac{\Gamma_h(z-\ii\bb;\bb)}{\Gamma_h(z;\bb)}
=2\cosh\bigl(\tfrac{\pi}{2\bb}(2z-\ii\bb^{-1})\bigr),
\qquad
\frac{\Gamma_h(z-\ii\bb^{-1};\bb)}{\Gamma_h(z;\bb)}
=2\cosh\bigl(\tfrac{\pi\bb}{2}(2z-\ii\bb)\bigr).
\end{split}
\end{equation}
Through the use of the latter equations, the hyperbolic gamma function \eqref{hgfdef} may be analytically continued to $\mathbb{C} - \ii\bb\mathbb{Z}_{\geq0} - \ii\bb^{-1}\mathbb{Z}_{\geq0} - \{\eta\}$ \cite{Ruijsenaars:1997:FOA}.

Define the following functions
\begin{equation}
\label{hypersols}
\begin{alignedat}{2}
s(\spn)&=\sqrt{2\sinh(2\pi x\bb)\sinh(2\pi x\bb^{-1})}, \qquad  & f_\theta(\sigma_i,\sigma_j)&=\EXP^{\frac{\pi\ii}{2}\bigl((\ii(\spn_i+\spn_j)+\theta)^2-2\theta^2\bigr)}, \\
W^{(1)}_\theta(\spn_i,\spn_j)&=
\frac{\Gamma_h(\spn_i\pm\spn_j+\ii\theta;\bb)}{\Gamma_h(\spn_i\pm\spn_j-\ii\theta;\bb)},\qquad 
& W^{(2)}_\theta(\spn_i,\spn_j)&=
\frac{\Gamma_h(\spn_i-\spn_j+\ii\theta;\bb)}{\Gamma_h(\spn_i-\spn_j-\ii\theta;\bb)},\\
\oW^{(1)}_\theta(\spn_i,\spn_j)&=
\frac{s(\spn_i)s(\spn_j)W^{(1)}_{\eta-\theta}(\spn_i,\spn_j)}{\Gamma_h\bigl(\ii(2\theta-\eta);\bb\bigr)}, \qquad 
& \oW^{(2)}_\theta(\spn_i,\spn_j)&=\frac{W^{(2)}_{\eta-\theta}(\spn_i,\spn_j)}{\Gamma_h\bigl(\ii(2\theta-\eta);\bb\bigr)}, \\
V^{(1)}_\theta(\spn_i,\spn_j)&=\Gamma_h(\spn_i\pm\spn_j+\ii\theta;\bb), \qquad 
& \oV^{(1)}_\theta(\spn_i,\spn_j)&=s(\spn_j)V^{(1)}_{\eta-\theta}(-\spn_i,\spn_j),\\
V^{(2)}_\theta(\spn_i,\spn_j)&=f_\theta(\sigma_i,\sigma_j)\Gamma_h(\spn_i+\spn_j+\ii\theta;\bb),\qquad 
& V^{(3)}_\theta(\spn_i,\spn_j)&=\EXP^{2\pi\ii\spn_i\spn_j},
\end{alignedat}
\end{equation}
where $\Gamma_h(x\pm y;\bb)$ indicates the product $\Gamma_h(x+y;\bb)\Gamma_h(x-y;\bb)$.  Each of these functions are symmetric under the exchange $\spn_i\leftrightarrow\spn_j$ (excluding $s(\spn)$), except for $V^{(1)}_\theta$ and $\oV^{(1)}_\theta$.

\begin{prop}
\label{thm:hypersols}
Let the functions \eqref{BWs} be given according to one of the five rows of Table \ref{tab:hyperBWcombos}. Then \eqref{BWs} is a solution of the star-triangle relations \eqref{STRdef} for $\spn_1,\spn_2,\spn_3\in\mathbb{R}$, $0\leq \theta<\eta$, and $C=\mathbb{R}$.

\begin{table}[htb!]
\centering
\begin{tabular}{c | c | c | c | c | c | c}
 Type & $W_\theta(\spn_i,\spn_j)$ & $\Wh_\theta(\spn_i,\spn_j)$  & $V_\theta(\spn_i,\spn_j)$ & $\oW_\theta(\spn_i,\spn_j)$ & $\oWh_\theta(\spn_i,\spn_j)$ & $\oV_\theta(\spn_i,\spn_j)$ 
 \\ \hline \\[-0.4cm]
I & \multicolumn{3}{c|}{$W^{(1)}_\theta(\spn_i,\spn_j)$} & \multicolumn{3}{c}{$\oW^{(1)}_{\theta}(\spn_i,\spn_j)$} 
\\ \hline \\[-0.4cm]
I & \multicolumn{3}{c|}{$W^{(2)}_\theta(\spn_i,\spn_j)$} & \multicolumn{3}{c}{$\oW^{(2)}_{\theta}(\spn_i,\spn_j)$}   
\\ \hline \\[-0.4cm]
III & $W^{(2)}_\theta(\spn_i,\spn_j)$ & $W^{(1)}_\theta(\spn_i,\spn_j)$ & $V^{(1)}_\theta(\spn_i,\spn_j)$ & $\oW^{(2)}_{\theta}(\spn_i,\spn_j)$ & $\oW^{(1)}_{\theta}(\spn_i,\spn_j)$ & $\oV^{(1)}_{\theta}(\spn_i,\spn_j)$
\\ \hline \\[-0.4cm]
II & \multicolumn{2}{c|}{$W^{(2)}_\theta(\spn_i,\spn_j)$} & $V^{(2)}_\theta(\spn_i,\spn_j)$ & \multicolumn{2}{c|}{$\oW^{(2)}_{\theta}(\spn_i,\spn_j)$} & $\bigl(V^{(2)}_{\eta+\theta}(\spn_i,\spn_j)\bigr)^{-1}$ 
\\ \hline \\[-0.4cm]
II & \multicolumn{2}{c|}{$W^{(2)}_\theta(\spn_i,\spn_j)$} & $V^{(3)}_\theta(\spn_i,\spn_j)$ & \multicolumn{2}{c|}{$\oW^{(2)}_{\theta}(\spn_i,\spn_j)$} & $V^{(3)}_{\theta}(-\spn_i,\spn_j)$ \\
\hline 
\end{tabular}
\caption{Assignment of the functions \eqref{hypersols} to \eqref{BWs} for five different solutions of the star-triangle relations \eqref{STRdef}.}
\label{tab:hyperBWcombos}
\end{table}

\end{prop}

\begin{remark}\label{rem:BWtypes}
In Table \ref{tab:hyperBWcombos}, the types are distinguished as follows.  Type I solutions are given in terms of up to two different functions $W$ and $\oW$, where $V=\Wh=W$ and $\oV=\Wh=\oW$.  Type II solutions are given in terms of up to four different functions $W,\oW,V,\oV$, where $W$ and $\oW$ are a type I solution, $\Wh=W$, and $\oWh=\oW$.  Type III solutions are given in terms of up to six different functions $V,\oV,\Wh,\oWh,W,\oW$, where $W$ and $\oW$, and $\Wh$ and $\oWh$, are two different type I solutions.  
\end{remark}

The second type I solution of Proposition \ref{thm:hypersols} is the star-triangle relation of the Faddeev-Volkov model introduced by Bazhanov, Mangazeev, and Sergeev \cite{Bazhanov:2007mh}.  The different type II and type III solutions of the star-triangle relations were obtained \cite{Kels:2018xge} as limits of the first type I solution, which is a star-triangle introduced by Spiridonov \cite{Spiridonov:2010em}.  The second type II solution is also equivalent to a self-duality relation introduced for the Faddeev-Volkov model \cite{Bazhanov:2007mh}.

\subsubsection{Quasi-classical limit and classical star-triangle relations}

The quasi-classical limit of the star-triangle relations \eqref{STRdef} involves taking some parameter $\hbar\to0$, such that they get the asymptotics of the form
\begin{equation}\label{STRqcl}
\begin{gathered}
\int_Cdx_d\, \EXP^{\hbar^{-1}\bigl(\olam_{v-w}(x_a,x_d)+\lam_{u-w}(x_b,x_d)+\olh_{u-v}(x_d,x_c)\bigr)+O(1)}=\EXP^{\hbar^{-1}\bigl(\lam_{v-w}(x_b,x_c)+\olam_{u-w}(x_a,x_c)+\lag_{u-v}(x_b,x_a)\bigr)}, \\
\int_Cdx_d\, \EXP^{\hbar^{-1}\bigl(\olam_{v-w}(x_d,x_a)+\lam_{u-w}(x_d,x_b)+\ol_{u-v}(x_c,x_d)\bigr)+O(1)}=\EXP^{\hbar^{-1}\bigl(\lam_{v-w}(x_c,x_b)+\olam_{u-w}(x_c,x_a)+\lagh_{u-v}(x_a,x_b)\bigr)},
\end{gathered}
\end{equation}
where each of 
\begin{equation}\label{CSTRsol}
\lag_{\alpha}(x_i,x_j),\quad\lagh_{\alpha}(x_i,x_j),\quad\ol_{\alpha}(x_i,x_j),\quad\olh_{\alpha}(x_i,x_j),\quad\lam_{\alpha}(x_i,x_j),\quad\olam_{\alpha}(x_i,x_j),
\end{equation}
are complex-valued functions of $x_i,x_j,\alpha$, which arise as the leading $O(\hbar^{-1})$ asymptotics of the respective Boltzmann weights in \eqref{BWs}.  The variables and parameters of \eqref{STRdef}  and \eqref{STRqcl} are related by
\begin{equation}\label{YBEcov}
(\spn_a,\spn_b,\spn_c,\spn_d)=\bigl(F_a(x_a),F_b(x_b),F_c(x_c),F_d(x_d)\bigr),\qquad (p,q,r)=\bigl(G(u),G(v),G(w)\bigr),
\end{equation}
where $F_I(z)$, $I\in\{a,b,c,d\}$, and $G(z)$ are some M\"obius transformations that depend on $\hbar$.  For example, for the star-triangle relations of Proposition \ref{thm:hypersols}, a quasi-classical limit can be taken by setting \cite{Bazhanov:2007mh,Bazhanov:2016ajm,Kels:2018xge}
\begin{equation}\label{hypqcl}
\spn_I=\frac{x_I}{\sqrt{2\pi\hbar}},\; I\in\{a,b,c,d\},\qquad
(p,q,r)=\Bigl(\frac{u}{\sqrt{2\pi\hbar}},\frac{v}{\sqrt{2\pi\hbar}},\frac{w}{\sqrt{2\pi\hbar}}\Bigr),
\end{equation}
where
\begin{equation}
\hbar=2\pi\bb^2,
\end{equation}
and taking $\hbar\to0$.

In general, identifying both sides of the star-triangle relations \eqref{STRqcl} at $O(\hbar^{-1})$ gives a classical star-triangle relation.  Specifically, defining
\begin{align}
\begin{split}
\mathcal{A}^{(STR)}_1(x_a,x_b,x_c,x_d;u,v,w)
=\olam_{v-w}(x_a,x_d)+\lam_{u-w}(x_b,x_d)+\olh_{u-v}(x_d,x_c)\phantom{,} \\
-\lam_{v-w}(x_b,x_c)-\olam_{u-w}(x_a,x_c)-\lag_{u-v}(x_b,x_a),
\end{split}
\\
\begin{split}
\mathcal{A}^{(STR)}_2(x_a,x_b,x_c,x_d;u,v,w)
=\olam_{v-w}(x_d,x_a)+\lam_{u-w}(x_d,x_b)+\ol_{u-v}(x_c,x_d)\phantom{,} \\
-\lam_{v-w}(x_c,x_b)-\olam_{u-w}(x_c,x_a)-\lagh_{u-v}(x_a,x_b),
\end{split}
\end{align}
the CSTRs are given by the equations
\begin{align}\label{CSTRdef1}
\mathcal{A}^{(STR)}_1(x_a,x_b,x_c,x_d;u,v,w)=0,
\\ \label{CSTRdef2}
\mathcal{A}^{(STR)}_2(x_a,x_b,x_c,x_d;u,v,w)=0,
\end{align}
where for \eqref{CSTRdef1} the variables $x_a,x_b,x_c,x_d,$ and parameters $u,v,w,$ are constrained to satisfy the equation for the saddle-point of the first of \eqref{STRqcl}
\begin{equation}\label{3legdef1}
\frac{\partial}{\partial x}
\Bigl(\mathcal{A}^{(STR)}_1(x_a,x_b,x_c,x;u,v,w)\Bigr)_{x=x_d}=0,
\end{equation}
and for \eqref{CSTRdef2} the variables $x_a,x_b,x_c,x_d,$ and parameters $u,v,w,$ are constrained to satisfy the equation for the saddle-point of the second of \eqref{STRqcl}
\begin{equation}\label{3legdef2}
\frac{\partial}{\partial x}
\Bigl(\mathcal{A}^{(STR)}_2(x_a,x_b,x_c,x;u,v,w)\Bigr)_{x=x_d}=0.
\end{equation}
Both of \eqref{CSTRdef1} and \eqref{CSTRdef2} depend on the four variables $x_a,x_b,x_c,x_d$ and three parameters $u,v,w$, subject to a single constraint \eqref{3legdef1} and \eqref{3legdef2}, respectively.  Although the same variable notations are used for both \eqref{CSTRdef1} and \eqref{CSTRdef2}, it should be clear that the first CSTR \eqref{CSTRdef1} and its constraint \eqref{3legdef1}, is a relation independent of the second CSTR \eqref{CSTRdef2} and its constraint \eqref{3legdef2}.  The solutions of the CSTRs obtained from the quasi-classical limit \eqref{hypqcl} of Proposition \ref{thm:hypersols} are given in Proposition \ref{prop:hyperlags}.

It is typically useful to work with a graphical representation of the star-triangle relations.  The different functions \eqref{CSTRsol} will be associated to the edges pictured in Figure \ref{fig:edges}, and in terms of these edges the CSTRs are pictured in Figure \ref{fig-CSTR}.

\begin{figure}[htb!]
\centering
\begin{tikzpicture}[scale=1.6]

\begin{scope}[yshift=57pt]
\draw[-] (0,-0.5)--(0,0.5);
\filldraw[fill=black,draw=black] (0,-0.5) circle (1.4pt)
node[below=4pt]{\color{black}\small $i$};
\filldraw[fill=black,draw=black] (0,0.5) circle (1.4pt)
node[above=4pt]{\color{black}\small $j$};

\fill (0,-0.9) circle(0.01pt)
node[below=0.05pt]{\color{black} $ \lag_{u-v}(\xv_i,\xv_j)$}
node[below=15.05pt]{\color{black} $ \lagh_{u-v}(\xv_i,\xv_j)$};
\end{scope}



\begin{scope}[xshift=100pt,yshift=57pt]
\draw[-,double] (0,-0.5)--(0,0.5);
\draw[->>>-,white] (0,-0.5)--(0,0.5);
\filldraw[fill=black,draw=black] (0,-0.5) circle (1.4pt)
node[below=4pt]{\color{black}\small $i$};
\filldraw[fill=black,draw=black] (0,0.5) circle (1.4pt)
node[above=4pt]{\color{black}\small $j$};

\fill (0,-0.9) circle(0.01pt)
node[below=7.55pt]{\color{black} $ \lam_{u-v}(\xv_i,\xv_j)$};
\end{scope}

\begin{scope}[xshift=50pt,yshift=57pt]
\draw[-|-] (0,-0.5)--(0,0.5);
\filldraw[fill=black,draw=black] (0,-0.5) circle (1.4pt)
node[below=4pt]{\color{black}\small $i$};
\filldraw[fill=black,draw=black] (0,0.5) circle (1.4pt)
node[above=4pt]{\color{black}\small $j$};

\fill (0,-0.9) circle(0.01pt)
node[below=0.05pt]{\color{black} $\ol_{u-v}(\xv_i,\xv_j)$}
node[below=15.05pt]{\color{black} $\olh_{u-v}(\xv_i,\xv_j)$};
\end{scope}



\begin{scope}[xshift=150pt,yshift=57pt]
\draw[-,double] (0,-0.5)--(0,0.5);
\draw[->>>>-,white] (0,-0.5)--(0,0.5);
\filldraw[fill=black,draw=black] (0,-0.5) circle (1.4pt)
node[below=4pt]{\color{black}\small $i$};
\filldraw[fill=black,draw=black] (0,0.5) circle (1.4pt)
node[above=4pt]{\color{black}\small $j$};

\fill (0,-0.9) circle(0.01pt)
node[below=7.55pt]{\color{black} $\olam_{u-v}(\xv_i,\xv_j)$};
\end{scope}

\end{tikzpicture}
\caption{Functions \eqref{CSTRsol} associated to four types of edges.  $\lag$ and $\lagh$ are associated to the same type of edge, and similarly for $\ol$ and $\olh$.}
\label{fig:edges}
\end{figure}

\begin{figure}[htb!]
\centering

\begin{tikzpicture}[scale=1.7]

\draw[white] (-1.9,0.5) circle (0.1pt)
node[right=2pt]{\color{black}$\olam_{v-w}$};
\draw[white] (-2.6,-0.2) circle (0.1pt)
node[above=2pt]{\color{black}$\ol_{u-v}$};
\draw[white] (-1.25,-0.3) circle (0.1pt)
node[above=2pt]{\color{black}$\lam_{u-w}$};

\draw[-|-] (-2,0)--(-2.87,-0.5);
\draw[-,double] (-2,0)--(-2,1);
\draw[-,double] (-2,0)--(-1.13,-0.5);
\draw[->>>>-,white] (-2,0)--(-2,1);
\draw[->>>-,white] (-2,0)--(-1.13,-0.5);

\fill (-2,0) circle (1.5pt)
node[below=3.5pt]{\color{black} $x_d$};
\filldraw[fill=black,draw=black] (-2,1) circle (1.5pt)
node[above=3.5pt] {\color{black} $x_a$};
\filldraw[fill=black,draw=black] (-2.87,-0.5) circle (1.5pt)
node[left=3.5pt] {\color{black} $x_c$};
\filldraw[fill=black,draw=black] (-1.13,-0.5) circle (1.5pt)
node[right=3.5pt] {\color{black} $x_b$};

\fill[white!] (-0.5,0.3) circle (0.01pt)
node[left=0.05pt] {\color{black}\Large $=$};

\begin{scope}[xshift=-30pt]

\draw[white] (2.0,-0.55) circle (0.1pt)
node[below=2pt]{\color{black}$\lam_{v-w}$};
\draw[white] (2.7,0.15) circle (0.1pt)
node[above=2pt]{\color{black}$\lagh_{u-v}$};
\draw[white] (1.15,0.15) circle (0.1pt)
node[above=2pt]{\color{black}$\olam_{u-w}$};

\draw[-,double] (1.13,-0.5)--(2,1);
\draw[-,double] (2.87,-0.5)--(1.13,-0.5);
\draw[->>>>-,white] (1.13,-0.5)--(2,1);
\draw[->>>-,white] (1.13,-0.5)--(2.87,-0.5);
\draw[-] (2.87,-0.5)--(2,1);

\filldraw[fill=black,draw=black] (2,1) circle (1.5pt)
node[above=3.5pt]{\color{black} $x_a$};
\filldraw[fill=black,draw=black] (1.13,-0.5) circle (1.5pt)
node[left=3.5pt]{\color{black} $x_c$};
\filldraw[fill=black,draw=black] (2.87,-0.5) circle (1.5pt)
node[right=3.5pt]{\color{black} $x_b$};

\end{scope}


\begin{scope}[yshift=70pt]

\draw[white] (-1.9,0.5) circle (0.1pt)
node[right=2pt]{\color{black}$\olam_{v-w}$};
\draw[white] (-2.6,-0.2) circle (0.1pt)
node[above=2pt]{\color{black}$\olh_{u-v}$};
\draw[white] (-1.25,-0.3) circle (0.1pt)
node[above=2pt]{\color{black}$\lam_{u-w}$};

\draw[-|-] (-2,0)--(-2.87,-0.5);
\draw[-,double] (-2,1)--(-2,0);
\draw[-,double] (-1.13,-0.5)--(-2,0);
\draw[->>>>-,white] (-2,1)--(-2,0);
\draw[->>>-,white] (-1.13,-0.5)--(-2,0);

\filldraw[fill=black,draw=black] (-2,0) circle (1.5pt)
node[below=3.5pt]{\color{black} $x_d$};
\filldraw[fill=black,draw=black] (-2,1) circle (1.5pt)
node[above=3.5pt] {\color{black} $x_a$};
\filldraw[fill=black,draw=black] (-2.87,-0.5) circle (1.5pt)
node[left=3.5pt] {\color{black} $x_c$};
\filldraw[fill=black,draw=black] (-1.13,-0.5) circle (1.5pt)
node[right=3.5pt] {\color{black} $x_b$};

\fill[white!] (-0.5,0.3) circle (0.01pt)
node[left=0.05pt] {\color{black}\Large $=$};

\begin{scope}[xshift=-30pt]

\draw[white] (2.0,-0.55) circle (0.1pt)
node[below=2pt]{\color{black}$\lam_{v-w}$};
\draw[white] (2.7,0.15) circle (0.1pt)
node[above=2pt]{\color{black}$\lag_{u-v}$};
\draw[white] (1.15,0.15) circle (0.1pt)
node[above=2pt]{\color{black}$\olam_{u-w}$};

\draw[-,double] (2,1)--(1.13,-0.5);
\draw[-,double] (2.87,-0.5)--(1.13,-0.5);
\draw[->>>>-,white] (2,1)--(1.13,-0.5);
\draw[->>>-,white] (2.87,-0.5)--(1.13,-0.5);
\draw[-] (2.87,-0.5)--(2,1);

\filldraw[fill=black,draw=black] (2,1) circle (1.5pt)
node[above=3.5pt]{\color{black} $x_a$};
\filldraw[fill=black,draw=black] (1.13,-0.5) circle (1.5pt)
node[left=3.5pt]{\color{black} $x_c$};
\filldraw[fill=black,draw=black] (2.87,-0.5) circle (1.5pt)
node[right=3.5pt]{\color{black} $x_b$};

\end{scope}

\end{scope}

\end{tikzpicture}

\caption{The CSTRs \eqref{CSTRdef1} (top) and \eqref{CSTRdef2} (bottom) in terms of the edges of Figure \ref{fig:edges}.}
\label{fig-CSTR}
\end{figure}

Generally, it is difficult to find non-trivial solutions to \eqref{CSTRdef1}--\eqref{3legdef2} that are valid for all complex $u,v,w,x_a,x_b,x_c,x_d\in\mathbb{C}$.  The main reason is that the interesting solutions typically depend on the complex logarithm (at least the solutions that are known to the author), and a solution that is valid in some region of the complex parameter space will not continue to hold when crossing a branch cut of the logarithm.

To account for this, it will be enough in this paper that the CSTRs \eqref{CSTRdef1} and \eqref{CSTRdef2} respectively hold up to terms of the form
\begin{equation}\label{addtermsCSTR}
    \sum_{I\in\{a,b,c,d\}}2\pi\ii k_Ix_I
    +C(u,v,w),
\end{equation}
for some integers $k_I\in\mathbb{Z}$ 
($I\in\{a,b,c,d\}$) dependent on $u,v,w,x_I$, and a constant $C(u,v,w)$ with respect to the variables $x_I$.  
In this case several interesting non-trivial solutions of the CSTRs can be found for all complex $u,v,w,x_I$, away from possible branch cuts of the functions \eqref{CSTRsol}.  For the purposes of this paper, the terms of the form \eqref{addtermsCSTR} can eventually be ignored, since the equations of interest for the Yang-Baxter maps given in Section \ref{sec:YBM} will involve exponentials of partial derivatives with respect to the variables $x_a,x_b,x_c,x_d$ of the CSTRs, to which the additional terms of the form \eqref{addtermsCSTR} make no contribution.

It will be assumed that the following symmetries are satisfied
\begin{equation}\label{addterms2}
 \lag_\alpha(x_i,x_j)=\lag_\alpha(x_j,x_i)  + C(\alpha),\qquad
 \ol_\alpha(x_i,x_j)=\ol_\alpha(x_j,x_i)  + C(\alpha), 
\end{equation}
and similarly for $\lagh_\alpha$ and $\olh_\alpha$, where $C(\alpha)$ is a constant with respect to $x_i$ and $x_j$.  Because of \eqref{addterms2}, the ordering of variables for the functions $\lag_\alpha,\ol_\alpha,\lagh_\alpha,\olh_\alpha$ can be ignored in the CSTRs \eqref{CSTRdef1}--\eqref{3legdef2}, since a change in ordering will only result in an additional term of the form \eqref{addtermsCSTR}.  On the other hand, the functions $\lam$ and $\olam$ are not assumed to satisfy symmetries of the form \eqref{addterms2}, and thus the ordering of variables for these functions should be kept as written in \eqref{CSTRdef1}--\eqref{3legdef2}.  This is the reason that these functions are associated to directed edges in Figure \ref{fig-CSTR}.

There is also another symmetry that is assumed to be satisfied
\begin{equation}\label{lagsyms}
\begin{split}
\lag_{-\alpha}(x_i,x_j)=&-\lag_{\alpha}(x_i,x_j)+2\pi\ii(k_ix_i+k_jx_j)+C(\alpha), \\ \ol_{-\alpha}(x_i,x_j)=&-\ol_{\alpha}(x_i,x_j)+2\pi\ii(k_ix_i+k_jx_j)+C(\alpha), 
\end{split}
\end{equation}
and similarly for $\lagh_\alpha$ and $\olh_\alpha$, where $k_i,k_j$ are some integers and $C(\alpha)$ is a constant with respect to $x_i$ and $x_j$.  The symmetries \eqref{lagsyms} won't be needed for this section, but will be useful for establishing certain properties of the Yang-Baxter maps that will be derived in Section \ref{sec:YBM}.  The analogue of the property \eqref{lagsyms} for Boltzmann weights is known as an inversion relation $W_{\theta}(\sigma_i,\sigma_j)W_{-\theta}(\sigma_i,\sigma_j)=1$ ({\it e.g.}, this inversion relation is satisfied by both $W^{(1)}$ and $W^{(2)}$ in \eqref{hypersols}).

Similarly to the examples of Proposition \ref{thm:hypersols}, the solutions \eqref{CSTRsol} of the CSTRs may be grouped into three different types that are distinguished as follows.  Type I solutions are given in terms of up to two different functions $\lag$ and $\ol$, where $\lam=\lagh=\lag$ and $\olam=\olh=\ol$.  Type II solutions are given in terms of up to four different functions $\lag$, $\ol$, $\lam$, $\olam$, where $\lag$ and $\ol$ are a type I solution, $\lagh=\lag$, and $\olh=\ol$.  Type III solutions are given in terms of up to six different functions $\lag$, $\ol$, $\lagh$, $\olh$, $\lam$, $\olam$, where $\lag$ and $\ol$, and $\lagh$ and $\olh$, are two different type I solutions.

Table \ref{tab:cases} summarises the functions used for the three different types of solutions to the CSTRs.

\begin{table}[htb!]
\centering
\begin{tabular}{c | c | c | c | c | c | c}
Type & $\lag$ & $\lagh$ & $\lam$  & $\ol$ & $\olh$ & $\olam$ \\[0.0cm]
\hline \\[-0.4cm]
I &  \multicolumn{3}{c|}{($\lam=\lagh=\lag$)} & \multicolumn{3}{|c}{($\olam=\olh=\ol$)} \\[0.1cm]
\hline \\[-0.4cm]
II & \multicolumn{2}{c|}{($\lagh=\lag$)} & $\lam$ & \multicolumn{2}{|c|}{($\olh=\ol$)}  &  $\olam$ \\[0.1cm]
\hline \\[-0.4cm]
III & $\lag$ & $\lagh$ & $\lam$  & $\ol$ & $\olh$ & $\olam$ \\[0.0cm]
\hline 
\end{tabular}
 \caption{Functions used for the three types of solutions \eqref{CSTRsol} of the CSTRs \eqref{CSTRdef1} and \eqref{CSTRdef2}}
\label{tab:cases}
\end{table}

\subsection{Classical Yang-Baxter equations}\label{sec:CYBE}

The classical Yang-Baxter equations considered here may be regarded as counterparts of quantum Yang-Baxter equations related to multicomponent hypergeometric integrals \cite{Bazhanov:2011mz,Bazhanov:2013bh}.  They involve two-component complex parameters
\begin{equation}\label{YBEpardef}
\bu=(\rua,\rub),\qquad \bv=(\rva,\rvb),\qquad \bw=(\rwa,\rwb).
\end{equation}
Introduce two complex-valued functions
\begin{equation}\label{CYBEsol}
R^{(1)}_{\bu\bv}(x_i,x_j;x'_i,x'_j),\quad R^{(2)}_{\bu\bv}(x_i,x_j;x'_i,x'_j), 
\end{equation}
where these functions depend on complex variables and parameters indicated in the arguments and subscripts, respectively.  Then defining 
\begin{equation}
\begin{split}
\mathcal{A}^{YB}
=&R^{(1)}_{\bu\bv}(x_i,x_j;\xli,\xlj)+R^{(2)}_{\bu\bw}(\xli,x_k;x''_i,\xlk)+R^{(2)}_{\bv\bw}(\xlj,\xlk;x''_j,x''_k)\phantom{,} \\
-&R^{(2)}_{\bv\bw}(x_j,x_k;\xrj,\xrk)-R^{(2)}_{\bu\bw}(x_i,\xrk;\xri,x''_k)-R^{(1)}_{\bu\bv}(\xri,\xrj;x''_i,x''_j),
\end{split}
\end{equation}
the CYBE is given by the expression
\begin{equation}
\label{CYBEdef}
\mathcal{A}^{YB}
=0,
\end{equation}
where the variables and parameters are constrained to satisfy the six equations
\begin{equation}\label{cuboc12}
\frac{\partial}{\partial X}\mathcal{A}^{YB}
=0,\qquad X\in\{\xli,\xlj,\xlk,\xri,\xrj,\xrk\}.
\end{equation}
The variables of the CYBE \eqref{CYBEdef} are shown in the diagram of Figure \ref{fig-CYBEdef}.

\begin{figure}[tbh!]
\centering
\begin{tikzpicture}[scale=1.4]

\begin{scope}[xshift=-140pt]



\draw (0,1.73)--(-1,0);
\draw (-1.5,0.88)--(0.5,0.88);
\draw (0.5,0.88)--(1.5,-0.88);
\draw (1.5,0.88)--(0.5,-0.88);
\draw (-1,0)--(0,-1.73);
\draw (0.5,-0.88)--(-1.5,-0.88);

\filldraw[fill=black,draw=black] (0,1.73) circle (0.016pt)
node[right=2pt]{\small $x''_j$};
\filldraw[fill=black,draw=black] (-1.5,0.88) circle (0.016pt)
node[left=2pt]{\small $x''_k$};
\filldraw[fill=black,draw=black] (-1.5,-0.88) circle (0.016pt)
node[below=2pt]{\small $x_i$};
\filldraw[fill=black,draw=black] (0,-1.73) circle (0.016pt)
node[right=2pt]{\small $x_j$};
\filldraw[fill=black,draw=black] (1.5,-0.88) circle (0.016pt)
node[right=2pt]{\small $x_k$};
\filldraw[fill=black,draw=black] (1.5,0.88) circle (0.016pt)
node[right=2pt]{\small $x''_i$};
\filldraw[fill=black,draw=black] (0.5,0.88) circle (0.016pt);
\filldraw[fill=black,draw=black] (0.6,0.88) circle (0.01pt)
node[above=3pt]{\small $\xlk$};
\filldraw[fill=black,draw=black] (-1,0) circle (0.016pt)
node[left=3pt]{\small $\xlj$};
\filldraw[fill=black,draw=black] (0.5,-0.88) circle (0.016pt);
\filldraw[fill=black,draw=black] (0.6,-0.88) circle (0.01pt)
node[below=2pt]{\small $\xli$};


\end{scope}



\draw (0,1.73)--(1,0);
\draw (1.5,0.88)--(-0.5,0.88);
\draw (-0.5,0.88)--(-1.5,-0.88);
\draw (-1.5,0.88)--(-0.5,-0.88);
\draw (1,0)--(0,-1.73);
\draw (-0.5,-0.88)--(1.5,-0.88);

\filldraw[fill=black,draw=black] (0,1.73) circle (0.016pt)
node[left=2pt]{\small $x''_j$};
\filldraw[fill=black,draw=black] (1.5,0.88) circle (0.016pt)
node[right=2pt]{\small $x''_i$};
\filldraw[fill=black,draw=black] (1.5,-0.88) circle (0.016pt)
node[right=2pt]{\small $x_k$};
\filldraw[fill=black,draw=black] (0,-1.73) circle (0.016pt)
node[left=2pt]{\small $x_j$};
\filldraw[fill=black,draw=black] (-1.5,-0.88) circle (0.016pt)
node[left=2pt]{\small $x_i$};
\filldraw[fill=black,draw=black] (-1.5,0.88) circle (0.016pt)
node[above=2pt]{\small $x''_k$};
\filldraw[fill=black,draw=black] (-0.5,0.88) circle (0.016pt);
\filldraw[fill=black,draw=black] (-0.6,0.88) circle (0.01pt)
node[above=2pt]{\small $\xri$};
\filldraw[fill=black,draw=black] (1,0) circle (0.016pt)
node[right=2pt]{\small $\xrj$};
\filldraw[fill=black,draw=black] (-0.5,-0.88) circle (0.016pt);
\filldraw[fill=black,draw=black] (-0.6,-0.88) circle (0.01pt)
node[below=2pt]{\small $\xrk$};


\draw[black!] (-2.3,0) circle (0.01pt)
node[left=1pt]{\color{black}\small $=$};

\end{tikzpicture}
\caption{The CYBE \eqref{CYBEdef}.}
\label{fig-CYBEdef}
\end{figure}

The CYBE \eqref{CYBEdef} is an equation involving twelve variables $x_I,x''_I,x^{(l)}_I,x^{(r)}_I$, $I\in\{i,j,k\}$, and three two-component parameters \eqref{YBEpardef}, subject to the six constraints \eqref{cuboc12}. In analogy with R-matrices that are solutions to the quantum Yang-Baxter equation with continuous spin variables \cite{Bazhanov:2011mz}, the solutions of the CYBE \eqref{CYBEdef}--\eqref{cuboc12} may naturally be referred to as classical R-matrices.  Similarly to the case for the CSTRs introduced in the previous subsection, it will be enough for our purposes that the CYBE is satisfied up to a term of the form ({\it c.f} \eqref{addtermsCSTR})
\begin{equation}\label{addtermsYBE}
\sum_{J\in\{i,j,k\}}2\pi\ii(k_Jx_J+k''_Jx''_J+k^{(l)}_Jx^{(l)}_J+k^{(r)}_Jx^{(r)}_J)
+C(\bu,\bv,\bw),
\end{equation}
for some integers $k_J,k''_J,k^{(l)}_J,k^{(r)}_J\in\mathbb{Z}$, $J\in\{i,j,k\}$, dependent on $x_J,x''_J,x^{(l)}_J,x^{(r)}_J$, 
and constant $C(\bu,\bv,\bw)$ with respect to the variables $x_J,x''_J,x^{(l)}_J,x^{(r)}_J$.  

Assuming that the CYBE \eqref{CYBEdef} is satisfied, in addition to the six equations \eqref{cuboc12} there are another six equations that are obtained by taking the partial derivatives of the CYBE with respect to each of the six variables $x_I,x''_I$, $I\in\{i,j,l\}$.  These six additional equations are
\begin{equation}\label{cuboc34}
\frac{\partial}{\partial x}\mathcal{A}^{YB}
=0, \qquad  x\in\{x_i,x_j,x_k,x''_i,x''_j,x''_k\}.
\end{equation}
The twelve equations for the partial derivatives \eqref{cuboc12} and \eqref{cuboc34} will be important for the construction of Yang-Baxter maps in the next section. While a solution of the CYBE \eqref{CYBEdef} holds up to possible terms of the form \eqref{addtermsYBE}, these additional terms give no contribution to the exponentials of the equations for the partial derivatives in \eqref{cuboc12} and \eqref{cuboc34}, which will be utilised in the next section.

\subsubsection{Classical R-matrices from solutions of the CSTRs}

It is known how to construct R-matrices for solutions of the Yang-Baxter equation through solutions of the star-triangle relation \cite{Bazhanov:1992jqa,KashaevStarSquare}.  This same idea may be used to construct solutions of the CYBE \eqref{CYBEdef} from the solutions of the CSTR \eqref{CSTRdef1} and \eqref{CSTRdef2}.

Introduce the following expressions for classical R-matrices given in terms of the functions \eqref{CSTRsol}
\begin{align}
\label{CRMAT}
\begin{split}
R_{\bu\bv}(x_i,x_j;x'_i,x'_j)&=
\ol_{\rua-\rva}(x'_i,x_j)+\ol_{\rub-\rvb}(x_i,x'_j)+\lag_{\rua-\rvb}(x_i,x_j)+\lag_{\rub-\rva}(x'_i,x'_j),
\end{split}
\\
\label{CRMAT2}
\begin{split}
\hat{R}_{\bu\bv}(x_i,x_j;x'_i,x'_j)&=
\olh_{\rua-\rva}(x_j,x_i')+\olh_{\rub-\rvb}(x_j',x_i)+\lagh_{\rua-\rvb}(x_j,x_i)+\lagh_{\rub-\rva}(x_j',x_i'),
\end{split}
\\
\label{CUMAT1}
\begin{split}
U_{\bu\bv}(x_i,x_j;x'_i,x'_j)&=
\olam_{\rua-\rva}(x'_i,x_j)+\olam_{\rub-\rvb}(x_i,x'_j)+\lam_{\rua-\rvb}(x_i,x_j)+\lam_{\rub-\rva}(x'_i,x'_j),
\end{split}
\\
\label{CUMAT2}
\begin{split}
\hat{U}_{\bu\bv}(x_i,x_j;x'_i,x'_j)&=
\olam_{\rua-\rva}(x_j,x_i')+\olam_{\rub-\rvb}(x_j',x_i)+\lam_{\rua-\rvb}(x_j,x_i)+\lam_{\rub-\rva}(x_j',x_i').
\end{split}
\end{align}
Using the assignment of functions to edges shown in Figure \ref{fig:edges}, the classical R-matrices  \eqref{CRMAT} and \eqref{CUMAT1} are pictured in Figure \ref{boxfig}.

\begin{figure}[htb!]
\centering
\begin{tikzpicture}[scale=1.8]

\begin{scope}[xshift=-100]

\draw[-] (-0.8,0)--(0,0.8)--(0.8,0)--(0,-0.8)--(-0.8,0);
\fill[black!] (-0.45,0.5) circle (0.01pt)
node[left=3.1pt]{\color{black}\small $\ol_{\rub-\rvb}$};
\fill[black!] (-1.1,-0.5) circle (0.01pt)
node[right=3.1pt]{\color{black}\small $\lag_{\rua-\rvb}$};
\fill[black!] (0.4,-0.5) circle (0.01pt)
node[right=3.1pt]{\color{black}\small $\ol_{\rua-\rva}$};
\fill[black!] (1.2,0.5) circle (0.01pt)
node[left=3.1pt]{\color{black}\small $\lag_{\rub-\rva}$};

\filldraw[fill=black,draw=black] (0,-0.8) circle (1.3pt)
node[below=3pt]{\color{black}\small $x_j$};
\filldraw[fill=black,draw=black] (0,0.8) circle (1.3pt)
node[above=3pt]{\color{black}\small $x_j'$};
\filldraw[fill=black,draw=black] (0.8,0) circle (1.3pt)
node[right=3pt]{\color{black}\small $x_i'$};
\filldraw[fill=black,draw=black] (-0.8,0) circle (1.3pt)
node[left=3pt]{\color{black}\small $x_i$};

\fill (0,-1.1) circle(0.01pt)
node[below=0.05pt]{\color{black} $R_{\bu\bv}\bigl(x_i,x_j;x_i',x_j'\bigr)$};

\end{scope}

\draw[-,double] (-0.8,0)--(0,0.8);
\draw[-,double] (0.8,0)--(0,0.8);
\draw[-,double] (0.8,0)--(0,-0.8);
\draw[-,double] (-0.8,0)--(0,-0.8);
\draw[->>>>-,white] (-0.8,0)--(0,0.8);
\draw[->>>-,white] (0.8,0)--(0,0.8);
\draw[->>>>-,white] (0.8,0)--(0,-0.8);
\draw[->>>-,white] (-0.8,0)--(0,-0.8);
\fill[black!] (-0.45,0.5) circle (0.01pt)
node[left=3.1pt]{\color{black}\small $\olam_{\rub-\rvb}$};
\fill[black!] (-1.15,-0.5) circle (0.01pt)
node[right=3.1pt]{\color{black}\small $\lam_{\rua-\rvb}$};
\fill[black!] (0.5,-0.5) circle (0.01pt)
node[right=3.1pt]{\color{black}\small $\olam_{\rua-\rva}$};
\fill[black!] (1.2,0.5) circle (0.01pt)
node[left=3.1pt]{\color{black}\small $\lam_{\rub-\rva}$};

\filldraw[fill=black,draw=black] (0,-0.8) circle (1.3pt)
node[below=3pt]{\color{black}\small $x_j$};
\filldraw[fill=black,draw=black] (0,0.8) circle (1.3pt)
node[above=3pt]{\color{black}\small $x_j'$};
\filldraw[fill=black,draw=black] (0.8,0) circle (1.3pt)
node[right=3pt]{\color{black}\small $x_i'$};
\filldraw[fill=black,draw=black] (-0.8,0) circle (1.3pt)
node[left=3pt]{\color{black}\small $x_i$};

\fill (0,-1.1) circle(0.01pt)
node[below=0.05pt]{\color{black} $U_{\bu\bv}\bigl(x_i,x_j;x_i',x_j'\bigr)$};

\end{tikzpicture}
\caption{Classical R-matrices \eqref{CRMAT} and \eqref{CUMAT1}.  
The classical R-matrix \eqref{CRMAT2} would also be represented by the diagram on the left but with ``hatted'' functions associated to each edge, while the classical R-matrix \eqref{CUMAT2} would be represented by the diagram on the right with the reverse orientation of all four edges.}
\label{boxfig}
\end{figure}

\begin{prop}\label{thm:CYBE}
If the six functions \eqref{CSTRsol} are a type I, type II, or type III solution of the CSTRs \eqref{CSTRdef1}--\eqref{3legdef2} 
up to respective terms of the form \eqref{addtermsCSTR}, then the following four choices of $R^{(1)}_{\bu\bv}$ and $R^{(2)}_{\bu\bv}$ given in terms of the classical R-matrices \eqref{CRMAT}--\eqref{CUMAT2}
\begin{align}\nonumber
\begin{split}
(1a) \qquad & R^{(1)}_{\bu\bv}=R^{(2)}_{\bu\bv}
=R_{\bu\bv},
\end{split}
\\ \nonumber
\begin{split}
(1b) \qquad & R^{(1)}_{\bu\bv}=R^{(2)}_{\bu\bv}
=\hat{R}_{\bu\bv},
\end{split}
\\ \nonumber
\begin{split}
(2a) \qquad & R^{(1)}_{\bu\bv}=R_{\bu\bv},\quad R^{(2)}_{\bu\bv}
=U_{\bu\bv},
\end{split}
\\ \nonumber
\begin{split}
(2b) \qquad & R^{(1)}_{\bu\bv}=\hat{R}_{\bu\bv},\quad R^{(2)}_{\bu\bv}
=\hat{U}_{\bu\bv},
\end{split}
\end{align}
are solutions to the CYBE \eqref{CYBEdef}--\eqref{cuboc12} up to a term of the form \eqref{addtermsYBE}.

\end{prop}

\begin{remark}
If \eqref{CSTRsol} is a type I solution of the CSTR, then the four cases (1a), (1b), (2a), (2b) are equivalent, while if \eqref{CSTRsol} is a type II solution of the CSTR, the cases (1a) and (1b) are equivalent, and the cases (2a) and (2b) are equivalent.
\end{remark}

\begin{proof}
A straightforward way to show Proposition \ref{thm:CYBE} is diagrammatically, where the CSTRs may be used to transform the left-hand side of the CYBE into the right-hand side, or vice versa.  The example of the CYBE for the case (2a) is shown in Figure \ref{fig-VYBE}. A sequence of transformations that shows that this CYBE is satisfied is shown in Figure \ref{fig-YBEtransform}.  The use of the CSTR is justified at each step, because each step involves either changing a triangle into a star which adds a variable at a new vertex whose value is determined by requiring the constraint \eqref{3legdef1} or \eqref{3legdef2} of the CSTR to be satisfied, or changing a star to a triangle where the constraint for the CSTR is already satisfied.  Each variable of the CYBE is involved in a transformation, and since the CSTRs are satisfied up to terms of the form \eqref{addtermsCSTR}, the CYBE must be satisfied up to terms of the form \eqref{addtermsYBE}.  The cases (1a), (1b), and (2b), follow from analogous transformations using the appropriate forms of the CSTRs for each case.  
\end{proof}

\begin{figure}[tbh!]
\centering
\begin{tikzpicture}[scale=1.4]

\begin{scope}[xshift=-140pt]


\draw[-] (0,-1.73)--(-1.5,-0.88)--(-1,0)--(0.5,-0.88)--(0,-1.73);
\draw[-,double] (0,1.73)--(-1.5,0.88);
\draw[-,double] (-1,0)--(-1.5,0.88);
\draw[-,double] (-1,0)--(0.5,0.88);
\draw[-,double] (0,1.73)--(0.5,0.88);
\draw[-,double] (1.5,0.88)--(0.5,0.88);
\draw[-,double] (1.5,0.88)--(1.5,-0.88);
\draw[-,double] (0.5,-0.88)--(1.5,-0.88);
\draw[-,double] (0.5,-0.88)--(0.5,0.88);
\draw[->>>-,white] (0,1.73)--(-1.5,0.88);
\draw[->>>>-,white] (-1,0)--(-1.5,0.88);
\draw[->>>-,white] (-1,0)--(0.5,0.88);
\draw[->>>>-,white] (0,1.73)--(0.5,0.88);
\draw[->>>-,white] (1.5,0.88)--(0.5,0.88);
\draw[->>>>-,white] (1.5,0.88)--(1.5,-0.88);
\draw[->>>-,white] (0.5,-0.88)--(1.5,-0.88);
\draw[->>>>-,white] (0.5,-0.88)--(0.5,0.88);

\filldraw[fill=black,draw=black] (0,1.73) circle (1.6pt)
node[right=2pt]{\small $x''_j$};
\filldraw[fill=black,draw=black] (-1.5,0.88) circle (1.6pt)
node[left=2pt]{\small $x''_k$};
\filldraw[fill=black,draw=black] (-1.5,-0.88) circle (1.6pt)
node[below=2pt]{\small $x_i$};
\filldraw[fill=black,draw=black] (0,-1.73) circle (1.6pt)
node[right=2pt]{\small $x_j$};
\filldraw[fill=black,draw=black] (1.5,-0.88) circle (1.6pt)
node[right=2pt]{\small $x_k$};
\filldraw[fill=black,draw=black] (1.5,0.88) circle (1.6pt)
node[right=2pt]{\small $x''_i$};
\filldraw[fill=black,draw=black] (0.5,0.88) circle (1.6pt);
\filldraw[fill=black,draw=black] (0.6,0.88) circle (0.01pt)
node[above=3pt]{\small $\xlk$};
\filldraw[fill=black,draw=black] (-1,0) circle (1.6pt)
node[left=3pt]{\small $\xlj$};
\filldraw[fill=black,draw=black] (0.5,-0.88) circle (1.6pt);
\filldraw[fill=black,draw=black] (0.6,-0.88) circle (0.01pt)
node[below=2pt]{\small $\xli$};

\fill[black] (-0.85,-0.85) circle (0.01pt)
node[right=1pt]{ $R_{\bu\bv}$};
\fill[black] (-0.85, 0.85) circle (0.01pt)
node[right=1pt]{ $U_{\bv\bw}$};
\fill[black] (1.4,-0.0) circle (0.01pt)
node[left=1pt]{ $U_{\bu\bw}$};

\end{scope}


\draw[-] (0,1.73)--(1.5,0.88)--(1,0)--(-0.5,0.88)--(0,1.73);
\draw[-,double] (0,-1.73)--(1.5,-0.88);
\draw[-,double] (1,0)--(1.5,-0.88);
\draw[-,double] (1,0)--(-0.5,-0.88);
\draw[-,double] (0,-1.73)--(-0.5,-0.88);
\draw[-,double] (-0.5,0.88)--(-1.5,0.88);
\draw[-,double] (-1.5,-0.88)--(-1.5,0.88);
\draw[-,double] (-1.5,-0.88)--(-0.5,-0.88);
\draw[-,double] (-0.5,0.88)--(-0.5,-0.88);
\draw[->>>-,white] (0,-1.73)--(1.5,-0.88);
\draw[->>>>-,white] (1,0)--(1.5,-0.88);
\draw[->>>-,white] (1,0)--(-0.5,-0.88);
\draw[->>>>-,white] (0,-1.73)--(-0.5,-0.88);
\draw[->>>-,white] (-0.5,0.88)--(-1.5,0.88);
\draw[->>>>-,white] (-1.5,-0.88)--(-1.5,0.88);
\draw[->>>-,white] (-1.5,-0.88)--(-0.5,-0.88);
\draw[->>>>-,white] (-0.5,0.88)--(-0.5,-0.88);

\filldraw[fill=black,draw=black] (0,1.73) circle (1.6pt)
node[left=2pt]{\small $x''_j$};
\filldraw[fill=black,draw=black] (1.5,0.88) circle (1.6pt)
node[right=2pt]{\small $x''_i$};
\filldraw[fill=black,draw=black] (1.5,-0.88) circle (1.6pt)
node[right=2pt]{\small $x_k$};
\filldraw[fill=black,draw=black] (0,-1.73) circle (1.6pt)
node[left=2pt]{\small $x_j$};
\filldraw[fill=black,draw=black] (-1.5,-0.88) circle (1.6pt)
node[left=2pt]{\small $x_i$};
\filldraw[fill=black,draw=black] (-1.5,0.88) circle (1.6pt)
node[above=2pt]{\small $x''_k$};
\filldraw[fill=black,draw=black] (-0.5,0.88) circle (1.6pt);
\filldraw[fill=black,draw=black] (-0.6,0.88) circle (0.01pt)
node[above=2pt]{\small $\xri$};
\filldraw[fill=black,draw=black] (1,0) circle (1.6pt)
node[right=2pt]{\small $\xrj$};
\filldraw[fill=black,draw=black] (-0.5,-0.88) circle (1.6pt);
\filldraw[fill=black,draw=black] (-0.6,-0.88) circle (0.01pt)
node[below=2pt]{\small $\xrk$};

\fill[black] (0.85, 0.85) circle (0.01pt)
node[left=1pt]{ $R_{\bu\bv}$};
\fill[black] (0.85,-0.85) circle (0.01pt)
node[left=1pt]{ $U_{\bv\bw}$};
\fill[black] (-1.4,-0.0) circle (0.01pt)
node[right=1pt]{ $U_{\bu\bw}$};

\draw[black!] (-2.3,0) circle (0.01pt)
node[left=1pt]{\color{black}\small $=$};

\end{tikzpicture}
\caption{CYBE for the case (2a) of Proposition \ref{thm:CYBE} using the graphical representation of Figure \ref{boxfig}.  The diagram for the case (2b) would be the same but with the reverse orientation of each directed edge.  
For the cases (1a) and (1b) there would be no directed edges.}
\label{fig-VYBE}
\end{figure}

\begin{figure}[tbh!]
\centering
\begin{tikzpicture}[scale=0.9]

\begin{scope}[xshift=-180pt]

\draw[-] (0,-1.73)--(-1.5,-0.88)--(-1,0)--(0.5,-0.88)--(0,-1.73);

\draw[-,double] (0,1.73)--(-1.5,0.88);
\draw[-,double] (-1,0)--(-1.5,0.88);
\draw[-,double] (-1,0)--(0.5,0.88);
\draw[-,double] (0,1.73)--(0.5,0.88);
\draw[-,double] (1.5,0.88)--(0.5,0.88);
\draw[-,double] (1.5,0.88)--(1.5,-0.88);
\draw[-,double] (0.5,-0.88)--(1.5,-0.88);
\draw[-,double] (0.5,-0.88)--(0.5,0.88);
\draw[->>>-,white] (0,1.73)--(-1.5,0.88);
\draw[->>>>-,white] (-1,0)--(-1.5,0.88);
\draw[->>>-,white] (-1,0)--(0.5,0.88);
\draw[->>>>-,white] (0,1.73)--(0.5,0.88);
\draw[->>>-,white] (1.5,0.88)--(0.5,0.88);
\draw[->>>>-,white] (1.5,0.88)--(1.5,-0.88);
\draw[->>>-,white] (0.5,-0.88)--(1.5,-0.88);
\draw[->>>>-,white] (0.5,-0.88)--(0.5,0.88);

\filldraw[fill=black,draw=black] (0,1.73) circle (2.2pt)
node[right=2pt]{\small $x''_j$};
\filldraw[fill=black,draw=black] (-1.5,0.88) circle (2.2pt)
node[left=2pt]{\small $x''_k$};
\filldraw[fill=black,draw=black] (-1.5,-0.88) circle (2.2pt)
node[below=2pt]{\small $x_i$};
\filldraw[fill=black,draw=black] (0,-1.73) circle (2.2pt)
node[below=2pt]{\small $x_j$};
\filldraw[fill=black,draw=black] (1.5,-0.88) circle (2.2pt)
node[right=2pt]{\small $x_k$};
\filldraw[fill=black,draw=black] (1.5,0.88) circle (2.2pt)
node[right=2pt]{\small $x''_i$};
\filldraw[fill=white,draw=black] (0.5,0.88) circle (2.2pt);
\filldraw[fill=white,draw=black] (-1,0) circle (2.2pt);
\filldraw[fill=white,draw=black] (0.5,-0.88) circle (2.2pt);

\end{scope}

\draw[-] (0,1.73)--(1.5,0.88)--(1,0)--(-0.5,0.88)--(0,1.73);
\draw[-,double] (0,-1.73)--(1.5,-0.88);
\draw[-,double] (1,0)--(1.5,-0.88);
\draw[-,double] (1,0)--(-0.5,-0.88);
\draw[-,double] (0,-1.73)--(-0.5,-0.88);
\draw[-,double] (-0.5,0.88)--(-1.5,0.88);
\draw[-,double] (-1.5,-0.88)--(-1.5,0.88);
\draw[-,double] (-1.5,-0.88)--(-0.5,-0.88);
\draw[-,double] (-0.5,0.88)--(-0.5,-0.88);
\draw[->>>-,white] (0,-1.73)--(1.5,-0.88);
\draw[->>>>-,white] (1,0)--(1.5,-0.88);
\draw[->>>-,white] (1,0)--(-0.5,-0.88);
\draw[->>>>-,white] (0,-1.73)--(-0.5,-0.88);
\draw[->>>-,white] (-0.5,0.88)--(-1.5,0.88);
\draw[->>>>-,white] (-1.5,-0.88)--(-1.5,0.88);
\draw[->>>-,white] (-1.5,-0.88)--(-0.5,-0.88);
\draw[->>>>-,white] (-0.5,0.88)--(-0.5,-0.88);

\filldraw[fill=black,draw=black] (0,1.73) circle (2.2pt)
node[left=2pt]{\small $x''_j$};
\filldraw[fill=black,draw=black] (1.5,0.88) circle (2.2pt)
node[right=2pt]{\small $x''_i$};
\filldraw[fill=black,draw=black] (1.5,-0.88) circle (2.2pt)
node[right=2pt]{\small $x_k$};
\filldraw[fill=black,draw=black] (0,-1.73) circle (2.2pt)
node[below=2pt]{\small $x_j$};
\filldraw[fill=black,draw=black] (-1.5,-0.88) circle (2.2pt)
node[left=2pt]{\small $x_i$};
\filldraw[fill=black,draw=black] (-1.5,0.88) circle (2.2pt)
node[above=2pt]{\small $x''_k$};
\filldraw[fill=white,draw=black] (-0.5,0.88) circle (2.2pt);
\filldraw[fill=white,draw=black] (1,0) circle (2.2pt);
\filldraw[fill=white,draw=black] (-0.5,-0.88) circle (2.2pt);

\draw[black!] (-2.9,0) circle (0.01pt)
node[left=1pt]{\color{black}\small $=$};

\draw[<->] (-7.7,-2)--(-7.7,-3);
\draw[<->] (1.3,-2)--(1.3,-3);

\begin{scope}[xshift=17,yshift=-20,scale=0.8]
\draw[<->] (-7.5,-7)--(-6.5,-8);
\end{scope}
\begin{scope}[xshift=-60,yshift=-20,scale=0.8]
\draw[<->] (1.3,-7)--(0.3,-8);
\end{scope}

\begin{scope}[yshift=-135pt]

\draw[-] (-0.5,0.88)--(0,1.73)--(1.5,0.88)--(1,0); \draw[-] (-0.5,-0.88)--(0,0);
\draw[-,double] (0,-1.73)--(1.5,-0.88);
\draw[-,double] (1,0)--(1.5,-0.88);
\draw[-,double] (1,0)--(0,0);
\draw[-,double] (0,-1.73)--(-0.5,-0.88);
\draw[-,double] (-0.5,0.88)--(-1.5,0.88);
\draw[-,double] (-1.5,-0.88)--(-1.5,0.88);
\draw[-,double] (-1.5,-0.88)--(-0.5,-0.88);
\draw[-,double] (-0.5,0.88)--(0,0);
\draw[->>>-,white] (0,-1.73)--(1.5,-0.88);
\draw[->>>>-,white] (1,0)--(1.5,-0.88);
\draw[->>>-,white] (1,0)--(0,0);
\draw[->>>>-,white] (0,-1.73)--(-0.5,-0.88);
\draw[->>>-,white] (-0.5,0.88)--(-1.5,0.88);
\draw[->>>>-,white] (-1.5,-0.88)--(-1.5,0.88);
\draw[->>>-,white] (-1.5,-0.88)--(-0.5,-0.88);
\draw[->>>>-,white] (-0.5,0.88)--(0,0);

\filldraw[fill=white,draw=black] (0,0) circle (2.2pt);

\filldraw[fill=black,draw=black] (0,1.73) circle (2.2pt)
node[left=2pt]{\small $x''_j$};
\filldraw[fill=black,draw=black] (1.5,0.88) circle (2.2pt)
node[right=2pt]{\small $x''_i$};
\filldraw[fill=black,draw=black] (1.5,-0.88) circle (2.2pt)
node[right=2pt]{\small $x_k$};
\filldraw[fill=black,draw=black] (0,-1.73) circle (2.2pt)
node[below=2pt]{\small $x_j$};
\filldraw[fill=black,draw=black] (-1.5,-0.88) circle (2.2pt)
node[left=2pt]{\small $x_i$};
\filldraw[fill=black,draw=black] (-1.5,0.88) circle (2.2pt)
node[above=2pt]{\small $x''_k$};
\filldraw[fill=white,draw=black] (-0.5,0.88) circle (2.2pt);
\filldraw[fill=white,draw=black] (1,0) circle (2.2pt);
\filldraw[fill=white,draw=black] (-0.5,-0.88) circle (2.2pt);

\end{scope}

\begin{scope}[xshift=-180pt,yshift=-135pt]

\draw[-] (0.5,-0.88)--(0,-1.73)--(-1.5,-0.88)--(-1,0);\draw[-] (0,0)--(0.5,0.88);
\draw[-,double] (0,1.73)--(-1.5,0.88);
\draw[-,double] (-1,0)--(-1.5,0.88);
\draw[-,double] (-1,0)--(0,0);
\draw[-,double] (0,1.73)--(0.5,0.88);
\draw[-,double] (1.5,0.88)--(0.5,0.88);
\draw[-,double] (1.5,0.88)--(1.5,-0.88);
\draw[-,double] (0.5,-0.88)--(1.5,-0.88);
\draw[-,double] (0.5,-0.88)--(0,0);
\draw[->>>-,white] (0,1.73)--(-1.5,0.88);
\draw[->>>>-,white] (-1,0)--(-1.5,0.88);
\draw[->>>-,white] (-1,0)--(0,0);
\draw[->>>>-,white] (0,1.73)--(0.5,0.88);
\draw[->>>-,white] (1.5,0.88)--(0.5,0.88);
\draw[->>>>-,white] (1.5,0.88)--(1.5,-0.88);
\draw[->>>-,white] (0.5,-0.88)--(1.5,-0.88);
\draw[->>>>-,white] (0.5,-0.88)--(0,0);

\filldraw[fill=white,draw=black] (0,0) circle (2.2pt);

\filldraw[fill=black,draw=black] (0,1.73) circle (2.2pt)
node[right=2pt]{\small $x''_j$};
\filldraw[fill=black,draw=black] (-1.5,0.88) circle (2.2pt)
node[left=2pt]{\small $x''_k$};
\filldraw[fill=black,draw=black] (-1.5,-0.88) circle (2.2pt)
node[below=2pt]{\small $x_i$};
\filldraw[fill=black,draw=black] (0,-1.73) circle (2.2pt)
node[below=2pt]{\small $x_j$};
\filldraw[fill=black,draw=black] (1.5,-0.88) circle (2.2pt)
node[right=2pt]{\small $x_k$};
\filldraw[fill=black,draw=black] (1.5,0.88) circle (2.2pt)
node[right=2pt]{\small $x''_i$};
\filldraw[fill=white,draw=black] (0.5,0.88) circle (2.2pt);
\filldraw[fill=white,draw=black] (-1,0) circle (2.2pt);
\filldraw[fill=white,draw=black] (0.5,-0.88) circle (2.2pt);

\end{scope}

\begin{scope}[xshift=-90pt,yshift=-250pt]

\draw[-] (0,-1.73)--(-1.5,-0.88);\draw[-] (0,1.73)--(1.5,0.88);
\draw[-] (0,0)--(1.5,-0.88); \draw[-] (0,0)--(-1.5,0.88);
\draw[-,double] (0,1.73)--(-1.5,0.88);
\draw[-,double] (-1.5,-0.88)--(-1.5,0.88);
\draw[-,double] (-1.5,-0.88)--(0,0);
\draw[-,double] (0,1.73)--(0,0);
\draw[-,double] (1.5,0.88)--(0,0);
\draw[-,double] (1.5,0.88)--(1.5,-0.88);
\draw[-,double] (0,-1.73)--(1.5,-0.88);
\draw[-,double] (0,-1.73)--(0,0);
\draw[->>>-,white] (0,1.73)--(-1.5,0.88);
\draw[->>>>-,white] (-1.5,-0.88)--(-1.5,0.88);
\draw[->>>-,white] (-1.5,-0.88)--(0,0);
\draw[->>>>-,white] (0,1.73)--(0,0);
\draw[->>>-,white] (1.5,0.88)--(0,0);
\draw[->>>>-,white] (1.5,0.88)--(1.5,-0.88);
\draw[->>>-,white] (0,-1.73)--(1.5,-0.88);
\draw[->>>>-,white] (0,-1.73)--(0,0);

\filldraw[fill=white,draw=black] (0,0) circle (2.2pt);

\filldraw[fill=black,draw=black] (0,1.73) circle (2.2pt)
node[above=2pt]{\small $x''_j$};
\filldraw[fill=black,draw=black] (-1.5,0.88) circle (2.2pt)
node[left=2pt]{\small $x''_k$};
\filldraw[fill=black,draw=black] (-1.5,-0.88) circle (2.2pt)
node[below=2pt]{\small $x_i$};
\filldraw[fill=black,draw=black] (0,-1.73) circle (2.2pt)
node[below=2pt]{\small $x_j$};
\filldraw[fill=black,draw=black] (1.5,-0.88) circle (2.2pt)
node[right=2pt]{\small $x_k$};
\filldraw[fill=black,draw=black] (1.5,0.88) circle (2.2pt)
node[right=2pt]{\small $x''_i$};

\end{scope}

\end{tikzpicture}
\caption{Deformations of the CYBE of Figure \ref{fig-VYBE}, with the use CSTR of Figure \ref{fig-CSTR}.  The deformations for the (2b) case of the CYBE are of the same form, but with the reverse orientation of each directed edge.  
For the cases (1a) and (1b) there would be no directed edges.  At each step, the partial derivatives with respect to the variables associated to white vertices vanish.}
\label{fig-YBEtransform}
\end{figure}


\begin{remark}
For the transformations used in Figure \ref{fig-YBEtransform}, each function of the type $\lag_{u-v}$ or  $\ol_{u-v}$ is transformed twice via the CSTRs.  They first get transformed to $\lagh_{u-v}$ or $\olh_{u-v}$, and then back to $\lag_{u-v}$ or $\ol_{u-v}$.  The effect of this is that the (2a) and (2b) cases only involve one of the classical R-matrices \eqref{CRMAT} or \eqref{CRMAT2}, and not both of them.  
\end{remark}

\section{From classical Yang-Baxter equation to Yang-Baxter maps}\label{sec:YBM}

The purpose of this section is to show how Yang-Baxter maps may be derived directly from the solutions of the classical Yang-Baxter equation of Proposition \ref{thm:CYBE}.

\subsection{Two-component Yang-Baxter maps}\label{sec:YBMdef}

Two different types of Yang-Baxter maps will be obtained, which will be denoted by
\begin{equation}
\label{YBMapdef}
R(\al,\bt)\colon X\times X\to X\times X,\qquad U(\al,\bt)\colon X\times X\to X\times X,
\end{equation}
where the parameters are $\al,\bt\in\mathbb{C}^2$.  For this paper, the set $X$ is always taken as
\begin{equation}
X=\cp\times\cp,
\end{equation}
and thus it is appropriate to refer to \eqref{YBMapdef} as two-component maps, as opposed to the more common expressions for Yang-Baxter maps where $X=\cp$.

The Yang-Baxter maps \eqref{YBMapdef} will provide solutions to two different forms of the functional Yang-Baxter equation  
given by
\begin{align}
\label{MYBE}
R_{jk}(\bt,\gm)\circ R_{ik}(\al,\gm)\circ R_{ij}(\al,\bt)= R_{ij}(\al,\bt)\circ R_{ik}(\al,\gm)\circ R_{jk}(\bt,\gm), \\
\label{MAYBE}
U_{jk}(\bt,\gm)\circ U_{ik}(\al,\gm)\circ R_{ij}(\al,\bt)= R_{ij}(\al,\bt)\circ U_{ik}(\al,\gm)\circ U_{jk}(\bt,\gm).\hspace{0.05cm}
\end{align}
These equations respectively express the equality of two different maps $X\times X\times X\to X\times X\times X$, where as usual, an individual map $R_{ij}$ or $U_{ij}$ only acts non-trivially on the $i$-th and $j$-th sets.  The first expression for the FYBE  \eqref{MYBE} is related to the more symmetric cases (1a) and (1b) of Proposition \ref{thm:CYBE}, and  
the second expression for the FYBE \eqref{MAYBE} is related to the (2a) and (2b) cases.  The first expression \eqref{MYBE} is the usual form of FYBE for Yang-Baxter maps, while the second expression \eqref{MAYBE} has been studied before as a case of the ``entwining'' Yang-Baxter equation \cite{KouloukasEntwining}. 

The Yang-Baxter maps can be given a graphical representation (which is independent of the diagrams that have been introduced in the previous section).  Defining the two-component variables 
\begin{equation}
\label{YBMvardef}
\bxi_i,\bxi_j\in\cp\times\cp,
\end{equation}
the Yang-Baxter maps \eqref{YBMapdef} may be visualised as having the variables and parameters on edges of a square as shown in Figure \ref{fig-YBMs}, with the arrow indicating the direction of the mapping.  The FYBE \eqref{MAYBE} is pictured in Figure \ref{fig-FYBE} (the FYBE \eqref{MYBE} is simply obtained from this figure by replacing each of the $U$ maps with $R$ maps).  The left-hand side of \eqref{MAYBE} maps $(\bxi_i,\bxi_j,\bxi_k)\mapsto(\bxi''^{(l)}_i,\bxi''^{(l)}_j,\bxi''^{(l)}_k)$, and the right-hand side of \eqref{MAYBE} maps $(\bxi_i,\bxi_j,\bxi_k)\mapsto(\bxi''^{(r)}_i,\bxi''^{(r)}_j,\bxi''^{(r)}_k)$, and the FYBE \eqref{MAYBE} states that these two different compositions of maps should be equivalent, such that $(\bxi''^{(l)}_i,\bxi''^{(l)}_j,\bxi''^{(l)}_k)= (\bxi''^{(r)}_i,\bxi''^{(r)}_j,\bxi''^{(r)}_k)$.    

\begin{figure}[htb!]
\centering
\begin{tikzpicture}[scale=1.4]

\begin{scope}[xshift=-150]

\draw[-] (-0.8,-0.8)--(0.8,-0.8)--(0.8,0.8)--(-0.8,0.8)--(-0.8,-0.8);

\draw[-Latex] (-0.4,-0.4)--(0.4,0.4);

\filldraw[fill=black,draw=black] (0,-0.85) circle (0.01pt)
node[below=1pt]{\color{black}\small $(\bxi_j, \bt)$};
\filldraw[fill=black,draw=black] (0,0.85) circle (0.01pt)
node[above=1pt]{\color{black}\small $( \bxi_j',\bt)$};
\filldraw[fill=black,draw=black] (0.85,0) circle (0.01pt)
node[right=1pt]{\color{black}\small $(\bxi_i',\al)$};
\filldraw[fill=black,draw=black] (-0.85,0) circle (0.01pt)
node[left=1pt]{\color{black}\small $(\bxi_i,\al)$};

\fill (0,-1.5) circle(0.01pt)
node[below=0.05pt]{\color{black} $R(\al,\bt)$};

\end{scope}

\draw[-] (-0.8,-0.8)--(0.8,-0.8)--(0.8,0.8)--(-0.8,0.8)--(-0.8,-0.8);

\draw[-Latex,double] (-0.4,-0.4)--(0.4,0.4);

\filldraw[fill=black,draw=black] (0,-0.85) circle (0.01pt)
node[below=1pt]{\color{black}\small $(\bxi_j, \bt)$};
\filldraw[fill=black,draw=black] (0,0.85) circle (0.01pt)
node[above=1pt]{\color{black}\small $( \bxi_j',\bt)$};
\filldraw[fill=black,draw=black] (0.85,0) circle (0.01pt)
node[right=1pt]{\color{black}\small $(\bxi_i',\al)$};
\filldraw[fill=black,draw=black] (-0.85,0) circle (0.01pt)
node[left=1pt]{\color{black}\small $(\bxi_i,\al)$};

\fill (0,-1.5) circle(0.01pt)
node[below=0.05pt]{\color{black} $U(\al,\bt)$};

\end{tikzpicture}
\caption{Yang-Baxter maps \eqref{YBMapdef} that have two-component variables $\bxi_i,\bxi_j$ and two-component parameters $\al,\bt$ on edges.  The single- and double-line arrows are used to distinguish the Yang-Baxter maps $R(\al,\bt)$ and $U(\al,\bt)$, respectively.  Parallel edges are assigned the same parameter.}  
\label{fig-YBMs}
\end{figure}

\begin{figure}[hbt!]
\centering
\begin{tikzpicture}[scale=0.85]
\begin{scope}[scale=0.8]

\draw[-] (-3,-1)--(-1,0)--(3,0)--(1.0,-1)--(-3,-1);
 \draw[-] (-3,-1)--(-3,3)--(-1,4)--(-1,0); 
 \draw[-] (3,0)--(3,4)--(-1,4);
 
 \draw[-Latex] (0.4,-0.8)--(-0.4,-0.2);
 \draw[-Latex,double] (-1.6,0.8)--(-2.4,2.2);
 \draw[-Latex,double] (1.6,1.4)--(0.4,2.6);

\filldraw[fill=black,draw=black] (-1,-1) circle (0.01pt)
node[below=1.5pt]{\color{black}\small $(\bxi_i,\al)$};
\fill[black!] (2.5,-0.5) circle (0.01pt)
node[below=0pt]{\color{black}\small $(\bxi_j,\bt)$};
\fill[black!] (-2,3.7) circle (0.01pt)
node[left=1.5pt]{\color{black}\small $(\bxi''^{(l)}_j,\bt)$};
\fill[black!] (-3,1.1) circle (0.01pt)
node[left=1.5pt]{\color{black}\small $(\bxi''^{(l)}_k,\gm)$};
\filldraw[fill=black,draw=black] (3,2.6) circle (0.01pt)
node[right=1.5pt]{\color{black}\small $(\bxi_k,\gm)$};
\filldraw[fill=black,draw=black] (1,4) circle (0.01pt)
node[above=1.5pt]{\color{black}\small $(\bxi''^{(l)}_i,\al)$};

\end{scope}

\draw[black] (4.5,1.35) circle (0.01pt)
node[below=1pt]{\color{black}$=$};

\begin{scope}[scale=0.8,xshift=332,yshift=0,rotate=0]

 \draw[-] (1,-1)--(1,3)--(3,4)--(3,0)--(1,-1);
 \draw[-] (1,3)--(-3,3)--(-3,-1)--(1,-1);
 \draw[-] (-3,3)--(-1,4)--(3,4);
 
 \draw[-Latex] (0.4,3.2)--(-0.4,3.8);
 \draw[-Latex,double] (2.4,0.8)--(1.6,2.2);
 \draw[-Latex,double] (-0.4,0.4)--(-1.6,1.6);

\fill[black!] (-2.8,3.3) circle (0.01pt)
node[above=1.5pt]{\color{black}\small $(\bxi''^{(r)}_j,\bt)$};
\filldraw[fill=black,draw=black] (1,4) circle (0.01pt)
node[above=1.5pt]{\color{black}\small $(\bxi''^{(r)}_i,\al)$};
\filldraw[fill=black,draw=black] (3,2) circle (0.01pt)
node[right=1.5pt]{\color{black}\small $(\bxi_k,\gm)$};
\fill[black!] (-3,0.5) circle (0.01pt)
node[left=1.5pt]{\color{black}\small $(\bxi''^{(r)}_k,\gm)$};
\filldraw[fill=black,draw=black] (-1,-1) circle (0.01pt)
node[below=1.5pt]{\color{black}\small $(\bxi_i,\al)$};
\fill[black!] (2.5,-0.5) circle (0.01pt)
node[below=0pt]{\color{black}\small $(\bxi_j,\bt)$};

\end{scope}

\end{tikzpicture}
\caption{FYBE \eqref{MAYBE} for the Yang-Baxter maps of Figure \ref{fig-YBMs}.}    
\label{fig-FYBE}
\end{figure}

\subsection{Classical R-matrix to Yang-Baxter map}\label{sec:YBMconversion}

In the following, details will be given for constructing the Yang-Baxter maps for the FYBE's \eqref{MYBE} and \eqref{MAYBE} from case (2a) of Proposition \ref{thm:CYBE}.  The other cases of Proposition \ref{thm:CYBE} follow from similar computations and so will not be considered here separately.

First, consider the classical R-matrix \eqref{CRMAT} and let the partial derivatives with respect to its four variables be denoted
\begin{equation}
\label{Rmatderivs}
\begin{gathered}
R_{\bu\bv,i}(x_j',x_j;x_i)=\frac{\partial R_{\bu\bv}\bigl(x_i,x_j;x_i',x_j'\bigr)}{\partial x_i}, \quad
R_{\bu\bv,i'}(x_j,x_j';x_i')=\frac{-\partial R_{\bu\bv}\bigl(x_i,x_j;x_i',x_j'\bigr)}{\partial x_i'}, \\
R_{\bu\bv,j}(x_i',x_i;x_j)=\frac{\partial R_{\bu\bv}\bigl(x_i,x_j;x_i',x_j'\bigr)}{\partial x_j}, \quad
R_{\bu\bv,j'}(x_i,x_i';x_j')=\frac{-\partial R_{\bu\bv}\bigl(x_i,x_j;x_i',x_j'\bigr)}{\partial x_j'}.
\end{gathered}
\end{equation}
From the definition \eqref{CRMAT}, each of the above partial derivatives are a sum of two terms that involve a partial derivative of $\lag_{u-v}(x_i,x_j)$ and a partial derivative of $\ol_{u-v}(x_i,x_j)$.

Similarly, the respective partial derivatives of the classical R-matrix \eqref{CUMAT1} will be denoted by
\begin{equation}
\label{Umatderivs}
\begin{gathered}
U_{\bu\bv,i}(x_j',x_j;x_i)=\frac{\partial U_{\bu\bv}\bigl(x_i,x_j;x_i',x_j'\bigr)}{\partial x_i}, \quad
U_{\bu\bv,i'}(x_j,x_j';x_i')=\frac{-\partial U_{\bu\bv}\bigl(x_i,x_j;x_i',x_j'\bigr)}{\partial x_i'}, \\
U_{\bu\bv,j}(x_i',x_i;x_j)=\frac{\partial U_{\bu\bv}\bigl(x_i,x_j;x_i',x_j'\bigr)}{\partial x_j}, \quad
U_{\bu\bv.j'}(x_i,x_i';x_j')=\frac{-\partial U_{\bu\bv}\bigl(x_i,x_j;x_i',x_j'\bigr)}{\partial x_j'}.
\end{gathered}
\end{equation}

Next, a change of variables is usually necessary to put the equations \eqref{Rmatderivs} and \eqref{Umatderivs} into a suitable form for the respective Yang-Baxter maps.  For \eqref{Rmatderivs}, this change of variables takes the following form
\begin{equation}
\label{RmattoYBMcov}
\begin{gathered}
\myj=f(x'_i),\qquad \myi=f(x'_j),\qquad \mxi=f(x_i),\qquad \mxj=f(x_j), \\[0.1cm]
\mpa=h(\rua),\qquad \mpb=h(\rub),\qquad \mqa=h(\rva), \qquad \mqb=h(\rvb),
\end{gathered}
\end{equation}
where $f(x)$ is chosen so that the exponentials of both $R_{\bu\bv,j}(x_i',x_i;x_j)$ and $R_{\bu\bv,i}(x_j',x_j;x_i)$ (or $R_{\bu\bv,j}(x_i',x_i;x_j)$ and $R_{\bu\bv,i}(x_j',x_j;x_i)$ themselves) depend linearly on $\myj$ and $\myi$, respectively.   Except for the elliptic case, it will turn out that the choices for $f(x)$ also make $R_{\bu\bv,j}(x_i',x_i;x_j)$ and $R_{\bu\bv,i}(x_j',x_j;x_i)$ have linear dependence on $\mxi$ and  $\mxj$, respectively.  
The function $h(x)$ is chosen so that the resulting expressions have an algebraic dependence on components of $\al$ and $\bt$. 

Similarly to \eqref{RmattoYBMcov}, the change of variables for \eqref{Umatderivs} takes the form
\begin{equation}
\label{UmattoYBMcov}
\begin{gathered}
\myj=f(x'_i),\qquad \myi=g(x'_j),\qquad \mxi=f(x_i),\qquad \mxj=g(x_j), \\[0.1cm]
\mpa=h(\rua),\qquad \mpb=h(\rub),\qquad \mqa=h(\rva), \qquad \mqb=h(\rvb),
\end{gathered}
\end{equation}
where $f(x)$ is the same function from \eqref{RmattoYBMcov}, and $g(x)$ is another function which is chosen so that the exponentials of $U_{\bu\bv,i}(x_j',x_j;x_i)$ and $U_{\bu\bv,j}(x_i',x_i;x_j)$ (or $U_{\bu\bv,i}(x_j',x_j;x_i)$ and $U_{\bu\bv,j}(x_i',x_i;x_j)$ themselves) depend linearly on $\myi$ and $\myj$, respectively.  For the explicit cases considered in this paper (besides the most complicated elliptic case), there may be found a choice of $f(x)$, $g(x)$, $h(x)$, which are compatible with the aforementioned requirements for transforming \eqref{Rmatderivs} and \eqref{Umatderivs}.

After the change of variables \eqref{RmattoYBMcov}, the exponentials of $R_{\bu\bv,i}$, $R_{\bu\bv,j}$, $R_{\bu\bv,i'}$, $R_{\bu\bv,j'}$ (or $R_{\bu\bv,i}$, $R_{\bu\bv,j}$, $R_{\bu\bv,i'}$, $R_{\bu\bv,j'}$ themselves) will each be rational bilinear functions of two out of the four variables $\mxi$, $\mxj$, $\myj$, $\myi$.  Let the latter functions be respectively denoted by $Y^{(R)}_{2,\al\bt}(\myi,\mxj;\mxi)$, $Y^{(R)}_{1,\al\bt}(\myj,\mxi;\mxj)$, $Z^{(R)}_{1,\al\bt}(\mxj,\myi;\myj)$, $Z^{(R)}_{2,\al\bt}(\mxi,\myj;\myi)$, where each function is rational bilinear in their first two arguments (the superscript $(R)$ refers to the classical R-matrix $R_{\bu\bv}\bigl(x_i,x_j;x_i',x_j'\bigr)$, while the superscript $(U)$ will be used to refer to $U_{\bu\bv}\bigl(x_i,x_j;x_i',x_j'\bigr)$).  The expressions $Y^{(R)}_{2,\al\bt}$, $Y^{(R)}_{1,\al\bt}$, $Z^{(R)}_{1,\al\bt}$, $Z^{(R)}_{2,\al\bt}$, will be components of the desired Yang-Baxter map, and these are labelled by the new variables $\mui$, $\muj$, $\mupi$, $\mupj$, as follows
\begin{equation}
\label{RmattoYBM}
\begin{gathered}
 \mui=Y^{(R)}_{2,\al\bt}(\myi,\mxj;\mxi)=\EXP^{R_{\bu\bv,i}(x_j',x_j;x_i)}, \qquad 
 \mupi=Z^{(R)}_{1,\al\bt}(\mxj,\myi;\myj)=\EXP^{R_{\bu\bv,i'}(x_j,x_j';x_i')}, \\
 \muj=Y^{(R)}_{1,\al\bt}(\myj,\mxi;\mxj)=\EXP^{R_{\bu\bv,j}(x_i',x_i;x_j)}, \qquad 
 \mupj=Z^{(R)}_{2,\al\bt}(\mxi,\myj;\myi)=\EXP^{R_{\bu\bv,j'}(x_i,x_i';x_j')}.
\end{gathered}
\end{equation}
Taking the exponentials is typically necessary to eliminate unwanted sums of logarithms, however, for some degenerate additive cases the partial derivatives of the classical R-matrix in \eqref{RmattoYBM} do not involve the complex logarithm, in which case no exponentials would be required.

The expressions \eqref{RmattoYBM} provide an implicit form of the desired Yang-Baxter map.  To arrive at the final explicit form, the two equations on the left-hand side of \eqref{RmattoYBM} can be uniquely solved (due to rational bilinearity) for the variables $\myj$ and $\myi$.  The resulting expressions for $\myj$ and $\myi$, will be written as 
\begin{equation}
\label{RmattoYBM2}
\myj=\Upsilon^{(R)}_{1,\al\bt}(\muj,\mxi;\mxj),\qquad \myi=\Upsilon^{(R)}_{2,\al\bt}(\mui,\mxj;\mxi),
\end{equation}
where $\Upsilon^{(R)}_{1,\al\bt}(x,y;z)$ and $\Upsilon^{(R)}_{2,\al\bt}(x,y;z)$ are both rational bilinear functions of $x$ and $y$.

Then the final expression for the Yang-Baxter map $R(\al,\bt)\colon (\bxi_i,\bxi_j)\mapsto(\bxi_i',\bxi_j')$ is given in terms of \eqref{RmattoYBM} and \eqref{RmattoYBM2} as
\begin{equation}
\label{finalYBM1}
\begin{gathered}
\myj=\Upsilon^{(R)}_{1,\al\bt}(\muj,\mxi;\mxj), \qquad \mupi=Z^{(R)}_{1,\al\bt}(\mxj,\myi;\myj), \\
\myi=\Upsilon^{(R)}_{2,\al\bt}(\mui,\mxj;\mxi), \qquad \mupj=Z^{(R)}_{2,\al\bt}(\mxi,\myj;\myi).
\end{gathered}
\end{equation}

The Yang-Baxter map \eqref{finalYBM1}, can be naturally split up into the following QRT-like sequence
\begin{equation}\label{QRT}
\begin{split}
(\bxi_i,\bxi_j)\mapsto(\bxi_i',\bxi_j')\colon\bigl((\mxi,\mui),(\mxj,\muj)\bigr)\mapsto\bigl((\myj,\mui),(\myi,\muj)\bigr)\mapsto\bigl((\myj,\mupi),(\myi,\mupj)\bigr),
\end{split}
\end{equation}
where the variables $y_i,y_j$ are mapped before the variables $z_i,z_j$.  However, unlike QRT maps, the separate maps for the $y_i,y_j$ and $z_i,z_j$ do not define involutions.

The expression for the Yang-Baxter map $U(\al,\bt)\colon (\bxi_i,\bxi_j)\mapsto(\bxi_i',\bxi_j')$ follows similarly.  That is, after the change of variables \eqref{RmattoYBMcov} and \eqref{UmattoYBMcov}, $\mui,\mupi,\muj,\mupj$ are defined as ({\it c.f.} \eqref{RmattoYBM})
\begin{equation}
\label{UmattoYBM}
\begin{split}
 \mui=Y^{(U)}_{2,\al\bt}(\myi,\mxj;\mxi)=\EXP^{U_{\bu\bv,i}(x_j',x_j;x_i)}, \qquad 
 \mupi=Z^{(U)}_{1,\al\bt}(\mxj,\myi;\myj)=\EXP^{U_{\bu\bv,i'}(x_j,x_j';x_i')}, \\
 \muj=Y^{(U)}_{1,\al\bt}(\myj,\mxi;\mxj)=\EXP^{U_{\bu\bv,j}(x_i',x_i;x_j)}, \qquad 
 \mupj=Z^{(U)}_{2,\al\bt}(\mxi,\myj;\myi)=\EXP^{U_{\bu\bv,j'}(x_i,x_i';x_j')},
\end{split}
\end{equation}
where $Y^{(U)}_{1,\al\bt}(x,y;z)$, $Y^{(U)}_{2,\al\bt}(x,y;z)$, $Z^{(U)}_{1,\al\bt}(x,y;z)$, $Z^{(U)}_{2,\al\bt}(x,y;z)$, are each rational bilinear functions of $x$ and $y$.  Rewriting in terms of the variables $\myj$, $\myi$, $\mupi$, $\mupj$, gives the final expression for the Yang-Baxter map $U(\al,\bt)\colon (\bxi_i,\bxi_j)\mapsto(\bxi_i',\bxi_j')$, as
\begin{equation}
\label{finalYBM2}
\begin{gathered}
\myj=\Upsilon^{(U)}_{1,\al\bt}(\muj,\mxi;\mxj), \qquad \mupi=Z^{(U)}_{1,\al\bt}(\mxj,\myi;\myj), \\
\myi=\Upsilon^{(U)}_{2,\al\bt}(\mui,\mxj;\mxi), \qquad \mupj=Z^{(U)}_{2,\al\bt}(\mxi,\myj;\myi),
\end{gathered}
\end{equation}
where $\Upsilon^{(R)}_{1,\al\bt}(x,y;z)$ and $\Upsilon^{(R)}_{2,\al\bt}(x,y;z)$ are rational bilinear functions of $x$ and $y$.  This map may also be split up in into the sequence \eqref{QRT}.

\subsection{Properties of Yang-Baxter maps and FYBE}\label{sec:FYBE}

The following property is a straightforward consequence of the rational bilinear expressions for the Yang-Baxter maps \eqref{finalYBM1} and \eqref{finalYBM2}.
\begin{prop}[Quadrirationality]\label{prop:quadri}
The maps \eqref{finalYBM1} and \eqref{finalYBM2} can be uniquely solved to define rational maps
\begin{equation}\label{quadri}
\begin{gathered}
(\bxi_i',\bxi_j')\mapsto(\bxi_i,\bxi_j)\colon\bigl((\myj,\mupi),(\myi,\mupj)\bigr)\mapsto\bigl((\mxi,\mupi),(\mxj,\mupj)\bigr)\mapsto\bigl((\mxi,\mui),(\mxj,\muj)\bigr), \\
(\bxi_i',\bxi_j)\mapsto(\bxi_i,\bxi_j')\colon\bigl((\myj,\mupi),(\mxj,\muj)\bigr)\mapsto\bigl((\mxi,\mupi),(\myi,\muj)\bigr)\mapsto\bigl((\mxi,\mui),(\myi,\mupj)\bigr), \\
(\bxi_i,\bxi_j')\mapsto(\bxi_i',\bxi_j)\colon\bigl((\mxi,\mui),(\myi,\mupj)\bigr)\mapsto\bigl((\myj,\mui),(\mxj,\mupj)\bigr)\mapsto\bigl((\myj,\mupi),(\mxj,\muj)\bigr).
\end{gathered}
\end{equation}
\end{prop}

\begin{remark}
The first of \eqref{quadri} is the inverse map, and the remaining two maps are sometimes referred to as companion maps.
\end{remark}

\begin{prop}[Reversibility]\label{prop:revers}
If \eqref{lagsyms} is satisfied then the map \eqref{finalYBM1} satisfies
\begin{equation}
\label{YBMreversibility}
 R_{ji}(\bt,\al)\circ R_{ij}(\al,\bt)=\mbox{Id},
\end{equation}
where $\mbox{Id}$ is the identity map.  
\end{prop}

The reversibility property \eqref{YBMreversibility} is depicted in Figure \ref{fig-reverse}.

\begin{figure}[htb!]
\centering
\begin{tikzpicture}[scale=1.3]

\begin{scope}[xshift=-120,rotate=-45]

\draw[-] (-0.8,-0.8)--(0.8,-0.8)--(0.8,0.8)--(-0.8,0.8)--(-0.8,-0.8);

\draw[-Latex] (-0.4,-0.4)--(0.4,0.4);

\filldraw[fill=black,draw=black] (0,-0.85) circle (0.01pt)
node[left=1pt]{\color{black}\small $(\bxi_j, \bt)$};
\filldraw[fill=black,draw=black] (0,0.85) circle (0.01pt)
node[right=1pt]{\color{black}\small $( \bxi_j',\bt)$};
\filldraw[fill=black,draw=black] (0.85,0) circle (0.01pt)
node[right=1pt]{\color{black}\small $(\bxi_i',\al)$};
\filldraw[fill=black,draw=black] (-0.85,0) circle (0.01pt)
node[left=1pt]{\color{black}\small $(\bxi_i,\al)$};


\end{scope}

\draw[->] (-2.5,0.6)--(-1.7,0.6);
\draw[->] (-2.5,-0.6)--(-1.7,-0.6);


\begin{scope}[rotate=45]

\draw[-] (-0.8,-0.8)--(0.8,-0.8)--(0.8,0.8)--(-0.8,0.8)--(-0.8,-0.8);

\draw[-Latex] (-0.4,0.4)--(0.4,-0.4);

\filldraw[fill=black,draw=black] (0,-0.85) circle (0.01pt)
node[right=1pt]{\color{black}\small $(\bxi_j'', \bt)$};
\filldraw[fill=black,draw=black] (0,0.85) circle (0.01pt)
node[left=1pt]{\color{black}\small $( \bxi_j',\bt)$};
\filldraw[fill=black,draw=black] (0.85,0) circle (0.01pt)
node[right=1pt]{\color{black}\small $(\bxi_i'',\al)$};
\filldraw[fill=black,draw=black] (-0.85,0) circle (0.01pt)
node[left=1pt]{\color{black}\small $(\bxi_i',\al)$};

\end{scope}


\end{tikzpicture}
\caption{The reversibility property \eqref{YBMreversibility} for the Yang-Baxter map on the left-hand side of Figure \ref{fig-YBMs}.  This property requires that $\bxi_i''=\bxi_i$ and $\bxi_j''=\bxi_j$.}
\label{fig-reverse}
\end{figure}

\begin{proof}

In terms of variables $\bxi_i=(y_i,z_i)$, $\bxi_j=(z_i,z_j)$, the left-hand side of \eqref{YBMreversibility} defines a map 
\begin{equation}
    R_{ji}\circ R_{ij}\colon (\bxi_i,\bxi_j)\mapsto(\bxi_i',\bxi_j')\mapsto(\bxi_i'',\bxi_j'').
\end{equation}
Thus \eqref{YBMreversibility} will hold if $\bxi_i''=\bxi_i$ and $\bxi_j''=\bxi_j$.

First it will be shown that the first components of $\bxi_i''$ and $\bxi_i$ are equal.  By the definitions \eqref{RmattoYBM}, \eqref{Rmatderivs}, and \eqref{CRMAT}, the first component $y_i''$ of the variable $\bxi_i''$ is determined from a change of variables \eqref{RmattoYBMcov} of $x_i''$ that appears in the equation
the equation
\begin{equation}
\label{req1}
z_j'=\exp\biggl(\frac{\partial}{\partial x_j'}\bigr(\ol_{v_2-u_2}(x_j',x_i'')+\lag_{v_2-u_1}(x_j',x_i')\bigr)\biggr).
\end{equation}
On the other hand, by \eqref{RmattoYBM}, \eqref{Rmatderivs}, and \eqref{CRMAT}, the first component $y_i$ of $\bxi_i$ is related by the change of variables \eqref{RmattoYBMcov} of $x_i$ which satisfies the equation 
\begin{equation}
\label{req2}
z_j'= \exp\biggl(-\frac{\partial}{\partial x_j'}\bigr(\ol_{u_2-v_2}(x_i,x_j')+\lag_{u_1-v_2}(x_j',x_i')\bigr)\biggr).
\end{equation}
Then equating \eqref{req1} and \eqref{req2} and using the relations \eqref{lagsyms} 
it follows that $x_i''=x_i$, implying that $y_i''=y_i$, {\it i.e.}, the first components of $\bxi_i''$ and $\bxi_i$ are equal. By symmetry, the first components of $\bxi_j''$ and  $\bxi_j$ will also be equal.

Then it remains to show that the second components of $\bxi_i''$ and $\bxi_j''$, are equal to the second components of $\bxi_i$ and $\bxi_j$, respectively.  By the definition \eqref{RmattoYBM}, \eqref{Rmatderivs}, and \eqref{CRMAT}, the second component $z_i''$ of the variable $\bxi_i''$ is up to the change of variables \eqref{RmattoYBMcov} given by
\begin{equation}
\label{req3}
z_i''=\exp\biggl(-\frac{\partial}{\partial x_i''}\bigl(\lag_{v_2-u_1}(x_j'',x_i'')+\ol_{v_2-u_2}(x_j',x_i'')\bigr)\biggr).
\end{equation}
On the other hand, by \eqref{RmattoYBM}, \eqref{Rmatderivs}, and \eqref{CRMAT}, the second component $z_i$ of the variable $\bxi_i$ is given by 
\begin{equation}
\label{req4}
z_i= \exp\biggl( \frac{\partial}{\partial x_i}\bigl(\lag_{u_1-v_2}(x_i,x_j)+\ol_{u_2-v_2}(x_i,x_j')\bigr)\biggr).
\end{equation}
By the relations \eqref{lagsyms}, and also using $x_j=x_j''$ (as was determined above), the equation \eqref{req4} is found to be equivalent to the equation \eqref{req3}.  This means that $z_i''=z_i$,  {\it i.e.}, the second components of $\bxi_i''$ and $\bxi_i$ are equal.  By symmetry, the second components of $\bxi_j''$ and $\bxi_j$ will also be equal.  

\end{proof}

\begin{remark}
The functions $\lam_{u-v}(x_i,x_j)$ and $\olam_{u-v}(x_i,x_j)$ are not assumed to satisfy analogues of the anti-symmetry relations \eqref{lagsyms}, and thus the reversibility property \eqref{YBMreversibility} can not be expected to hold for the Yang-Baxter map $U(\al,\bt)$ in \eqref{finalYBM2}.
\end{remark}

The main result of this section is the following relation between the CYBE \eqref{CYBEdef}--\eqref{cuboc12} and the FYBE's \eqref{MYBE} and \eqref{MAYBE}.

\begin{theorem}\label{thm:maintheorem}
The maps \eqref{finalYBM1} and \eqref{finalYBM2} satisfy the FYBE's \eqref{MYBE} and \eqref{MAYBE}
if and only if the classical R-matrices \eqref{CRMAT} and \eqref{CUMAT1} are type-(1a) and type-(2a) solutions of the CYBE \eqref{CYBEdef}--\eqref{cuboc12}.
\end{theorem}

\begin{proof}

Only the details for the case of the FYBE \eqref{MAYBE} will be considered (corresponding to a type-(2a) solution of the CYBE), since \eqref{MYBE} follows from the case of \eqref{MAYBE} when all Yang-Baxter maps are of the same type (corresponding to a type-(1a) solution of the CYBE).

For the first part of the proof it will be shown that if the FYBE \eqref{MAYBE} is satisfied by the maps \eqref{finalYBM1} and \eqref{finalYBM2}, then this implies that exponentials of the equations for the twelve partial derivatives \eqref{cuboc12} and \eqref{cuboc34} are satisfied by the classical R-matrices \eqref{CRMAT} and \eqref{CUMAT1}.  
This in turn implies that the CYBE holds up to a term of the form \eqref{addtermsYBE}.

Consider the two mappings
\begin{align}\label{mapcomp1}
U_{jk}(\bt,\gm)\circ U_{ik}(\al,\gm)\circ R_{ij}(\al,\bt)\colon & (\bxi_i,\bxi_j,\bxi_k)\mapsto(\bxi'^{(l)}_i,\bxi'^{(l)}_j,\bxi'^{(l)}_k)\mapsto(\bxi''^{(l)}_i,\bxi''^{(l)}_j,\bxi''^{(l)}_k),
\\ \label{mapcomp2}
R_{ij}(\al,\bt)\circ U_{ik}(\al,\gm)\circ U_{jk}(\bt,\gm)\colon & (\bxi_i,\bxi_j,\bxi_k)\mapsto(\bxi'^{(r)}_i,\bxi'^{(r)}_j,\bxi'^{(r)}_k)\mapsto(\bxi''^{(r)}_i,\bxi''^{(r)}_j,\bxi''^{(r)}_k).
\end{align}
where the variables are 
\begin{equation}\label{FYBEcomponents}
\bxi_J=(y_J,z_J),\; \bxi'^{(l)}_J=(y'^{(l)}_J,z'^{(l)}_J),\; \bxi''^{(l)}_J=(y''^{(l)}_J,z''^{(l)}_J),\; \bxi'^{(r)}_J=(y'^{(r)}_J,z'^{(r)}_J),\; \bxi''^{(r)}_J=(y''^{(r)}_J,z''^{(r)}_J),
\end{equation}
for $J\in\{i,j,k\}$.

Assume now that the FYBE \eqref{MAYBE} is satisfied such that \eqref{mapcomp1} and \eqref{mapcomp2} are equivalent, and let three variables $\bxi''_i,\bxi''_j,\bxi''_k$ be defined in terms of the variables on the right hand side as
\begin{equation}\label{finalvariabledef}
\bxi''_i=\bxi''^{(l)}_i=\bxi''^{(r)}_i,\quad \bxi''_j=\bxi''^{(l)}_j=\bxi''^{(r)}_j,\quad \bxi''_k=\bxi''^{(l)}_k=\bxi''^{(r)}_k.
\end{equation}
The twelve partial derivatives \eqref{cuboc12} and \eqref{cuboc34} are simply a consequence of the equivalence of initial variables $\bxi_i,\bxi_j,\bxi_k$, and final variables $\bxi''_i,\bxi''_j,\bxi''_k$, on both sides of the FYBE, as well as the definition of the compositions of maps \eqref{mapcomp1} and \eqref{mapcomp2}, which can be seen as follows.

The second components $z_i,z_j,z_k$ of $\bxi_i,\bxi_j,\bxi_k$, are defined differently on the left and right-hand sides of the FYBE according to \eqref{RmattoYBM}, \eqref{UmattoYBM}, \eqref{RmattoYBMcov}, and \eqref{UmattoYBMcov}.  Since the initial variables are the same on both sides of the FYBE (regardless of whether the FYBE is satisfied or not), these different definitions of $z_i,z_j,z_k$ should be equivalent, which implies the following three equalities
\begin{equation}\label{cubocequivfirst}
\begin{split}
\exp\biggl(\frac{\partial}{\partial x_i}R_{\bu\bv}(x_i,x_j,x'^{(l)}_i,x'^{(l)}_j)\biggr)
&=\exp\biggl(\frac{\partial}{\partial x_i}U_{\bu\bw}(x_i,x'^{(r)}_k,x'^{(r)}_i,x''_k)\biggr), \\
\exp\biggl(\frac{\partial}{\partial x_j}R_{\bu\bv}(x_i,x_j,x'^{(l)}_i,x'^{(l)}_j)\biggr)
&=\exp\biggl(\frac{\partial}{\partial x_j}U_{\bv\bw}(x_j,x_k,x'^{(r)}_j,x'^{(r)}_k)\biggr), \\
\exp\biggl(\frac{\partial}{\partial x_k}U_{\bu\bw}(x'^{(l)}_i,x_k,x''_i,x'^{(l)}_k)\biggr)
&=\exp\biggl(\frac{\partial}{\partial x_k}U_{\bv\bw}(x_j,x_k,x'^{(r)}_j,x'^{(r)}_k)\biggr),
\end{split}
\end{equation}
where according to \eqref{RmattoYBMcov} and \eqref{UmattoYBMcov} the variables in \eqref{cubocequivfirst} are related to the first components of the variables \eqref{FYBEcomponents} as
\begin{equation}\label{proofcov}
\begin{gathered}
y_J=f(x_J),\;y'^{(l)}_J=f(x'^{(l)}_J),\;y'^{(r)}_J=f(x'^{(r)}_J),\;y''_J=f(x''_J),\quad J\in\{i,j\}, \\
y_k=g(x_k),\;y'^{(l)}_k=g(x'^{(r)}_k),\;y'^{(r)}_k=g(x'^{(r)}_k),\;y''_J=g(x''_k).
\end{gathered}
\end{equation}
Equations \eqref{cubocequivfirst} are seen to be equivalent (mod $2\pi\ii$) to three equations for the partial derivatives given in \eqref{cuboc34} after identifying the following variables 
\begin{equation}\label{cubocidentification}
(x'^{(l)}_i,x'^{(l)}_j,x'^{(l)}_k,x'^{(r)}_i,x'^{(r)}_j,x'^{(r)}_k)=(\xli,\xlj,\xlk,\xri,\xrj,\xrk),
\end{equation}
where the six variables on the right hand side are those that are used for the partial derivatives of the CYBE.

Similarly, the second components $z''_i,z''_j,z''_k$ of $\bxi''_i,\bxi''_j,\bxi''_k$, are defined differently on the left and right-hand sides of the FYBE according to \eqref{RmattoYBM}, \eqref{UmattoYBM}, \eqref{RmattoYBMcov}, and \eqref{UmattoYBMcov}.  Since the final variables are the same on both sides of the FYBE (by the assumption that FYBE is satisfed), these different definitions of $z''_i,z''_j,z''_k$ should be equivalent, which implies the following three equalities

\begin{equation}\label{cubocequiv2}
\begin{split}
\exp\biggl(\frac{\partial}{\partial x''_i}U_{\bu\bw}(x'^{(l)}_i,x_k,x''_i,x'^{(l)}_k)\biggr)
&=\exp\biggl(\frac{\partial}{\partial x''_i}R_{\bu\bv}(x'^{(r)}_i,x'^{(r)}_j,x''_i,x''_j)\biggr), \\
\exp\biggl(\frac{\partial}{\partial x''_j}U_{\bv\bw}(x'^{(l)}_j,x'^{(l)}_k,x''_j,x''_k)\biggr)
&=\exp\biggl(\frac{\partial}{\partial x''_j}R_{\bu\bv}(x'^{(r)}_i,x'^{(r)}_j,x''_i,x''_j)\biggr), \\
\exp\biggl(\frac{\partial}{\partial x''_k}U_{\bv\bw}(x'^{(l)}_j,x'^{(l)}_k,x''_j,x''_k)\biggr)
&=\exp\biggl(\frac{\partial}{\partial x''_k}U_{\bu\bw}(x_i,x'^{(r)}_k,x'^{(r)}_i,x''_k)\biggr),
\end{split}
\end{equation}
where according to \eqref{RmattoYBMcov} and \eqref{UmattoYBMcov} the variables in \eqref{cubocequiv2} are related to the first components of the variables \eqref{FYBEcomponents} as given in \eqref{proofcov}.  Equations \eqref{cubocequiv2} are seen to be equivalent (mod $2\pi\ii$) to the remaining three equations for the partial derivatives given in \eqref{cuboc34} after using the identification given in \eqref{cubocidentification}.

It remains to account for the six equations for the partial derivatives in \eqref{cuboc12}.  Consider the composition of maps \eqref{mapcomp1} for the left-hand side of the FYBE \eqref{MAYBE}.  The map $R_{ij}(\al,\bt)\colon (\bxi_i,\bxi_j)\mapsto(\bxi'^{(l)}_i,\bxi'^{(l)}_j)$ determines $\bxi'^{(l)}_i$ and $\bxi'^{(l)}_j$, which are subsequently mapped by $U_{ik}(\al,\gm)$ and $U_{jk}(\bt,\gm)$, respectively.  From the definitions \eqref{RmattoYBM} and \eqref{UmattoYBM}, the variables $\bxi'^{(l)}_i$ and $\bxi'^{(l)}_j$ that come from the map $R_{ij}(\al,\bt)$ are defined differently from the corresponding variables that are subsequently mapped by $U_{ik}(\al,\gm)$ and $U_{jk}(\bt,\gm)$.  However, the act of composition of the respective maps requires that the two different definitions are equated, implying the following two equations
\begin{equation}\label{cubocequiv3a}
 \begin{split}
\exp\biggl(-\frac{\partial}{\partial x'^{(l)}_i}R_{\bu\bv}(x_i,x_j,x'^{(l)}_i,x'^{(l)}_j)\biggr)
&=\exp\biggl(\frac{\partial}{\partial x'^{(l)}_i}U_{\bu\bw}(x'^{(l)}_i,x_k,x''_i,x'^{(l)}_k)\biggr), \\
\exp\biggl(-\frac{\partial}{\partial x'^{(l)}_j}R_{\bu\bv}(x_i,x_j,x'^{(l)}_i,x'^{(l)}_j)\biggr)
&=\exp\biggl(\frac{\partial}{\partial x'^{(l)}_j}U_{\bv\bw}(x'^{(l)}_j,x'^{(l)}_k,x''_j,x''_k)\biggr),
 \end{split}
\end{equation}
where the variables of \eqref{cubocequiv3a} are related to the first components of the variables \eqref{FYBEcomponents} as given in \eqref{proofcov}.

Similarly, the map $U_{ik}(\al,\gm)\colon (\bxi'^{(l)}_i,\bxi_k)\mapsto(\bxi''^{(l)}_i,\bxi'^{(l)}_k)$ determines $\bxi'^{(l)}_k$, which is subsequently mapped by $U_{jk}(\bt,\gm)$.  By the definition of the second component of $\bxi'^{(l)}_k$, the composition of the two maps requires that the following equation is satisfied
\begin{equation}\label{cubocequiv3b}
\begin{split}
\exp\biggl(-\frac{\partial}{\partial x'^{(l)}_k}U_{\bu\bw}(x'^{(l)}_i,x_k,x''_i,x'^{(l)}_k)\biggr)
&=\exp\biggl(\frac{\partial}{\partial x'^{(l)}_k}U_{\bv\bw}(x'^{(l)}_j,x'^{(l)}_k,x''_j,x''_k)\biggr).
\end{split}
\end{equation}
The three equations \eqref{cubocequiv3a}--\eqref{cubocequiv3b} are seen to be equivalent (mod $2\pi\ii$) to three equations for the partial derivatives in \eqref{cuboc12} after using the identification \eqref{cubocidentification}.

In an analogous way, the composition of maps \eqref{mapcomp2} for the right-hand side of the FYBE \eqref{MAYBE} can be shown to imply the following three equations
\begin{equation}\label{cubocequivlast}
\begin{split}
\exp\biggl(-\frac{\partial}{\partial x'^{(r)}_i}U_{\bu\bw}(x_i,x'^{(r)}_k,x'^{(r)}_i,x''_k)\biggr)
&=\exp\biggl(\frac{\partial}{\partial x'^{(r)}_i}R_{\bu\bv}(x'^{(r)}_i,x'^{(r)}_j,x''_i,x''_j)\biggr), \\
\exp\biggl(-\frac{\partial}{\partial x'^{(r)}_j}U_{\bv\bw}(x_j,x_k,x'^{(r)}_j,x'^{(r)}_k)\biggr)
&=\exp\biggl(\frac{\partial}{\partial x'^{(r)}_j}R_{\bu\bv}(x'^{(r)}_i,x'^{(r)}_j,x''_i,x''_j)\biggr), \\
\exp\biggl(-\frac{\partial}{\partial x'^{(r)}_k}U_{\bv\bw}(x_j,x_k,x'^{(r)}_j,x'^{(r)}_k)\biggr)
&=\exp\biggl(\frac{\partial}{\partial x'^{(r)}_k}U_{\bu\bw}(x_i,x'^{(r)}_k,x'^{(r)}_i,x''_k)\biggr),
\end{split}
\end{equation}
which are equivalent (mod $2\pi\ii$) to the remaining equations for the three partial derivatives in \eqref{cuboc12}.  Then since the twelve equations for the partial derivatives with respect to the twelve variables of the CYBE in \eqref{cuboc12} and \eqref{cuboc34} are satisfied in the forms \eqref{cubocequivfirst}--\eqref{cubocequivlast}, it follows that the classical R-matrices \eqref{CRMAT} and \eqref{CUMAT1} are solutions to the CYBE \eqref{CYBEdef} up to the additional terms of the form \eqref{addtermsYBE}.

\medskip

For the second part of the proof it will be shown if the classical R-matrices \eqref{CRMAT} and \eqref{CUMAT1} are a type-(2a) solution of the CYBE \eqref{CYBEdef}--\eqref{cuboc12}, then this implies that the FYBE \eqref{MAYBE} is satisfied by the maps \eqref{finalYBM1} and \eqref{finalYBM2}.

Since the FYBE \eqref{MAYBE} is no longer assumed to be satisfied, the equalities \eqref{finalvariabledef} are no longer assumed and the variables $\bxi''_i,\bxi''_j,\bxi''_k$ will no longer be used.  It was shown above for \eqref{cubocequiv3a} and \eqref{cubocequiv3b} that the composition of maps \eqref{mapcomp1} for the left-hand side of the FYBE \eqref{MAYBE} implies the following equalities
\begin{equation}\label{comp0}
 \begin{split}
\exp\biggl(-\frac{\partial}{\partial x'^{(l)}_i}R_{\bu\bv}(x_i,x_j,x'^{(l)}_i,x'^{(l)}_j)\biggr)
&=\exp\biggl(\frac{\partial}{\partial x'^{(l)}_i}U_{\bu\bw}(x'^{(l)}_i,x_k,x''^{(l)}_i,x'^{(l)}_k)\biggr), \\
\exp\biggl(-\frac{\partial}{\partial x'^{(l)}_j}R_{\bu\bv}(x_i,x_j,x'^{(l)}_i,x'^{(l)}_j)\biggr)
&=\exp\biggl(\frac{\partial}{\partial x'^{(l)}_j}U_{\bv\bw}(x'^{(l)}_j,x'^{(l)}_k,x''^{(l)}_j,x''^{(l)}_k)\biggr), \\
\exp\biggl(-\frac{\partial}{\partial x'^{(l)}_k}U_{\bu\bw}(x'^{(l)}_i,x_k,x''^{(l)}_i,x'^{(l)}_k)\biggr)
&=\exp\biggl(\frac{\partial}{\partial x'^{(l)}_k}U_{\bv\bw}(x'^{(l)}_j,x'^{(l)}_k,x''^{(l)}_j,x''^{(l)}_k)\biggr),
\end{split}
\end{equation}
where according to \eqref{RmattoYBMcov} and \eqref{UmattoYBMcov} the variables in \eqref{comp0} are related to the first components of the variables \eqref{FYBEcomponents} as
\begin{equation}\label{proofcov2}
\begin{gathered}
y_J=f(x_J),\;y'^{(l)}_J=f(x'^{(l)}_J),\;y'^{(r)}_J=f(x'^{(r)}_J),\;y''^{(l)}_J=f(x''^{(l)}_J),\;y''^{(r)}_J=f(x''^{(r)}_J),\quad J\in\{i,j\}, \\
y_k=g(x_k),\;y'^{(l)}_k=g(x'^{(r)}_k),\;y'^{(r)}_k=g(x'^{(r)}_k),\;y''^{(l)}_J=g(x''^{(l)}_k),\;y''^{(r)}_J=g(x''^{(r)}_k).
\end{gathered}
\end{equation}
Equations \eqref{comp0} are seen to be equivalent to three equations for the  partial derivatives in \eqref{cuboc12} if the following variables are identified
\begin{equation}\label{cubocidentification2}
(x'^{(l)}_i,x'^{(l)}_j,x'^{(l)}_k,x''^{(l)}_i,x''^{(l)}_j,x''^{(l)}_k)=(\xli,\xlj,\xlk,x''_i,x''_j,x''_k),
\end{equation}
where the six variables on the right are those that are used for the partial derivatives of the CYBE.  Note that the equations \eqref{comp0} hold regardless of whether the FYBE \eqref{MAYBE} is satisfied or not, as it is simply a consequence of the definition of the maps and their composition.

Assume that the two classical R-matrices $R_{\bu\bv}$ and $U_{\bu\bv}$ involved in the equations \eqref{comp0} are a solution to the CYBE \eqref{CYBEdef}--\eqref{cuboc12}.  Then the remaining nine partial derivatives in \eqref{cuboc12} and \eqref{cuboc34} that are implied by this CYBE may be used to show that the FYBE \eqref{MAYBE} is satisfied as follows.

The first map $U_{jk}(\bt,\gm)\colon (\bxi_j,\bxi_k)\mapsto(\bxi'^{(r)}_j,\bxi'^{(r)}_k)$ in \eqref{mapcomp2} for the RHS of the FYBE \eqref{MAYBE} determines $\bxi'^{(r)}_j$ and $\bxi'^{(r)}_k$.  By the definition \eqref{UmattoYBM}, the first components $y'^{(r)}_j$ and $y'^{(r)}_k$ of these variables are respectively related through \eqref{proofcov2} to two variables $x'^{(r)}_j$ and $x'^{(r)}_k$ which satisfy
\begin{equation}
\label{comp3}
\begin{gathered}
z_k=\exp\biggl(\frac{\partial}{\partial x_k}U_{\bv\bw}(x_j,x_k,x'^{(r)}_j,x'^{(r)}_k)\biggr), \quad
z_j=\exp\biggl(\frac{\partial}{\partial x_j}U_{\bv\bw}(x_j,x_k,x'^{(r)}_j,x'^{(r)}_k)\biggr),
\end{gathered}
\end{equation}
where the $z_k$ and $z_j$ are respectively the second components of $\bxi_k$ and $\bxi_j$ (note that $z_k$ is independent of $x'^{(r)}_k$ and that $z_j$ is independent of $x'^{(r)}_j$).  On the other hand, the definitions of $z_k$ and $z_j$ in \eqref{mapcomp1} for the LHS of the FYBE \eqref{MAYBE} are different, and are given by
\begin{equation}
\label{comp4}
\begin{gathered}
z_k=\exp\biggl(\frac{\partial}{\partial x_k}U_{\bu\bw}(x'^{(l)}_i,x_k,x''^{(l)}_i,x'^{(l)}_k)\biggr), \quad
z_j=\exp\biggl(\frac{\partial}{\partial x_j}R_{\bu\bv}(x_i,x_j,x'^{(l)}_i,x'^{(l)}_j)\biggr).
\end{gathered}
\end{equation}
Then after equating \eqref{comp3} and \eqref{comp4}, and recalling \eqref{cubocidentification2}, it is seen that $x'^{(r)}_j$ and $x'^{(r)}_k$ are respectively solutions (mod $2\pi\ii$) to the equations for the partial derivatives with respect to $x_k$ and $x_j$ given in \eqref{cuboc34}. This means that the variables $\xrj$ and $\xrk$ of the CYBE may be respectively identified with the variables $x'^{(r)}_j$ and $x'^{(r)}_k$ in \eqref{proofcov2}.

Next, the map $U_{ik}(\al,\gm)\colon(\bxi_i,\bxi'^{(r)}_k)\mapsto (\bxi'^{(r)}_i,\bxi''^{(r)}_k)$ in \eqref{mapcomp2} for the RHS of the FYBE \eqref{MAYBE} determines $\bxi'^{(r)}_i$ and $\bxi''^{(r)}_k$.  By the definition \eqref{UmattoYBM}, the first components $y'^{(r)}_k$ and $y''^{(r)}_k$ of these variables are respectively related through \eqref{proofcov2} to two variables $x'^{(r)}_i$ and $x''^{(r)}_k$ which satisfy
\begin{equation}
\label{comp5}
\begin{gathered}
z'^{(r)}_k=\exp\biggl(\frac{\partial}{\partial x'^{(r)}_k}U_{\bu\bw}(x_i,x'^{(r)}_k,x'^{(r)}_i,x''^{(r)}_k)\biggr), \quad
z_i=\exp\biggl(\frac{\partial}{\partial x_i}U_{\bu\bw}(x_i,x'^{(r)}_k,x'^{(r)}_i,x''^{(r)}_k)\biggr),
\end{gathered}
\end{equation}
where the $z'^{(r)}_k$ and $z_i$ are respectively the second components of $\bxi'^{(r)}_k$ and $\bxi_i$ (note that $z'^{(r)}_k$ is independent of $x''^{(r)}_k$ and that $z_i$ is independent of $x'^{(r)}_i$).  On the other hand, the map $U_{jk}(\bt,\gm)$ in \eqref{mapcomp2} for the RHS of the FYBE \eqref{MAYBE}, and the map $R_{ij}(\al,\bt)$ in \eqref{mapcomp1} for the LHS of the FYBE \eqref{MAYBE}, provide different expressions for $z'^{(r)}_k$ and $z_i$ respectively, which are given by
\begin{equation}
\label{comp6}
\begin{gathered}
z'^{(r)}_k=\exp\biggl(-\frac{\partial}{\partial x'^{(r)}_k}U_{\bv\bw}(x_j,x_k,x'^{(r)}_j,x'^{(r)}_k)\biggr), \quad
z_i=\exp\biggl(\frac{\partial}{\partial x_i}R_{\bu\bv}(x_i,x_j,x'^{(l)}_i,x'^{(l)}_j)\biggr).
\end{gathered}
\end{equation}
Then after equating \eqref{comp5} and \eqref{comp6}, and using $\xrj=x'^{(r)}_j$ and $\xrk=x'^{(r)}_k$ (shown in the previous step), it is seen that $x'^{(r)}_i$ and $x''^{(r)}_k$ are respectively solutions (mod $2\pi\ii$) to the equation for the partial derivative with respect to $x^{(r)}_k$ given in \eqref{cuboc12} and the equation for the partial derivative with respect to $x_i$ given in \eqref{cuboc34} (recalling also \eqref{cubocidentification2}). This means that the variables $\xri$ and $x''_k$ of the CYBE may be respectively identified with the variables $x'^{(r)}_i$ and $x''^{(r)}_k$ in \eqref{proofcov2}.  
Since the variable $x''_k$ of the CYBE was previously identified with $x''^{(l)}_k$ in \eqref{cubocidentification2}, this in turn implies that $x''^{(l)}_k=x''^{(r)}_k$, and thus $y''^{(l)}_k=y''^{(r)}_k$ through the relations \eqref{proofcov2}.

Finally, the map $R_{ij}(\al,\bt)\colon(\bxi'^{(r)}_i,\bxi'^{(r)}_j)\mapsto(\bxi''^{(r)}_i,\bxi''^{(r)}_j)$ in \eqref{mapcomp2} for the RHS of the FYBE \eqref{MAYBE} determines $\bxi''^{(r)}_i$ and $\bxi''^{(r)}_j$.  By the definition \eqref{RmattoYBM} and \eqref{UmattoYBM}, the first components of these variables $y''^{(r)}_i$ and $y''^{(r)}_j$ are respectively related through \eqref{proofcov2} to two variables $x''^{(r)}_i$ and $x''^{(r)}_j$ which satisfy
\begin{equation}
\label{comp7}
\begin{gathered}
z'^{(r)}_j=\exp\biggl(\frac{\partial}{\partial x'^{(r)}_j}R_{\bu\bv}(x'^{(r)}_i,x'^{(r)}_j,x''^{(r)}_i,x''^{(r)}_j)\biggr), \quad
z'^{(r)}_i=\exp\biggl(\frac{\partial}{\partial x'^{(r)}_i}R_{\bu\bv}(x'^{(r)}_i,x'^{(r)}_j,x''^{(r)}_i,x''^{(r)}_j)\biggr),
\end{gathered}
\end{equation}
where the $z'^{(r)}_j$ and $z'^{(r)}_i$ are respectively the second components of $\bxi'^{(r)}_j$ and $\bxi'^{(r)}_i$ (note that $z'^{(r)}_j$ is independent of $x''^{(r)}_j$ and that $z'^{(r)}_i$ is independent of $x''^{(r)}_i$).  On the other hand, the latter variables were first mapped by $U_{jk}(\bt,\gm)$ and $U_{ik}(\al,\gm)$, where they were defined through the different equations
\begin{equation}
\label{comp8}
\begin{gathered}
z'^{(r)}_j=\exp\biggl(-\frac{\partial}{\partial x'^{(r)}_j}U_{\bv\bw}(x_j,x_k,x'^{(r)}_j,x'^{(r)}_k)\biggr), \quad
z'^{(r)}_i=\exp\biggl(-\frac{\partial}{\partial x'^{(r)}_i}U_{\bu\bw}(x_i,x'^{(r)}_k,x'^{(r)}_i,x''^{(r)}_k)\biggr).
\end{gathered}
\end{equation}
Then after equating \eqref{comp7} and \eqref{comp8}, and using $\xrj=x'^{(r)}_j$, $\xrk=x'^{(r)}_k$, $\xri=x'^{(r)}_i$, and $x''^{(l)}_k=x''^{(r)}_k$ (shown in the previous two steps), it is seen that $x''^{(r)}_i$ and $x''^{(r)}_j$ are respectively solutions (mod $2\pi\ii$) to the equations for the partial derivatives with respect to $x^{(r)}_j$ and $x^{(r)}_i$ given in \eqref{cuboc12}. This means that the variables $x'_i$ and $x'_j$ of the CYBE may be respectively identified with the variables $x''^{(r)}_i$ and $x''^{(r)}_j$ in \eqref{proofcov2}.  
Since the variables $x'_i$ and $x'_j$ of the CYBE were previously respectively identified with the variables $x''^{(l)}_i$ and $x''^{(l)}_j$ in \eqref{cubocidentification2}, this in turn implies that $x''^{(l)}_i=x''^{(r)}_i$ and $x''^{(l)}_j=x''^{(r)}_j$, and thus $y''^{(l)}_i=y''^{(r)}_i$ and $y''^{(l)}_j=y''^{(r)}_j$ through the relations \eqref{proofcov2}.

Thus it has been shown that for the FYBE \eqref{MAYBE} with the maps \eqref{finalYBM1} and \eqref{finalYBM2}, the first components $y''^{(l)}_i$, $y''^{(l)}_j$, $y''^{(l)}_k$ of the variables $\bxi''^{(l)}_i$, $\bxi''^{(l)}_j$, $\bxi''^{(l)}_k$, are equal to the first components $y''^{(r)}_i$, $y''^{(r)}_j$, $y''^{(r)}_k$ of the variables $\bxi''^{(r)}_i$, $\bxi''^{(r)}_j$, $\bxi''^{(r)}_k$, respectively.

It remains to show that the second components $z''^{(l)}_i$, $z''^{(l)}_j$, $z''^{(l)}_k$ of the variables $\bxi''^{(l)}_i$, $\bxi''^{(l)}_j$, $\bxi''^{(l)}_k$, are equal to the second components $z''^{(r)}_i$, $z''^{(r)}_j$, $z''^{(r)}_k$ of the variables $\bxi''^{(r)}_i$, $\bxi''^{(r)}_j$, $\bxi''^{(r)}_k$, respectively.

As was determined above, the first components of the variables involved on both sides of the FYBE \eqref{MAYBE} are related through a change of variables \eqref{RmattoYBMcov} and \eqref{UmattoYBMcov} to the variables of the CYBE \eqref{CYBEdef}.
This means that the three equations for the partial derivatives with respect to $x''_i$, $x''_j$, and $x''_k$, given in \eqref{cuboc34}, hold also for the corresponding variables of the FYBE.  Then using the definitions \eqref{RmattoYBM}, \eqref{UmattoYBM}, and \eqref{Rmatderivs}, \eqref{Umatderivs}, these three equations are seen to be equivalent to the following three equations
\begin{equation}
z''^{(l)}_i=z''^{(r)}_i, \qquad  z''^{(l)}_j=z''^{(r)}_j, \qquad z''^{(l)}_k=z''^{(r)}_k.
\end{equation}
Therefore $\bxi''^{(l)}_i=\bxi''^{(r)}_i$, $\bxi''^{(l)}_j=\bxi''^{(r)}_j$, $\bxi''^{(l)}_k=\bxi''^{(r)}_k$, so the maps \eqref{finalYBM1} and \eqref{finalYBM2} satisfy the FYBE \eqref{MAYBE}.  

\end{proof}

There are some simple transformations of the Yang-Baxter maps \eqref{finalYBM1} and \eqref{finalYBM2} under which the FYBE's \eqref{MYBE} and \eqref{MAYBE} remain satisfied.  First, the FYBE's will still hold if the opposite signs are chosen for each of the expressions appearing on the right-hand sides of \eqref{Rmatderivs} and \eqref{Umatderivs}. 
According to \eqref{RmattoYBM} and \eqref{UmattoYBM}, this corresponds to setting  $\mui\to(\mui)^{-1}$, $\muj\to(\muj)^{-1}$, $\mupi\to(\mupi)^{-1}$, $\mupj\to(\mupj)^{-1}$, in the expressions for the Yang-Baxter maps \eqref{finalYBM1} and \eqref{finalYBM2}, respectively (or $\mui\to-\mui$, $\muj\to-\muj$, $\mupi\to-\mupi$, $\mupj\to-\mupj$, for degenerate additive cases).  The computations for the proof of Theorem \ref{thm:maintheorem} are invariant under this change in sign.


The second transformation is less trivial.  Let $\phi_{u,v}(x)$ be a function satisfying
\begin{equation}
\phi_{u,v}(x)+\phi_{-u,-v}(x)=0.
\end{equation}
Theorem \ref{thm:maintheorem} will still hold under the following gauge transformations for \eqref{Rmatderivs}
\begin{equation}\label{gauge1}
\begin{split}
R_{\bu\bv,i}(x_j',x_j;x_i)\to & R_{\bu\bv,i}(x_j',x_j;x_i) + \phi_{\rua,\rub}(x_i), \\
R_{\bu\bv,i'}(x_j,x_j';x_i') \to & R_{\bu\bv,i'}(x_j,x_j';x_i') + \phi_{\rub,\rua}(x_i'), \\
R_{\bu\bv,j}(x_i',x_i;x_j)\to & R_{\bu\bv,j}(x_i',x_i;x_j) + \phi_{\rva,\rvb}(x_j), \\
R_{\bu\bv,j'}(x_i,x_i';x_j')\to & R_{\bu\bv,j'}(x_i,x_i';x_j') +\phi_{\rvb,\rva}(x_j'),
\end{split}
\end{equation}
with identical transformations applied to $U_{\bu\bv,i}(x_j',x_j;x_i)$, $U_{\bu\bv,i'}(x_j,x_j';x_i')$, $U_{\bu\bv,j}(x_i',x_i;x_j)$, and $U_{\bu\bv,j'}(x_i,x_i';x_j')$, respectively.

According to \eqref{RmattoYBM} and \eqref{UmattoYBM}, this is equivalent to a change in the variables for the FYBE's \eqref{MYBE} and \eqref{MAYBE} of the form
\begin{equation}\label{gauge2}
\begin{gathered}
\mui\to \mui\Phi_{\mpa,\mpb}(\mxi),\qquad \mupi\to \mupi\Phi_{\mpb,\mpa}(\myj), \\
\muj\to \muj\Phi_{\mqa,\mqb}(\mxj),\qquad \mupj\to \mupj\Phi_{\mqb,\mqa}(\myi),
\end{gathered}
\end{equation}
where $\Phi_{\alpha,\beta}(x)$ satisfies
\begin{equation}
\Phi_{\alpha,\beta}(x)\Phi_{-\alpha,-\beta}(x)=1.
\end{equation}


Note that for the degenerate algebraic (additive) cases the changes would instead be of the form
\begin{equation}\label{gauge3}
\begin{gathered}
\mui\to \mui+\phi_{\mpa,\mpb}(\mxi),\qquad \mupi\to \mupi+\phi_{\mpb,\mpa}(\myj), \\
\muj\to \muj+\phi_{\mqa,\mqb}(\mxj),\qquad \mupj\to \mupj+\phi_{\mqb,\mqa}(\myi).
\end{gathered}
\end{equation}

\section{Classical star-triangle relations and Yang-Baxter maps}\label{sec:examples}

Using different solutions of the CSTRs \eqref{CSTRdef1}--\eqref{3legdef2}, the method introduced in the previous section will be used to derive explicit two-component Yang-Baxter maps.

Throughout this section the symbol $\pm$ will be used to indicate a sum of two terms involving $+$ and $-$ respectively, {\it e.g.},
\begin{equation}
\label{comnot}
\begin{gathered}
f(x\pm y)=f(x+y)+f(x-y), \\
f(\pm x \pm y)=f(x+y)+f(x-y)+f(-x+y)+f(-x-y).
\end{gathered}
\end{equation}

\subsection{Solutions to the CSTRs}
\label{sec:CSTRlist}

The solutions to the CSTRs in the elliptic and hyperbolic cases will be given in terms of the dilogarithm function, defined for $z\in\mathbb{C}\setminus [1,\infty)$ by
\begin{equation}\label{dilogdef}
\lie(z)=-\int^z_0\frac{\Log(1-t)}{t}dt.
\end{equation}
The solutions to the CSTRs in the rational and algebraic cases will be given in terms of a function $\gamma(z)$, which is defined in terms of the complex logarithm as
\begin{equation}
\label{gamdef}
\gamma(z)=\ii z\Log(\ii z),\qquad \bigl|\textrm{Arg}(\ii z)\bigr|<\pi.
\end{equation}

For the elliptic case the Weierstrass functions will also be used, where $\sigma(z)$ denotes the Weierstrass sigma function and $\zeta(z)$ denotes the Weierstrass zeta function, which both depend on the elliptic invariants $g_2,g_3$ or associated half-periods $\omega_1,\omega_2$ \cite{WW}.  These functions are related to each other and to the Weierstrass elliptic function $\wp(z)$, by 
\begin{equation}
\label{WeierstrassRels}
\frac{\partial}{\partial z}\Log\sigma(z)=\zeta(z),\qquad \frac{\partial}{\partial z}\zeta(z)=-\wp(z).
\end{equation}

\subsubsection{Elliptic case}

Elliptic solutions of the CSTRs may be derived from the quasi-classical limit of (quantum) star-triangle relations that are equivalent to elliptic hypergeometric integrals \cite{Bazhanov:2010kz,Kels:2017vbc}.  In these cases the solutions to the CSTRs arise from the leading order quasi-classical asymptotics of the elliptic gamma function \cite{Ruijsenaars:1997:FOA}, and may be written in terms of
\begin{equation}
\label{lagq4}
\begin{split}
\ds\mathcal{L}^{(1)}_{\alpha}(x_i,x_j)=&\frac{-\alpha\bigl((2x-\frac{\pi}{2})^2+(2y-\frac{\pi}{2})^2\bigr)}{\pi\tau}
-\phi(2\ii\alpha;2\tau)+\phi(\pm x_i\pm x_j-\alpha;\tau), \\
\ol^{(1)}_{\alpha}(x_i,x_j)=&\lag_{-\frac{\pi\tau}{2}-\alpha}(x_i,x_j),
\end{split}
\end{equation}
where $\tau$ is a complex parameter satisfying $\im(\tau)>0$, and 
\begin{equation}\label{q4asymptotics}
\phi(z)=\frac{1}{2}\sum_{j=0}^\infty\Bigl(\lie\bigl(-\EXP^{2\ii z}\EXP^{\pi\ii\tau(2j+1)}\bigr)-
                                              \lie\bigl(-\EXP^{-2\ii z}\EXP^{\pi\ii\tau(2j+1)}\bigr)\Bigr).
\end{equation}

\begin{prop}\label{prop:ellipticCSTR}
Let the six functions \eqref{CSTRsol} be defined in terms of \eqref{lagq4} as
\begin{equation}\label{elllagcombos}
\begin{split}
\lag_\alpha(x_i,x_j)=\lagh_\alpha(x_i,x_j)=\lam_\alpha(x_i,x_j)=\lag^{(1)}_\alpha(x_i,x_j), \\
\olh_\alpha(x_i,x_j)=\ol_\alpha(x_i,x_j)=\olam_\alpha(x_i,x_j)=\ol^{(1)}_{\alpha}(x_i,x_j).
\end{split}
\end{equation}
Then \eqref{CSTRsol} is a solution to the CSTRs \eqref{CSTRdef1}--\eqref{3legdef2} up to terms of the form \eqref{addtermsCSTR}.  
\end{prop}

\begin{proof}
The equations \eqref{CSTRdef1} and \eqref{CSTRdef2} defined with the combination of functions given in \eqref{elllagcombos} are equivalent, and thus only one of these relations needs to be shown.  Let $\mathcal{A}^{(STR)}_1(x_a,x_b,x_c,x_d;\,u,v,w)$ be the left-hand side of the CSTR \eqref{CSTRdef1} defined with \eqref{elllagcombos}.  If the exponential of the equation \eqref{3legdef1} and the following three equations 
\begin{equation}\label{4ellderivs}
\exp\Bigl(\ii\frac{\partial}{\partial x_I}\mathcal{A}^{(STR)}_1(x_a,x_b,x_c,x_d;\,u,v,w)\Bigr)=1,\qquad I\in\{a,b,c\},
\end{equation}
are each satisfied, then the CSTR \eqref{CSTRdef1} will be satisfied up to the terms of the form \eqref{addtermsCSTR} (since these equations effectively state that each of the exponentials of the partial derivatives of $A^{(STR)}$ with respect to the four variables $x_a,x_b,x_c,x_d$ vanish mod $2\pi\ii$).  The respective partial derivatives are explicitly given by
\begin{equation}\label{4ellderivsexplicit}
\begin{split}
\exp\Bigl(\ii\frac{\partial}{\partial x_d}\mathcal{A}^{(STR)}_1(x_a,x_b,x_c,x_d;\,u,v,w)\Bigr)=&
\frac{\psi(x_d,x_a,\alpha_1)\psi(x_d,x_c,\alpha_3)}{\psi(x_d,x_b-\tfrac{\pi\tau}{2},\alpha_1+\alpha_3)}, \\
\exp\Bigl(\ii\frac{\partial}{\partial x_a}\mathcal{A}^{(STR)}_1(x_a,x_b,x_c,x_d;\,u,v,w)\Bigr)=&
\frac{\psi(x_a,x_d,\alpha_1)\psi(x_a,x_b-\tfrac{\pi\tau}{2},\alpha_3)}{\psi(x_a,x_c,\alpha_1+\alpha_3)}, \\
\exp\Bigl(\ii\frac{\partial}{\partial x_b}\mathcal{A}^{(STR)}_1(x_a,x_b,x_c,x_d;\,u,v,w)\Bigr)=&
\frac{\psi(x_b-\tfrac{\pi\tau}{2},x_c,\alpha_1)\psi(x_b-\tfrac{\pi\tau}{2},x_a,\alpha_3)}{\psi(x_b-\tfrac{\pi\tau}{2},x_d,\alpha_1+\alpha_3)}, \\
\exp\Bigl(\ii\frac{\partial}{\partial x_c}\mathcal{A}^{(STR)}_1(x_a,x_b,x_c,x_d;\,u,v,w)\Bigr)=&
\frac{\psi(x_c,x_d,\alpha_3)\psi(x_c,x_b-\tfrac{\pi\tau}{2},\alpha_1)}{\psi(x_c,x_a,\alpha_1+\alpha_3)},
\end{split}
\end{equation}
where $\alpha_1=v-w$ and $\alpha_3=u-v$, and $\psi(x_i,x_j,\alpha)$ is defined in terms of $\sigma(z)$ by
\begin{equation}
\psi(x_i,x_j,\alpha)=
\frac{\sigma(\frac{2\omega_1(x_i+x_j+\alpha)}{\pi})\sigma(\frac{2\omega_1(x_i-x_j+\alpha)}{\pi})}
     {\sigma(\frac{2\omega_1(x_i+x_j-\alpha)}{\pi})\sigma(\frac{2\omega_1(x_i-x_j-\alpha)}{\pi})}.
\end{equation}
The Weierstrass half-periods implicitly involved in \eqref{4ellderivs} and \eqref{4ellderivsexplicit} are related to the parameter $\tau$ in \eqref{lagq4} by $\tau=\frac{\omega_2}{\omega_1}$.

From \eqref{4ellderivsexplicit}, the exponential of \eqref{3legdef1} and the three equations \eqref{4ellderivs} may be recognised as four equivalent versions of the three-leg form for the elliptic discrete integrable quad equation known as $Q4$ \cite{ABS,AdlerSurisQ4}.  More specifically, under the following change of variables
\begin{equation}\label{ellcov}
\begin{gathered}
  X_a=\wp\bigl(\tfrac{2\omega_1 x_a}{\pi}\bigr),\; X_b=\wp\bigl(\tfrac{2\omega_1 x_b}{\pi}-\omega_2\bigr),\; X_c=\wp\bigl(\tfrac{2\omega_1 x_c}{\pi}\bigr),\; X_d=\wp\bigl(\tfrac{2\omega_1 x_d}{\pi}\bigr), \\
  A_1=\wp\bigl(\tfrac{2\omega_1 \alpha_1}{\pi}\bigr),\; A_3=\wp\bigl(\tfrac{2\omega_1 \alpha_3}{\pi}\bigr),
  \end{gathered}
\end{equation}
the exponential of \eqref{3legdef1} and the three equations \eqref{4ellderivs} are each equivalently written as the following multilinear polynomial equation
\begin{multline}\label{q4def}
\kappa_0X_aX_bX_cX_d
+\kappa_1(X_aX_bX_c+X_aX_bX_d+X_aX_cX_d+X_bX_cX_d)
+\kappa_2(X_aX_c+X_bX_d) \\
+\kappa_3(X_aX_d+X_bX_c)
+\kappa_4(X_aX_b+X_cX_d)
+\kappa_5(X_a+X_b+X_c+X_d)
+\kappa_6=0,
\end{multline}
where the coefficients are given by
\begin{equation}
\begin{gathered}
\kappa_0=A'_1-A'_3,\quad
\kappa_1=A_1A'_3-A_3A'_1,\quad
\kappa_2=A_3^2A'_1-A_1^2A'_3,\\
\kappa_3=\frac{A'_1A'_3(A'_1-A'_3)}{2(A_3-A_1)}+A_3^2A'_1+\Bigl(2A_1^2-\frac{g_2}{4}\Bigr)A'_3,\\
\kappa_4=\frac{A'_1A'_3(A'_1-A'_3)}{2(A_1-A_3)}-A_1^2A'_3-\Bigl(2A_3^2-\frac{g_2}{4}\Bigr)A'_1,\\
\kappa_5=\frac{g_3\kappa_0}{2}-\frac{g_2\kappa_1}{4},\quad
\kappa_6=\frac{g_2^2\kappa_0}{16}-g_3\kappa_1,
\end{gathered}
\end{equation}
and $A'_1$ and $A'_3$ denote derivatives of the Weierstrass functions involved in the change of variables \eqref{ellcov}, {\it i.e.},
\begin{equation}
A'_1=\wp'\bigl(\tfrac{2\omega_1 \alpha_1}{\pi}\bigr),\qquad A'_3=\wp'\bigl(\tfrac{2\omega_1 \alpha_3}{\pi}\bigr).
\end{equation}

This implies that the CSTR \eqref{CSTRdef1} holds up to a term of the form \eqref{addtermsCSTR} on solutions of the $Q4$ integrable quad equation in the three-leg forms \eqref{4ellderivs} and \eqref{3legdef1}.

\end{proof}

\begin{remark}
The equations \eqref{3legdef1} and \eqref{4ellderivs} may be written in terms of a simpler form of $Q4$ than \eqref{q4def} through the use of Jacobi functions instead of Weierstrass functions in \eqref{ellcov} \cite{Bazhanov:2010kz}.
\end{remark}

\subsubsection{Hyperbolic cases}

Solutions to the CSTRs may be obtained using the quasi-classical asymptotics of the hyperbolic gamma function (or non-compact quantum dilogarithm), which parameterises the Boltzmann weights of Proposition \ref{thm:hypersols}.  Such asymptotics are given in terms of the dilogarithm \eqref{dilogdef}.  The resulting solutions of the CSTRs may be given in terms of the following functions
\begin{equation}\label{hyperlags}
\begin{split}
\lag^{(1)}_\alpha(x_i,x_j)&=\lie(-\EXP^{\pm x_i\pm x_j+\ii\alpha}) -2\lie(-\EXP^{\ii\alpha})+x_i^2+x_j^2-\tfrac{\alpha^2}{2}+\tfrac{\pi^2}{6}, \\
\lag^{(2)}_\alpha(x_i,x_j)&=\lie(-\EXP^{\pm(x_i-x_j)+\ii\alpha})-2\lie(-\EXP^{\ii\alpha})+\tfrac{(x_i-x_j)^2}{2}, \\
\lam^{(1)}_\alpha(x_i,x_j)&=\lie(-\EXP^{x_i\pm x_j+\ii\alpha})+\tfrac{(x_i+\ii\alpha)^2+x_j^2}{2}, \\
\lam^{(2)}_\alpha(x_i,x_j)&=\lie(-\EXP^{x_i+x_j+\ii\alpha})+x_i^2+x_j^2-\tfrac{\alpha^2}{2}+\ii\alpha(x_i+x_j)+\tfrac{\pi^2}{6}, \\
\lam^{(3)}_\alpha(x_i,x_j)&=-x_ix_j.
\end{split}
\end{equation}

\begin{prop} \label{prop:hyperlags}
Let the functions \eqref{CSTRsol} be given in terms of \eqref{hyperlags} according to one of the five rows of Table \ref{tab:hyperlagcombos}. Then \eqref{CSTRsol} is a solution to the CSTRs \eqref{CSTRdef1}--\eqref{3legdef2} up to terms of the form \eqref{addtermsCSTR}. 

\begin{table}[htb!]
\centering
\begin{tabular}{c | c | c | c | c | c | c}
Type & $\lag_\alpha(x_i,x_j)$  & $\lagh_\alpha(x_i,x_j)$  & $\lam_\alpha(x_i,x_j)$  & $\ol_\alpha(x_i,x_j)$ & $\olh_\alpha(x_i,x_j)$ & $\olam_\alpha(x_i,x_j)$  
\\ \hline \\[-0.4cm]
I & \multicolumn{3}{c|}{$\lag^{(1)}_\alpha(x_i,x_j)$} & \multicolumn{3}{c}{$\lag^{(1)}_{\pi-\alpha}(x_i,x_j)$} 
\\ \hline \\[-0.4cm]
I & \multicolumn{3}{c|}{$\lag^{(2)}_\alpha(x_i,x_j)$} & \multicolumn{3}{c}{$\lag^{(2)}_{\pi-\alpha}(x_i,x_j)$}   
\\ \hline \\[-0.4cm]
III & $\lag^{(2)}_\alpha(x_i,x_j)$ & $\lag^{(1)}_\alpha(x_i,x_j)$ & $\lam^{(1)}_\alpha(x_i,x_j)$ & $\lag^{(2)}_{\pi-\alpha}(x_i,x_j)$ & $\lag^{(1)}_{\pi-\alpha}(x_i,x_j)$ & $\lam^{(1)}_{\pi-\alpha}(-x_i,x_j)$ 
\\ \hline \\[-0.4cm]
II & \multicolumn{2}{c|}{$\lag^{(2)}_\alpha(x_i,x_j)$} & $\lam^{(2)}_\alpha(x_i,x_j)$ & \multicolumn{2}{c|}{$\lag^{(2)}_{\pi-\alpha}(x_i,x_j)$} & $-\lam^{(2)}_{\pi+\alpha}(x_i,x_j)$ 
\\ \hline \\[-0.4cm]
II & \multicolumn{2}{c|}{$\lag^{(2)}_\alpha(x_i,x_j)$} & $\lam^{(3)}_\alpha(x_i,x_j)$ & \multicolumn{2}{c|}{$\lag^{(2)}_{\pi-\alpha}(x_i,x_j)$} & $\lam^{(3)}_{\pi-\alpha}(-x_i,x_j)$ \\
\hline 
\end{tabular}
 \caption{Assignment of the functions \eqref{hyperlags} to \eqref{CSTRsol} for five different solutions of the CSTRs \eqref{CSTRdef1}--\eqref{3legdef2}.}
\label{tab:hyperlagcombos}
\end{table}

\end{prop}

The proof is basically the same as for (but simpler than) Proposition \ref{prop:ellipticCSTR} and can be shown case-by-case.  The type I, II, or III, is also given in Table \ref{tab:hyperlagcombos} according to the remarks at the end of Section \ref{sec:CSTR}. 

The functions \eqref{hyperlags} originate from the quasi-classical limit of star-triangle relations \cite{Bazhanov:2016ajm,Kels:2018xge} that are equivalent to hyperbolic hypergeometric integrals, which have been given in Proposition \ref{thm:hypersols}. 
Namely, from top to bottom row the functions arise from asymptotics of the hyperbolic beta function \cite{StokmanHyperbolic}, hyperbolic Saalsch\"utz integral \cite{BultThesis}, hyperbolic Askey-Wilson \cite{StokmanHyperbolic,Ruijsenaars2003} and Saalsch\"utz integrals, hyperbolic Barnes's first lemma \cite{Kels:2018xge}, and hyperbolic Barnes's $_2F_1$ integral \cite{Kels:2018xge}.

Also following from the results of \cite{Bazhanov:2016ajm,Kels:2018xge}, for each of the five solutions of the CSTRs given in Table \ref{tab:hyperlagcombos}, the equations \eqref{3legdef1} and \eqref{3legdef2} are equivalent up to a point transformation of variables to an integrable quad equation in the Adler-Bobenko-Suris (ABS) list \cite{ABS,ABS2}.  Namely, from top to bottom row the equations \eqref{3legdef1} and \eqref{3legdef2} both correspond to the equations known as $Q3_{(\delta=1)}$, $Q3_{(\delta=0)}$, $H3_{(\delta=1;\,\varepsilon=1)}$, $H3_{(\delta=0,1;\,\varepsilon=1-\delta)}$, and $H3_{(\delta=0;\,\varepsilon=0)}$.

\subsubsection{Rational cases}

Solutions to the CSTRs in these cases may be obtained using the quasi-classical asymptotics of the regular gamma function, which parameterise Boltzmann weights of the star-triangle relations \cite{Bazhanov:2016ajm,Kels:2018xge}.  Such asymptotics are given by the well-known Stirling formula involving the complex logarithm.  
The resulting solutions of the CSTRs \eqref{CSTRdef1} and \eqref{CSTRdef2} may be given in terms of the following functions
\begin{equation}
\begin{alignedat}{2}\label{ratlags}
&\lag^{(1)}_\alpha(x_i,x_j)=\gamma(\pm x_i\pm x_j+\ii\alpha),
 \qquad 
&&\ol^{(1)}_\alpha(x_i,x_j)=\lag^{(1)}_{-\alpha}(x_i,x_j)+\gamma(2\ii\alpha),
\\
&\lag^{(2)}_\alpha(x_i,x_j)=\gamma(x_i-x_j-\ii\alpha)-\gamma(x_i-x_j+\ii\alpha),
\qquad
&&\ol^{(2)}_\alpha(x_i,x_j)=\lag^{(2)}_{-\alpha}(x_i,x_j)+\gamma(2\ii\alpha),
\\
&\lam^{(1)}_\alpha(x_i,x_j)=\gamma(x_i\pm x_j+\ii\alpha),
\qquad
&&\lam^{(2)}_\alpha(x_i,x_j)=\gamma(x_i+x_j+\ii\alpha).
\end{alignedat}
\end{equation}

\begin{prop} 
Let the functions \eqref{CSTRsol} be given in terms of \eqref{ratlags} according to one of the four rows of Table \ref{tab:ratlagcombos}. Then \eqref{CSTRsol} is a solution to the CSTRs \eqref{CSTRdef1}--\eqref{3legdef2} up to terms of the form \eqref{addtermsCSTR}.


\begin{table}[htb!]
\centering
\begin{tabular}{c | c | c | c | c | c | c}
Type & $\lag_\alpha(x_i,x_j)$  & $\lagh_\alpha(x_i,x_j)$  & $\lam_\alpha(x_i,x_j)$  & $\ol_\alpha(x_i,x_j)$ & $\olh_\alpha(x_i,x_j)$ & $\olam_\alpha(x_i,x_j)$  
\\ \hline \\[-0.4cm]
I & \multicolumn{3}{c|}{$\lag^{(1)}_\alpha(x_i,x_j)$} & \multicolumn{3}{c}{$\ol^{(1)}_\alpha(x_i,x_j)$} 
\\ \hline \\[-0.4cm]
I & \multicolumn{3}{c|}{$\lag^{(2)}_\alpha(x_i,x_j)$} & \multicolumn{3}{c}{$\ol^{(2)}_\alpha(x_i,x_j)$} 
\\ \hline \\[-0.4cm]
III & $\lag^{(2)}_\alpha(x_i,x_j)$ & $\lag^{(1)}_\alpha(x_i,x_j)$ & $\lam^{(1)}_\alpha(x_i,x_j)$ & $\ol^{(2)}_\alpha(x_i,x_j)$ & $\ol^{(1)}_\alpha(x_i,x_j)$ & $\lam^{(1)}_{-\alpha}(-x_i,x_j)$ 
\\ \hline \\[-0.4cm]
II & \multicolumn{2}{c|}{$\lag^{(2)}_\alpha(x_i,x_j)$} & $\lam^{(2)}_\alpha(x_i,x_j)$ & \multicolumn{2}{c|}{$\ol^{(2)}_\alpha(x_i,x_j)$} & $-\lam^{(2)}_{\alpha}(x_i,x_j)$ \\
\hline 
\end{tabular}
 \caption{Assignment of the functions \eqref{ratlags} to \eqref{CSTRsol} for four different solutions of the CSTRs \eqref{CSTRdef1}--\eqref{3legdef2}.}
\label{tab:ratlagcombos}
\end{table}

\end{prop}

The functions \eqref{ratlags} originate from the quasi-classical limit of star-triangle relations \cite{Bazhanov:2016ajm,Kels:2018xge} that are equivalent to rational hypergeometric integrals, the most well-known examples of which are probably the Barnes Lemmas \cite{Barnes1908,Barnes1910}.  Namely, from top to bottom row the functions arise from asymptotics of the rational (Askey) beta function \cite{Askey1989}, Barnes's second lemma \cite{Barnes1910}, de Branges-Wilson integral \cite{DeBranges1972,Wilson1980} and Barnes's second lemma, and Barnes's first lemma \cite{Barnes1908}.

Also following from the results of \cite{Bazhanov:2016ajm,Kels:2018xge}, for each of the four solutions of the CSTRs given in Table \ref{tab:ratlagcombos}, the equations \eqref{3legdef1} and \eqref{3legdef2} are equivalent up to a point transformation of variables to an integrable quad equation in the ABS list \cite{ABS,ABS2}.  Namely, from top to bottom row the equations \eqref{3legdef1} and \eqref{3legdef2} both correspond to the equations known as $Q2$, $Q1_{(\delta=1)}$, $H2_{(\varepsilon=1)}$, and $H2_{(\varepsilon=0)}$.

\subsubsection{Algebraic cases}

Solutions of the CSTRs in these cases don't arise from the asymptotics of some specific special function, and this makes taking the quasi-classical expansion less systematic than in the previous cases, as may be seen from the cases derived in \cite{Kels:2018xge}.  The resulting functions that solve the CSTRs \eqref{CSTRdef1} and \eqref{CSTRdef2} may be written in terms of the following 
\begin{equation}
\begin{alignedat}{2}\label{alglags}
&\lag^{(1)}_\alpha(x_i,x_j)=\gamma(x_i-x_j-\ii\alpha)-\gamma(x_i-x_j+\ii\alpha),
 \qquad 
&&\ol^{(1)}_\alpha(x_i,x_j)=\lag^{(1)}_{-\alpha}(x_i,x_j)+\gamma(2\ii\alpha),
\\
&\lag^{(2)}_\alpha(x_i,x_j)=-2\alpha\Log(x-y),
\qquad
&&\ol^{(2)}_\alpha(x_i,x_j)=\lag^{(2)}_{-\alpha}(x_i,x_j)+\gamma(2\ii\alpha),
\\
&\lam^{(1)}_\alpha(x_i,x_j)=(\ii x_j-\alpha)\Log(x_i),
\qquad
&&\lam^{(2)}_\alpha(x_i,x_j)=x_ix_j+\alpha.
\end{alignedat}
\end{equation}
Note that $\lag_\alpha^{(1)}$ and $\ol_\alpha^{(1)}$ in \eqref{alglags} are respectively equivalent to $\lag_\alpha^{(2)}$ and $\ol_\alpha^{(2)}$ in \eqref{ratlags}.

\begin{prop} 
Let the functions \eqref{CSTRsol} be given in terms of \eqref{alglags} according to one of the three rows of Table \ref{tab:alglagcombos}. Then \eqref{CSTRsol} is a solution to the CSTRs \eqref{CSTRdef1}--\eqref{3legdef2} up to terms of the form \eqref{addtermsCSTR}.



\begin{table}[htb!]
\centering
\begin{tabular}{c | c | c | c | c | c | c}
Type & $\lag_\alpha(x_i,x_j)$  & $\lagh_\alpha(x_i,x_j)$  & $\lam_\alpha(x_i,x_j)$  & $\ol_\alpha(x_i,x_j)$ & $\olh_\alpha(x_i,x_j)$ & $\olam_\alpha(x_i,x_j)$  
\\ \hline \\[-0.4cm]
I & \multicolumn{3}{c|}{$\lag^{(2)}_\alpha(x_i,x_j)$} & \multicolumn{3}{c}{$\ol^{(2)}_\alpha(x_i,x_j)$}
\\ \hline \\[-0.4cm]
III & $\lag^{(2)}_\alpha(x_i,x_j)$ & $\lag^{(1)}_\alpha(x_i,x_j)$ & $\lam^{(1)}_\alpha(x_i,x_j)$ & $\ol^{(2)}_\alpha(x_i,x_j)$ & $\ol^{(1)}_\alpha(x_i,x_j)$ & $-\lam^{(1)}_{\alpha}(x_i,x_j)$
\\ \hline \\[-0.4cm]
II & \multicolumn{2}{c|}{$\lag^{(2)}_\alpha(x_i,x_j)$} & $\lam^{(2)}_\alpha(x_i,x_j)$ & \multicolumn{2}{c|}{$\ol^{(2)}_\alpha(x_i,x_j)$} &  $\lam^{(2)}_{-\alpha}(-x_i,x_j)$ \\
\hline 
\end{tabular}
 \caption{Assignment of the functions \eqref{alglags} to \eqref{CSTRsol} for three different solutions of the CSTRs \eqref{CSTRdef1}--\eqref{3legdef2}.}
\label{tab:alglagcombos}
\end{table}

\end{prop}

The functions \eqref{alglags} originate from the quasi-classical limit of star-triangle relations \cite{Bazhanov:2016ajm,Kels:2018xge} that are equivalent to classical hypergeometric integrals, the most well-known examples of which are probably the Euler beta function and the integral representation of Gauss's hypergeometric function.   Namely, from top to bottom row the functions arise from asymptotics of a Selberg-type integral \cite{Selberg,Rains2009}, a special case of Barnes's $_2F_1$ formula\footnote{\label{footnote:unknown}Only for the CSTR \eqref{CSTRdef1}.  This is because the CSTR \eqref{CSTRdef2} didn't arise from the analysis of \cite{Kels:2018xge} and its hypergeometric/quantum counterpart is presently unknown.} \cite{Barnes1908,WW}, and Euler beta function.  

Also following from the results of \cite{Bazhanov:2016ajm,Kels:2018xge}, for each of the three solutions of the CSTRs given in Table \ref{tab:alglagcombos}, the equations \eqref{3legdef1} and \eqref{3legdef2} are equivalent up to a point transformation of variables to an integrable quad equation in the ABS list \cite{ABS,ABS2}.  Namely, from top to bottom row the equations \eqref{3legdef1} and \eqref{3legdef2} both correspond to the equations known as $Q1_{(\delta=0)}$, $H1_{(\varepsilon=1)}$, and $H1_{(\varepsilon=0)}$.

\subsection{Explicit Yang-Baxter maps}
\label{sec:YBMlist}

The method to derive Yang-Baxter maps given in Section \ref{sec:YBMconversion} will be used for each of the solutions of the CSTRs given in Section \ref{sec:CSTRlist}.  Apart from the elliptic case, only type-II and type-III solutions of the CSTRs need to be considered, because they can be used to obtain the same Yang-Baxter maps as type-I solutions of the CSTRs.  

A list of Yang-Baxter maps derived in this section is presented in Appendix \ref{app:YBMlist}.

\subsubsection{Elliptic case}

In the following, the notation $\dot{x}$, and $\breve{x}$, will be used to respectively denote
\begin{equation}\label{ellnot}
\dot{x}^2=4x^3-g_2x-g_3, \qquad \breve{x}=\frac{\dot{x}^2}{4(x-\me)^2}-x-\me,
\end{equation}
where $x$ is a variable or parameter, and $\wp(x)$ is the Weierstrass elliptic function with elliptic invariants $g_2,g_3$ or associated half-periods $\omega_1,\omega_2$.  The expressions \eqref{ellnot} originate from computations involving the well-known relation between $\wp(z)$ and its derivative, and an addition formula for $\wp(z)$.

The classical R-matrix for this case is given by \eqref{CRMAT}, with the functions of \eqref{lagq4}.  The expression for  $R_{\bu\bv,i}(x_j',x_j;x_i)$ in \eqref{Rmatderivs} is obtained as a partial derivative of the latter classical R-matrix with respect to $x_i$, resulting in
\begin{multline}\label{q4ex1}
R_{\bu\bv,i}(x_j',x_j;x_i)=-\ii\pi+\frac{16x_i(\rua-\rub)\zeta(\omega_1)\omega_1}{\pi^2}+\frac{2\ii\omega_1(\rua-\rub)(\pi-4x_i)}{\pi\omega_2} \\  
\begin{aligned}
  + \Log\frac{\sigma\bigl(\frac{2\omega_1}{\pi}(x_i-x_j'-\rub+\rvb)\bigr)}{\sigma\bigl(\frac{2\omega_1}{\pi}(x_i-x_j'+\rub-\rvb)\bigr)} 
 + \Log\frac{\sigma\bigl(\frac{2\omega_1}{\pi}(x_i-x_j+\rua-\rvb)+\omega_2\bigr)}{\sigma\bigl(\frac{2\omega_1}{\pi}(x_i-x_j-\rua+\rvb)+\omega_2\bigr)}\phantom{,} \\
   + \Log\frac{\sigma\bigl(\frac{2\omega_1}{\pi}(x_i+x_j'-\rub+\rvb)\bigr)}{\sigma\bigl(\frac{2\omega_1}{\pi}(x_i+x_j'+\rub-\rvb)\bigr)} 
 + \Log\frac{\sigma\bigl(\frac{2\omega_1}{\pi}(x_i+x_j+\rua-\rvb)-\omega_2\bigr)}{\sigma\bigl(\frac{2\omega_1}{\pi}(x_i+x_j-\rua+\rvb)-\omega_2\bigr)},
\end{aligned}
\end{multline}
where $\sigma(z)$ and $\zeta(z)$ are the Weierstrass sigma and zeta functions respectively (see \eqref{WeierstrassRels}).  The other three expressions in \eqref{Rmatderivs} may also be obtained from \eqref{q4ex1} by comparing the respective dependencies on the variables.  For example, $R_{\bu\bv,j}(x_i',x_i;x_j)$ is equivalent to the right-hand side of \eqref{q4ex1} with
\begin{equation}\label{q4cov1}
x_i\leftrightarrow x_j,\quad x_j'\to x_i', \qquad \rub\to-\rva,\quad \rua\to-\rvb,\quad \rvb\to-\rua,
\end{equation}
$R_{\bu\bv,i'}(x_j,x_j';x_i')$ is obtained from \eqref{q4ex1} with
\begin{equation}\label{q4cov2}
x_i\to x_i',\quad x_j\leftrightarrow x_j', \qquad \rua\leftrightarrow\rub, \quad \rvb\to\rva,
\end{equation}
and $R_{\bu\bv,j'}(x_i,x_i';x_j')$ is obtained from \eqref{q4ex1} with
\begin{equation}\label{q4cov3}
x_i\leftrightarrow x_j',\quad x_j\to x_i', \qquad \rua\to-\rva, \quad \rub\to-\rvb,\quad \rvb\to-\rub.
\end{equation}

The terms on the first line in \eqref{q4ex1} can be ignored, since by \eqref{q4cov1}, \eqref{q4cov2}, and \eqref{q4cov3}, the first term will only contribute an overall sign to each Yang-Baxter map, and the second and third terms will only appear in the form of the gauge transformation given in \eqref{gauge1}.

The change of variables according to \eqref{RmattoYBMcov} is given by
\begin{equation}\label{q4cov}
f(x)=\wp(2\omega_1 x\pi^{-1}), \qquad h(x)=\wp(2\omega_1 x\pi^{-1}).
\end{equation}

Then using the change of variables \eqref{q4cov} for $R_{\bu\bv,i}(x_j',x_j;x_i)$ in \eqref{q4ex1} gives an expression for $\mui$ in \eqref{RmattoYBM} as
\begin{equation}\label{q4ex2}
\mui =\left(\frac{\sigma(x_i+\rua)\sigma(x_i-\rub)}{\sigma(x_i-\rua)\sigma(x_i+\rub)}\right)^2E(\mxj,\mqa,\mqb,\mpa)^2
\frac{G(\mxi,\breve{\mxj},\mpa,\mqb)G(\mxi,\myi,\mqa,\mpa)}{G(\mxi,\breve{\mxj},\mqb,\mpa)G(\mxi,\myi,\mpa,\mqa)}, 
\end{equation}
where
\begin{equation}
\begin{gathered}
A(\mpc,\mqc)=\frac{(\tmp+\tmq)^2}{4(\mpc-\mqc)^2}-\mpc-\mqc, \quad
B(x,\mpc,\mqc)=\tmp\mqc+\tmq\mpc-(\tmp+\tmq)x, \\
E(x,\mpc,\mqc,\gamma)=\frac{1}{E(x,\mqc,\mpc,\gamma)}=\frac{(B(x,\mpc,\gamma)+(\mpc-\gamma)\dot{x})(B(x,\mqc,\gamma)-(\mqc-\gamma)\dot{x})}{(B(x,\mpc,\gamma)-(\mpc-\gamma)\dot{x})(B(x,\mqc,\gamma)+(\mqc-\gamma)\dot{x})}, \\[0.2cm]
F(x,\mpc,\mqc)=2\dot{x}\frac{B\bigl(A(\mpc,\mqc),\mpc,\mqc\bigr)}{\mpc-\mqc}+2g_3+(A(\mpc,\mqc)+x)(g_2-4xA(\mpc,\mqc)), \\
G(x,y,\mpc,\mqc)=F(x,\mpc,\mqc)+4y(A(\mpc,\mqc)-x)^2.
\end{gathered}
\end{equation}
The expressions for $\muj$, $\mupi$, and $\mupj$ in \eqref{RmattoYBM} take a similar form (for $\mupi$ and $\mupj$ see \eqref{q4map}).  Note that from \eqref{q4cov1}, \eqref{q4cov2}, and \eqref{q4cov3}, the factor involving Weierstrass sigma functions in \eqref{q4ex2} is in the form of the gauge transformation of \eqref{gauge2} and thus can  effectively be set to $1$. Also note that $z_i$ is a ratio of polynomials of degree one in $\myi$, but degree two in $\mxj$.  Similarly the expressions  for $\muj$, $\mupi$, and $\mupj$, from \eqref{RmattoYBM}, will be ratios of polynomials of degree 1 and degree 2 in their first and second arguments respectively, instead of both degree 1.  The consequence of this is that the resulting Yang-Baxter map for the elliptic case is birational rather than quadriational.  Such a quadratic dependence for these variables will only appear for this elliptic case.\footnote{The author expects that a simpler quadrirational expression should exist but was not able to obtain it.}

Finally, let $\Upsilon_1(\muj,\mxi;\mxj)$, $\Upsilon_2(\mui,\mxj;\mxi)$, be the following expressions 
\begin{equation}\label{ellipticupsilondef}
\begin{split}
\Upsilon_1 &=  
\frac{\muj G(\mxj,\mxi,\mqb,\mpa)F\bigl(\mxj,\mpa,\mqa) E(\mxj,\mqa,\mqb,\mpa)^2
     -     G(\mxj,\mxi,\mpa,\mqb)F\bigl(\mxj,\mqa,\mpa)}
     {4\bigl(\mxj-A(\mpa,\mqa)\bigr)^2\bigl(G(\mxj,\mxi,\mpa,\mpb)-\muj G(\mxj,\mxi,\mqb,\mpa) E(\mxj,\mqa,\mqb,\mpa)^2\bigr)}, \\
\Upsilon_2 &=  
\frac{\mui G(\mxi,\mxj,\mqb,\mpa)F\bigl(\mxi,\mpb,\mqb) E(\mxi,\mpa,\mpb,\mqb)^2
     -     G(\mxi,\mxj,\mpa,\mqb)F\bigl(\mxi,\mqb,\mpb)}
     {4\bigl(\mxi-A(\mpb,\mqb)\bigr)^2\bigl(G(\mxi,\mxj,\mpa,\mpb)-\mui G(\mxi,\mxj,\mqb,\mpa) E(\mxi,\mpa,\mpb,\mqb)^2\bigr)},
\end{split}
\end{equation}
which are obtained by solving $\muj$, and $\mui$ in \eqref{q4ex2}, for $\myj$ and $\myi$, respectively.

Then the Yang-Baxter map $R(\al,\bt)\colon\bigl((y_i,y_j),(z_i,z_j)\bigr)\mapsto\bigl((y_i',y_j'),(z_i',z_j')\bigr)$ is given by
\begin{equation}
\label{q4map}
\begin{split}
\myj=\ds \Upsilon_1, \qquad
\mupi&=\ds E(\Upsilon_1,\mpb,\mpa,\mqa)^2 
\frac{G(\Upsilon_1,\mxj,\mpa,\mqa)G(\Upsilon_1,\breve{\Upsilon}_2,\mqa,\mpb)}
     {G(\Upsilon_1,\mxj,\mqa,\mpa)G(\Upsilon_1,\breve{\Upsilon}_2,\mpb,\mqa)}, \\
\myi=\ds \Upsilon_2, \qquad
\mupj&=\ds E(\Upsilon_2,\mqb,\mqa,\mpb)^2 
\frac{G(\Upsilon_2,\mxi,\mpb,\mqb)G(\Upsilon_2,\breve{\Upsilon}_1,\mqa,\mpb)}
     {G(\Upsilon_2,\mxi,\mqb,\mpb)G(\Upsilon_2,\breve{\Upsilon}_1,\mpb,\mqa)}.
\end{split}
\end{equation}
According to Theorem \ref{thm:maintheorem}, this is a solution of the FYBE \eqref{MYBE}.

\subsubsection{Hyperbolic cases}

In the following, for any variable or parameter $x$, the notation $\overline{x}$ will be used to denote
\begin{equation}\label{coshnot}
\overline{x}=x+\sqrt{x^2-1}.
\end{equation}
This comes from the use of the expression for the inverse of the cosh function.

  \tocless\paragraph{Case 1.}
  

This case results in two different solutions of the FYBE \eqref{MAYBE} that are obtained from the type-III solution of the CSTR in Table \ref{tab:hyperlagcombos}.   The latter solution also leads to the Yang-Baxter maps that would be obtained from the two type-I solutions given in Table \ref{tab:hyperlagcombos}.  
The first set of classical R-matrices are given by \eqref{CRMAT} and \eqref{CUMAT1}, and the second set of classical R-matrices for this case are given by \eqref{CRMAT2} and \eqref{CUMAT2}.

First, let $R_{\bu\bv}$ and $U_{\bu\bv}$ denote the classical R-matrices \eqref{CRMAT} and \eqref{CUMAT1} that are constructed from the functions for the type-III solution of the CSTR from Table \ref{tab:hyperlagcombos}.  The change of variables for \eqref{RmattoYBMcov} and \eqref{UmattoYBMcov} will be
\begin{equation}\label{q3d0cov}
f(x)=\EXP^{x},\qquad g(x)=\cosh(x),\qquad h(x)=\EXP^{\ii x}.
\end{equation}

For \eqref{CRMAT}, setting $z_i=\EXP^{R_{\bu\bv,i}(x_j',x_j;x_i)}$ and $z_j=\EXP^{R_{\bu\bv,j}(x_i',x_i;x_j)}$ as in \eqref{RmattoYBM} and applying the change of variables \eqref{q3d0cov} according to \eqref{RmattoYBMcov}, allows to solve for $y'_i$ and $y'_j$ as $y'_i=\Upsilon_1(\muj,\mxi;\mxj)$ and $y'_j=\Upsilon_2(\mui,\mxj;\mxi)$, where
\begin{equation}
\Upsilon_1 =  \mxj\frac{\mpa(\mpa\mxi + \mqb\mxj) + \mqa\muj(\mpa\mxj + \mqb\mxi) }{ \mqa(\mpa\mxi+\mqb\mxj) + \mpa\muj(\mpa\mxj + \mqb\mxi)}, \quad
\Upsilon_2 = \mxi\frac{\mpb(\mpa\mxj + \mqb\mxi) + \mqb\mui(\mpa\mxi + \mqb\mxj) }{\mqb(\mpa\mxj + \mqb\mxi) + \mpb\mui(\mpa\mxi+\mqb\mxj)}.
\end{equation}
Also setting $z'_i=\EXP^{R_{\bu\bv,i'}(x_j,x_j';x_i')}$ and $z'_j=\EXP^{R_{\bu\bv,j'}(x_i,x_i';x_j')}$ as in \eqref{RmattoYBM} and applying the change of variables \eqref{q3d0cov} according to \eqref{RmattoYBMcov}, then gives the desired expression for the Yang-Baxter map $R(\al,\bt)\colon \bigl((y_i,y_j),(z_i,z_j)\bigr)\mapsto\bigl((y_i',y_j'),(z_i',z_j')\bigr)$ as 
%
\begin{empheq}{equation}
\label{q3d0map}
\begin{gathered}
\myj=\ds \Upsilon_1, \qquad 
\mupi=\ds \frac{(\mpa\mxj-\mqa \Upsilon_1)( \mpb \Upsilon_1 + \mqa \Upsilon_2)}{(\mpa \Upsilon_1-\mqa\mxj)( \mpb \Upsilon_2 + \mqa \Upsilon_1)}, \\
\myi=\ds \Upsilon_2, \qquad 
\mupj=\ds \frac{(\mpb \mxi - \mqb \Upsilon_2)(\mpb \Upsilon_2+\mqa \Upsilon_1  )}{(\mpb \Upsilon_2 - \mqb\mxi )(\mpb \Upsilon_1 + \mqa \Upsilon_2)},
\end{gathered}
\end{empheq}
According to Theorem \ref{thm:maintheorem}, this is a solution of the FYBE \eqref{MYBE}.

Similarly for \eqref{CUMAT1}, setting $z_i=\EXP^{U_{\bu\bv,i}(x_j',x_j;x_i)}$ and $z_j=\EXP^{U_{\bu\bv,j}(x_i',x_i;x_j)}$ as in \eqref{UmattoYBM} and applying the change of variables \eqref{q3d0cov} according to \eqref{UmattoYBMcov}, allows to solve for $y'_i$ and $y'_j$ as $y'_i=\Upsilon_3(\muj,\mxi;\mxj)$ and $y'_j=\Upsilon_4(\mui,\mxj;\mxi)$, where
\begin{equation}
\Upsilon_3 =  \frac{\mqa}{\mpa}\frac{\mqb\omxj(\muj-1) + \mpa\mxi(\muj\omxj^2-1)}{\mpa\mxi\omxj(\muj-1) + \mqb(\muj - \omxj^2)}, \quad
\Upsilon_4 = {\mxj}{\mui} + \mxi\frac{\mpb+\mpa\mui}{2\mqb} + \mqb\frac{\mpa+\mpb\mui}{2\mpa\mpb\mxi}.
\end{equation}
Also setting $z'_i=\EXP^{U_{\bu\bv,i'}(x_j,x_j';x_i')}$ and $z'_j=\EXP^{U_{\bu\bv,j'}(x_i,x_i';x_j')}$ as in \eqref{UmattoYBM} and applying the change of variables \eqref{q3d0cov} according to \eqref{UmattoYBMcov}, then gives the desired expression for the Yang-Baxter map $U(\al,\bt)\colon \bigl((y_i,y_j),(z_i,z_j)\bigr)\mapsto\bigl((y_i',y_j'),(z_i',z_j')\bigr)$ as 
\begin{empheq}{equation}
\label{h3d1altmapb}
\begin{gathered}
\myj=\ds  \Upsilon_3, \qquad 
\mupi=\ds -\frac{\mpa}{\mpb}\frac{\mqa^2  + \mpb^2 \Upsilon_3^2 + 2\mpb\mqa \Upsilon_3\Upsilon_4}
                                 {\mqa^2 + \mpa^2\Upsilon_3^2 - 2\mpa\mqa\mxj \Upsilon_3} \\
\myi=\ds \Upsilon_4, \qquad 
\mupj=\ds \frac{(\mpb\mxi-\mqb\overline{\Upsilon}_4)(\mqa + \mpb \Upsilon_3\overline{\Upsilon}_4)}
               {(\mpb \Upsilon_3+\mqa\overline{\Upsilon}_4)(\mpb\mxi\overline{\Upsilon}_4-\mqb)}.
\end{gathered}
\end{empheq}
According to Theorem \ref{thm:maintheorem}, the Yang-Baxter maps \eqref{q3d0map} and \eqref{h3d1altmapb} together are a solution of the FYBE \eqref{MAYBE}.

Next, let $\hat{R}_{\bu\bv}$ and $\hat{U}_{\bu\bv}$ denote the classical R-matrices \eqref{CRMAT2} and \eqref{CUMAT2} that are constructed from the functions for the type-III solution of the CSTR from Table \ref{tab:hyperlagcombos}.  The change of variables for \eqref{RmattoYBMcov} and \eqref{UmattoYBMcov} will be
\begin{equation}\label{q3d1cov}
f(x)=\cosh(x),\qquad g(x)=\EXP^{x},\qquad h(x)=\EXP^{\ii x}.
\end{equation}

For \eqref{CRMAT2}, setting $z_i=\EXP^{\hat{R}_{\bu\bv,i}(x_j',x_j;x_i)}$ and $z_j=\EXP^{\hat{R}_{\bu\bv,j}(x_i',x_i;x_j)}$ as in \eqref{RmattoYBM} (with $R$ replaced by $\hat{R}$) and applying the change of variables \eqref{q3d1cov} according to \eqref{RmattoYBMcov}, allows to solve for $y'_i$ and $y'_j$ as $y'_i=\Upsilon_1(\muj,\mxi;\mxj)$ and $y'_j=\Upsilon_2(\mui,\mxj;\mxi)$, where
\begin{equation}
\begin{gathered}
\Upsilon_1 = \frac{\omxj^2(1-\muj)\bigl(\mpa^2 + \frac{\mqa^2\mqb^2}{\mpa^2}\bigr) + (\omxj^4-\muj)\mqb^2 + (1-\muj\omxj^4)\mqa^2 - 2\mpa\mqb\mxi\omxj\bigl(\muj - \omxj^2+ (\muj\omxj^2-1)\frac{\mqa^2}{\mpa^2} \bigr)}{2\frac{\mqa}{\mpa}\omxj\bigl(\mqb^2(\omxj^2-\muj) + \mpa^2(1-\muj\omxj^2) - 2\mpa\mqb\mxi\omxj(\muj-1) \bigr)}, \\[0.1cm]
\Upsilon_2 = \frac{\omxi^2(1-\mui)\bigl(\mqb^2 + \frac{\mpa^2\mpb^2}{\mqb^2}\bigr)+ (\omxi^4-\mui)\mpb^2  + (1 - \mui\omxi^4)\mpa^2 -             2\mpa\mqb\omxi\mxj\bigl((\mui - \omxi^2)\frac{\mpb^2}{\mqb^2}  + \mui\omxi^2-1\bigr) }{  2\frac{\mpb}{\mqb}\omxi\bigl(\mqb^2(\omxi^2-\mui) + \mpa^2(1 - \mui\omxi^2) - 2\mpa\mqb\omxi\mxj( \mui-1)\bigr)}.
\end{gathered}
\end{equation}
Also setting $z'_i=\EXP^{\hat{R}_{\bu\bv,i'}(x_j,x_j';x_i')}$ and $z'_j=\EXP^{\hat{R}_{\bu\bv,j'}(x_i,x_i';x_j')}$ as in \eqref{RmattoYBM} (with $R$ replaced by $\hat{R}$) and applying the change of variables \eqref{q3d1cov} according to \eqref{RmattoYBMcov}, then gives the desired expression for the Yang-Baxter map $R(\al,\bt)\colon \bigl((y_i,y_j),(z_i,z_j)\bigr)\mapsto\bigl((y_i',y_j'),(z_i',z_j')\bigr)$ as 
\begin{empheq}{equation}
\label{q3d1map}
\begin{gathered}
\myj=\ds \Upsilon_1, \qquad 
\mupi=\ds \frac{(\mpa^2 + \mqa^2\overline{\Upsilon}_1^2 - 2\mpa\mqa\overline{\Upsilon}_1\mxj)(\mqa^2 + \mpb^2\overline{\Upsilon}_1^2 + 2\mpb\mqa\overline{\Upsilon}_1\Upsilon_2 )} 
  {(\mqa^2 + \mpa^2\overline{\Upsilon}_1^2 - 2\mpa\mqa\overline{\Upsilon}_1\mxj)(\mpb^2 + \mqa^2\overline{\Upsilon}_1^2 + 2\mpb\mqa\overline{\Upsilon}_1\Upsilon_2)}, \\
\myi=\ds \Upsilon_2, \qquad 
\mupj=\ds \frac{(\mpb^2 + \mqb^2\overline{\Upsilon}_2^2 - 2\mpb\mqb\overline{\Upsilon}_2\mxi)(\mqa^2 + \mpb^2\overline{\Upsilon}_2^2 + 2\mpb\mqa\overline{\Upsilon}_2 \Upsilon_1)}
    {(\mqb^2 + \mpb^2\overline{\Upsilon}_2^2 - 2\mpb\mqb\overline{\Upsilon}_2\mxi)(\mpb^2 + \mqa^2\overline{\Upsilon}_2^2 + 2\mpb\mqa\overline{\Upsilon}_2 \Upsilon_1)},
\end{gathered}
\end{empheq}
According to Theorem \ref{thm:maintheorem}, this is a solution of the FYBE \eqref{MYBE}.

Similarly for \eqref{CUMAT2}, setting $z_i=\EXP^{\hat{U}_{\bu\bv,i}(x_j',x_j;x_i)}$ and $z_j=\EXP^{\hat{U}_{\bu\bv,j}(x_i',x_i;x_j)}$ as in \eqref{UmattoYBM} (with $U$ replaced by $\hat{U}$) and applying the change of variables \eqref{q3d1cov} according to \eqref{UmattoYBMcov}, allows to solve for $y'_i$ and $y'_j$ as $y'_i=\Upsilon_3(\muj,\mxi;\mxj)$ and $y'_j=\Upsilon_4(\mui,\mxj;\mxi)$, where
\begin{equation}
\Upsilon_3 = \mxi\muj + \frac{\mqa + \mqb\muj}{2\mpa\mxj} + \mpa\mxj\frac{\mqb+\mqa\muj}{2\mqa\mqb}, \quad
\Upsilon_4 = \frac{\mqb}{\mpb}\frac{\mqb\omxi(\mui-1) + \mpa\mxj(\mui\omxi^2-1)}{\mpa\omxi\mxj(\mui-1) + \mqb(\mui - \omxi^2)}.
\end{equation}
Also setting $z'_i=\EXP^{\hat{U}_{\bu\bv,i'}(x_j,x_j';x_i')}$ and $z'_j=\EXP^{\hat{U}_{\bu\bv,j'}(x_i,x_i';x_j')}$ as in \eqref{UmattoYBM} (with $U$ replaced by $\hat{U}$) and applying the change of variables \eqref{q3d1cov} according to \eqref{UmattoYBMcov}, then gives the desired expression for the Yang-Baxter map $U(\al,\bt)\colon \bigl((y_i,y_j),(z_i,z_j)\bigr)\mapsto\bigl((y_i',y_j'),(z_i',z_j')\bigr)$ as 
\begin{empheq}{equation}
\label{h3d1mapb}
\begin{split}
\myj=\ds  \Upsilon_3, \qquad 
\mupi=&\ds \frac{(\mpa\mxj - \mqa\overline{\Upsilon}_3)(\mqa + \mpb \Upsilon_4\overline{\Upsilon}_3)}{(\mpb \Upsilon_4 + \mqa\overline{\Upsilon}_3)(\mpa\mxj\overline{\Upsilon}_3-\mqa)}, \\
\myi=\ds \Upsilon_4, \qquad 
\mupj=&\ds -\frac{\mqb}{\mqa}\frac{\mqa^2 + \mpb^2\Upsilon_4^2 + 2\mpb\mqa\Upsilon_4\Upsilon_3}
                                  {\mqb^2 + \mpb^2\Upsilon_4^2 - 2\mpb\mqb\mxi \Upsilon_4}.
\end{split}
\end{empheq}
According to Theorem \ref{thm:maintheorem}, the Yang-Baxter maps \eqref{q3d1map} and \eqref{h3d1mapb} together are a solution of the FYBE \eqref{MAYBE}.

\tocless\paragraph{Case 2.}

The classical R-matrices for this case are given by \eqref{CRMAT} and \eqref{CUMAT1} using the functions for the first type-II solution of the CSTR given in Table \ref{tab:hyperlagcombos}.  The Yang-Baxter map for \eqref{CRMAT} was already derived in \eqref{q3d0map}, so it remains to derive the Yang-Baxter map for \eqref{CUMAT1}.

The classical R-matrix \eqref{CUMAT1} that is constructed from the functions for the first type-II solution of the CSTR from Table \ref{tab:hyperlagcombos} is given by
\begin{equation}
\label{h3d01rmat}
\begin{split}
U_{\bu\bv}\bigl(x_i,x_j;x_i',x_j'\bigr)=&\ii\left(\pi(x_i+x_j+x_i'+x_j')+(x_i-x_i')(\rua-\rub)+(x_j-x_j')(\rva-\rvb)\right) \\[0.1cm]
&+\lie(-\EXP^{x_i+x_j+\ii(\rua-\rvb)})-\lie(\EXP^{x_i+x_j'+\ii(\rub-\rvb)}) \\[0.1cm]
&+\lie(-\EXP^{x_i'+x_j'+\ii(\rub-\rva)})-\lie(\EXP^{x_i'+x_j+\ii(\rua-\rva)})+r(\bu,\bv). 
\end{split}
\end{equation}
where $r(\bu,\bv)$ represents terms that are constant with respect to $x_i,x_j,x'_i,x'_j$.  The change of variables used in \eqref{UmattoYBMcov} is given by
\begin{equation}\label{h3d01cov}
f(x)=\EXP^{x},\qquad g(x)=\EXP^{x},\qquad h(x)=\EXP^{\ii x}.
\end{equation}
Using the expressions \eqref{UmattoYBM} formed from the classical R-matrix \eqref{h3d0rmat} and applying the change of variables \eqref{h3d01cov} according to \eqref{UmattoYBMcov}, then gives the desired expression for the Yang-Baxter map $U(\al,\bt)\colon \bigl((y_i,y_j),(z_i,z_j)\bigr)\mapsto\bigl((y_i',y_j'),(z_i',z_j')\bigr)$ as 
\begin{empheq}{equation}
\label{h3d01mapb}
\begin{split}
\myj=\ds  \mxi\muj + \frac{\mqa + \mqb\muj}{\mpa\mxj}, \qquad 
\mupi=&\ds \mui + \frac{(\mpa+\mpb\mui)(\mqa+\mqb\muj)}{\mpa\mpb\mxi\mxj\muj}, \\
\myi=\ds \mxj\mui + \mqb\frac{\mpa + \mpb\mui}{\mpa\mpb\mxi}, \qquad 
\mupj=&\ds \muj + \frac{(\mpa + \mpb\mui)(\mqa + \mqb\muj)}{\mpa\mpb\mxi\mxj\mui}.
\end{split}
\end{empheq}
According to Theorem \ref{thm:maintheorem}, \eqref{q3d0map} and \eqref{h3d01mapb} together are a solution of the FYBE \eqref{MAYBE}.

\tocless\paragraph{Case 3.}

The classical R-matrices for this case are given by \eqref{CRMAT} and \eqref{CUMAT1} using the functions for the second type-II solution of the CSTR given in Table \ref{tab:hyperlagcombos}.   The Yang-Baxter map for \eqref{CRMAT} was already derived in \eqref{q3d0map}, so it remains to derive the Yang-Baxter map for \eqref{CUMAT1}.

The classical R-matrix \eqref{CUMAT1} that is constructed from the functions for the second type-II solution of the CSTR from Table \ref{tab:hyperlagcombos} is given by
\begin{equation}
\begin{split}\label{h3d0rmat}
U_{\bu\bv}\bigl(x_i,x_j;x_i',x_j'\bigr)=&(x_i-x_i')(x_j'-x_j).
\end{split}
\end{equation}
The change of variables used in \eqref{UmattoYBMcov} is given by
\begin{equation}\label{h3d0cov}
f(x)=\EXP^{x},\qquad g(x)=\EXP^{x},\qquad h(x)=\EXP^{\ii x}.
\end{equation}
Using the expressions \eqref{UmattoYBM} formed from the classical R-matrix \eqref{h3d0rmat} and applying the change of variables \eqref{h3d0cov} according to \eqref{UmattoYBMcov}, then gives the desired expression for the Yang-Baxter map $U(\al,\bt)\colon \bigl((y_i,y_j),(z_i,z_j)\bigr)\mapsto\bigl((y_i',y_j'),(z_i',z_j')\bigr)$ as 
\begin{empheq}{equation}
\label{h3d0mapb}
\begin{split}
\myj=&\ds  \muj\mxi, \qquad 
\mupi=\ds \mui, \\
\myi=&\ds \mui\mxj, \qquad 
\mupj=\ds \muj.
\end{split}
\end{empheq}
According to Theorem \ref{thm:maintheorem}, \eqref{q3d0map} and \eqref{h3d0mapb} together are a solution of the FYBE \eqref{MAYBE}.

\subsubsection{Rational cases}

\tocless\paragraph{Case 1.}

This case results in two different solutions of the FYBE \eqref{MAYBE} that are obtained from the type-III solution of the CSTR in Table \ref{tab:ratlagcombos}.  The latter solution also leads to the Yang-Baxter maps that would be obtained from the two type-I solutions given in Table \ref{tab:ratlagcombos}.  The first set of classical R-matrices are given by \eqref{CRMAT} and \eqref{CUMAT1}, and the second set of classical R-matrices are given by \eqref{CRMAT2} and \eqref{CUMAT2}.

First, let $R_{\bu\bv}$ and $U_{\bu\bv}$ denote the classical R-matrices \eqref{CRMAT} and \eqref{CUMAT1} that are constructed from the functions for the type-III solution of the CSTR from Table \ref{tab:ratlagcombos}.  The change of variables for \eqref{RmattoYBMcov} and \eqref{UmattoYBMcov} will be
\begin{equation}\label{q2cov}
f(x)=x,\qquad g(x)=x^2,\qquad h(x)=\ii x.
\end{equation}

For \eqref{CRMAT}, setting $z_i=\EXP^{R_{\bu\bv,i}(x_j',x_j;x_i)}$ and $z_j=\EXP^{R_{\bu\bv,j}(x_i',x_i;x_j)}$ as in \eqref{RmattoYBM} and applying the change of variables \eqref{q2cov} according to \eqref{RmattoYBMcov}, allows to solve for $y'_i$ and $y'_j$ as $y'_i=\Upsilon_1(\muj,\mxi;\mxj)$ and $y'_j=\Upsilon_2(\mui,\mxj;\mxi)$, where
\begin{equation}
\begin{gathered}
\Upsilon_1 =  \mxj + (\mpa - \mqa)\frac{(\muj+1)(\mxi-\mxj) + (1-\muj)(\mpa - \mqb)}{(1-\muj)(\mxi - \mxj) + (\muj+1)(\mpa-\mqb)}, \\
\Upsilon_2 = \mxi + (\mpb - \mqb)\frac{(\mui+1)(\mxi- \mxj) + (\mui-1)(\mpa - \mqb) }{(1-\mui)(\mxi - \mxj) - (\mui+1)(\mpa-\mqb)}.
\end{gathered}
\end{equation}
Also setting $z'_i=\EXP^{R_{\bu\bv,i'}(x_j,x_j';x_i')}$ and $z'_j=\EXP^{R_{\bu\bv,j'}(x_i,x_i';x_j')}$ as in \eqref{RmattoYBM} and applying the change of variables \eqref{q2cov} according to \eqref{RmattoYBMcov}, then gives the desired expression for the Yang-Baxter map $R(\al,\bt)\colon \bigl((y_i,y_j),(z_i,z_j)\bigr)\mapsto\bigl((y_i',y_j'),(z_i',z_j')\bigr)$ as 
\begin{empheq}{equation}
\label{q1d1map}
\begin{gathered}
\myj=\ds \Upsilon_1, \qquad
\mupi=\ds \frac{(\mpa - \mqa + (\mxj - \Upsilon_1))( \mpb-\mqa + (\Upsilon_1 - \Upsilon_2))}
               {(\mpa - \mqa - (\mxj - \Upsilon_1))( \mpb-\mqa - (\Upsilon_1 - \Upsilon_2))}, \\
\myi=\ds \Upsilon_2, \qquad
\mupj=\ds \frac{(\mpb - \mqb + (\mxi - \Upsilon_2))(\mpb - \mqa - (\Upsilon_1 - \Upsilon_2))}
               {(\mpb - \mqb - (\mxi - \Upsilon_2))(\mpb - \mqa + (\Upsilon_1 - \Upsilon_2))},
\end{gathered}
\end{empheq}
According to Theorem \ref{thm:maintheorem}, this is a solution of the FYBE \eqref{MYBE}.

Similarly for \eqref{CUMAT1}, setting $z_i=\EXP^{U_{\bu\bv,i}(x_j',x_j;x_i)}$ and $z_j=\EXP^{U_{\bu\bv,j}(x_i',x_i;x_j)}$ as in \eqref{UmattoYBM} and applying the change of variables \eqref{q2cov} according to \eqref{UmattoYBMcov}, allows to solve for $y'_i$ and $y'_j$ as $y'_i=\Upsilon_3(\muj,\mxi;\mxj)$ and $y'_j=\Upsilon_4(\mui,\mxj;\mxi)$, where
\begin{equation}
\begin{split}
\Upsilon_3 =& (\mqa-\mpa) + \frac{(\muj-1)\mxj + (\muj+1)(\mxi + \mpa - \mqb)\sqrt{\mxj}}{(\muj-1)(\mxi +\mpa - \mqb ) + (\muj+1)\sqrt{\mxj}}, \\
\Upsilon_4 =& ( \mpb-\mqb + \mxi)^2 + \mui\bigl(\mxj - (\mpa-\mqb + \mxi)^2\bigr).
\end{split}
\end{equation}
Also setting $z'_i=\EXP^{U_{\bu\bv,i'}(x_j,x_j';x_i')}$ and $z'_j=\EXP^{U_{\bu\bv,j'}(x_i,x_i';x_j')}$ as in \eqref{UmattoYBM} and applying the change of variables \eqref{q2cov} according to \eqref{UmattoYBMcov}, then gives the desired expression for the Yang-Baxter map $U(\al,\bt)\colon \bigl((y_i,y_j),(z_i,z_j)\bigr)\mapsto\bigl((y_i',y_j'),(z_i',z_j')\bigr)$ as 
\begin{empheq}{equation}
\label{h2d1mapb}
\begin{split}
\myj=\ds  \Upsilon_3, \qquad 
\mupi=&\ds \frac{\Upsilon_4 - (\mpb-\mqa + \Upsilon_3)^2}{\mxj - (\mpa-\mqa + \Upsilon_3)^2}, \\
\myi=\ds \Upsilon_4, \qquad 
\mupj=&\ds \frac{(\mpb - \mqb + \mxi - \sqrt{\Upsilon_4})(\mpb-\mqa + \Upsilon_3 + \sqrt{\Upsilon_4})}
                {(\mpb - \mqb + \mxi + \sqrt{\Upsilon_4})(\mpb-\mqa + \Upsilon_3 - \sqrt{\Upsilon_4})}.
\end{split}
\end{empheq}
According to Theorem \ref{thm:maintheorem}, the Yang-Baxter maps \eqref{q1d1map} and \eqref{h2d1mapb} together are a solution of the FYBE \eqref{MAYBE}.

Next, let $\hat{R}_{\bu\bv}$ and $\hat{U}_{\bu\bv}$ denote the classical R-matrices \eqref{CRMAT2} and \eqref{CUMAT2} that are constructed from the functions for the type-III solution of the CSTR from Table \ref{tab:ratlagcombos}.  The change of variables for \eqref{RmattoYBMcov} and \eqref{UmattoYBMcov} will be
\begin{equation}\label{q1d1cov}
f(x)=x^2,\qquad g(x)=x,\qquad h(x)=\ii x.
\end{equation}

For \eqref{CRMAT2}, setting $z_i=\EXP^{\hat{R}_{\bu\bv,i}(x_j',x_j;x_i)}$ and $z_j=\EXP^{\hat{R}_{\bu\bv,j}(x_i',x_i;x_j)}$ as in \eqref{RmattoYBM} (with $R$ replaced by $\hat{R}$) and applying the change of variables \eqref{q1d1cov} according to \eqref{RmattoYBMcov}, allows to solve for $y'_i$ and $y'_j$ as $y'_i=\Upsilon_1(\muj,\mxi;\mxj)$ and $y'_j=\Upsilon_2(\mui,\mxj;\mxi)$, where
\begin{equation}
\begin{gathered}
\Upsilon_1 = \mxj + (\mpa-\mqa)^2\Bigl(1 - \frac{2\sqrt{\mxj}}{\mpa-\mqa}
\frac{(\mxj-\mxi+(\mpa-\mqb)^2)(\muj+1) + 2(\mpa-\mqb)(\muj-1)\sqrt{\mxj}}
     {(\mxj-\mxi+(\mpa-\mqb)^2)(\muj-1) + 2(\mpa-\mqb)(\muj+1)\sqrt{\mxj}}\Bigr)  , \\
\Upsilon_2 = \mxi + (\mpb-\mqb)^2\Bigl(1 - \frac{2\sqrt{\mxi}}{\mpb-\mqb}
\frac{(\mxi-\mxj+(\mpa-\mqb)^2)(\mui+1) + 2(\mpa-\mqb)(\mui-1)\sqrt{\mxi}}
     {(\mxi-\mxj+(\mpa-\mqb)^2)(\mui-1) + 2(\mpa-\mqb)(\mui+1)\sqrt{\mxi}}\Bigr).
\end{gathered}
\end{equation}
Also setting $z'_i=\EXP^{\hat{R}_{\bu\bv,i'}(x_j,x_j';x_i')}$ and $z'_j=\EXP^{\hat{R}_{\bu\bv,j'}(x_i,x_i';x_j')}$ as in \eqref{RmattoYBM} (with $R$ replaced by $\hat{R}$) and applying the change of variables \eqref{q1d1cov} according to \eqref{RmattoYBMcov}, then gives the desired expression for the Yang-Baxter map $R(\al,\bt)\colon \bigl((y_i,y_j),(z_i,z_j)\bigr)\mapsto\bigl((y_i',y_j'),(z_i',z_j')\bigr)$ as 
\begin{empheq}{equation}
\label{q2map}
\begin{gathered}
\myj=\ds \Upsilon_1, \qquad
\mupi=\ds \frac{(\mxj - ( \mpa-\mqa - \sqrt{\Upsilon_1})^2)(\Upsilon_2 - (\mpb - \mqa + \sqrt{\Upsilon_1})^2)}
               {(\mxj - (\mpa - \mqa + \sqrt{\Upsilon_1})^2)(\Upsilon_2 - ( \mpb-\mqa - \sqrt{\Upsilon_1})^2)}, \\
\myi=\ds \Upsilon_2, \qquad
\mupj=\ds \frac{(\mxi - ( \mpb-\mqb - \sqrt{\Upsilon_2})^2)(\Upsilon_1- (\mpb - \mqa + \sqrt{\Upsilon_2})^2)}
               {(\mxi - ( \mpb - \mqb + \sqrt{\Upsilon_2})^2)(\Upsilon_1 - (\mpb-\mqa - \sqrt{\Upsilon_2})^2 )}.
\end{gathered}
\end{empheq}
According to Theorem \ref{thm:maintheorem}, this is a solution of the FYBE \eqref{MYBE}.

Similarly for \eqref{CUMAT2}, setting $z_i=\EXP^{\hat{U}_{\bu\bv,i}(x_j',x_j;x_i)}$ and $z_j=\EXP^{\hat{U}_{\bu\bv,j}(x_i',x_i;x_j)}$ as in \eqref{UmattoYBM} (with $U$ replaced by $\hat{U}$) and applying the change of variables \eqref{q3d1cov} according to \eqref{UmattoYBMcov}, allows to solve for $y'_i$ and $y'_j$ as $y'_i=\Upsilon_3(\muj,\mxi;\mxj)$ and $y'_j=\Upsilon_4(\mui,\mxj;\mxi)$, where
\begin{equation}
\begin{split}
\Upsilon_3 =& (\mpa - \mqa + \mxj)^2 + \muj\bigl(\mxi - (\mpa - \mqb + \mxj)^2\bigr), \\
\Upsilon_4 =& (\mqb-\mpb) + \frac{(\mui-1)\mxi+(\mui+1)(\mxj+\mpa-\mqb)\sqrt{\mxi}}{(\mui-1)(\mxj+\mpa-\mqb)+(\mui+1)\sqrt{\mxi}}.
\end{split}
\end{equation}
Also setting $z'_i=\EXP^{\hat{U}_{\bu\bv,i'}(x_j,x_j';x_i')}$ and $z'_j=\EXP^{\hat{U}_{\bu\bv,j'}(x_i,x_i';x_j')}$ as in \eqref{UmattoYBM} (with $U$ replaced by $\hat{U}$) and applying the change of variables \eqref{q3d1cov} according to \eqref{UmattoYBMcov}, then gives the desired expression for the Yang-Baxter map $U(\al,\bt)\colon \bigl((y_i,y_j),(z_i,z_j)\bigr)\mapsto\bigl((y_i',y_j'),(z_i',z_j')\bigr)$ as 
\begin{empheq}{equation}
\label{h2d1altmapb}
\begin{split}
\myj=\ds  \Upsilon_3, \qquad 
\mupi=&\ds \frac{(\mpa - \mqa + \mxj - \sqrt{\Upsilon_3})(\mpb-\mqa + \Upsilon_4 + \sqrt{\Upsilon_3})}
{(\mpa - \mqa + \mxj + \sqrt{\Upsilon_3})(\mpb-\mqa + \Upsilon_4 - \sqrt{\Upsilon_3})}, \\
\myi=\ds \Upsilon_4, \qquad 
\mupj=&\ds \frac{\Upsilon_3-(\mpb - \mqa + \Upsilon_4)^2}{\mxi - (\mpb-\mqb + \Upsilon_4)^2}.
\end{split}
\end{empheq}
According to Theorem \ref{thm:maintheorem}, the Yang-Baxter maps \eqref{q2map} and \eqref{h2d1altmapb} together are a solution of the FYBE \eqref{MAYBE}.

\tocless\paragraph{Case 2.}

The classical R-matrices for this case are given by \eqref{CRMAT} and \eqref{CUMAT1} using the functions for the type-II solution of the CSTR given in Table \ref{tab:ratlagcombos}.  The Yang-Baxter map for \eqref{CRMAT} was already derived in \eqref{q1d1map}, so it remains to derive the Yang-Baxter map for \eqref{CUMAT1}.

The classical R-matrix \eqref{CUMAT1} that is constructed from the functions for the first type-II solution of the CSTR from Table \ref{tab:hyperlagcombos} is given by (recall that $\gamma(z)$ is defined in \eqref{gamdef})
\begin{equation}
\begin{split}
\label{h2d0rmat}
U_{\bu\bv}\bigl(x_i,x_j;x_i',x_j'\bigr)=
\gamma\bigl(x_i+x_j+\ii(\rua-\rvb)\bigr)+\gamma\bigl(-x_i-x_j'-\ii(\rub-\rvb)\bigr)\phantom{.} \\
+ \gamma\bigl(x_i'+x_j'+\ii(\rub-\rva)\bigr)+\gamma\bigl(-x_i'-x_j-\ii(\rua-\rva)\bigr).
\end{split}
\end{equation}
The change of variables used in \eqref{UmattoYBMcov} is given by
\begin{equation}\label{h2d0cov}
f(x)={x},\qquad g(x)={x},\qquad h(x)={\ii x}.
\end{equation}
Using the expressions \eqref{UmattoYBM} formed from the classical R-matrix \eqref{h2d0rmat} and applying the change of variables \eqref{h2d0cov} according to \eqref{UmattoYBMcov}, then gives the desired expression for the Yang-Baxter map $U(\al,\bt)\colon \bigl((y_i,y_j),(z_i,z_j)\bigr)\mapsto\bigl((y_i',y_j'),(z_i',z_j')\bigr)$ as 
\begin{empheq}{equation}
\label{h2d0mapb}
\begin{gathered}
\myj=\ds  -\mxj-(\mpa-\mqa) + \muj\bigl( \mpa - \mqb + \mxi + \mxj\bigr), \qquad 
\mupi=\ds \muj^{-1}(\mui+\muj-1), \\[0.1cm] 
\myi=\ds -\mxi-(\mpb-\mqb) + \mui\bigl( \mpa - \mqb + \mxi + \mxj\bigr), \qquad 
\mupj=\ds \mui^{-1}(\mui+\muj-1).
\end{gathered}
\end{empheq}
According to Theorem \ref{thm:maintheorem}, the Yang-Baxter maps \eqref{h2d0mapb} and \eqref{q1d1map} together are a solution of the FYBE \eqref{MAYBE}.

\subsubsection{Algebraic cases}

\tocless\paragraph{Case 1.}

This case results in two different solutions of the FYBE \eqref{MAYBE} that are obtained from the type-III solution of the CSTR in Table \ref{tab:alglagcombos}.  
The first set of classical R-matrices for this case are given by \eqref{CRMAT} and \eqref{CUMAT1}, and the second set of classical R-matrices for this case are given by \eqref{CRMAT2} and \eqref{CUMAT2}.  The Yang-Baxter map for \eqref{CRMAT2} has already been derived in \eqref{q1d1map}.

First, let $R_{\bu\bv}$ and $U_{\bu\bv}$ denote the classical R-matrices \eqref{CRMAT} and \eqref{CUMAT1} that are constructed from the functions for the type-III solution of the CSTR from Table \ref{tab:alglagcombos}.  The change of variables for \eqref{RmattoYBMcov} and \eqref{UmattoYBMcov} will be
\begin{equation}\label{q1cov}
f(x)=x,\qquad g(x)=x,\qquad h(x)= \ii x.
\end{equation}

For \eqref{CRMAT}, setting $z_i=R_{\bu\bv,i}(x_j',x_j;x_i)$ and $z_j=R_{\bu\bv,j}(x_i',x_i;x_j)$ (the exponentials in \eqref{RmattoYBM} are not needed) and applying the change of variables \eqref{q1cov} according to \eqref{RmattoYBMcov}, allows to solve for $y'_i$ and $y'_j$ as $y'_i=\Upsilon_1(\muj,\mxi;\mxj)$ and $y'_j=\Upsilon_2(\mui,\mxj;\mxi)$, where
\begin{equation}
\begin{gathered}
\Upsilon_1= \mxj + \frac{(\mxi-\mxj)(\mpa-\mqa)}{\muj(\mxi-\mxj)+(\mpa-\mqb)}, \qquad
\Upsilon_2= \mxi + \frac{(\mxi-\mxj)(\mpb-\mqb)}{\mui(\mxi-\mxj)-(\mpa-\mqb)}.
\end{gathered}
\end{equation}
Also setting $z'_i={R_{\bu\bv,i'}(x_j,x_j';x_i')}$ and $z'_j={R_{\bu\bv,j'}(x_i,x_i';x_j')}$ (the exponentials in \eqref{RmattoYBM} are not needed) and applying the change of variables \eqref{q1cov} according to \eqref{RmattoYBMcov}, then gives the desired expression for the Yang-Baxter map $R(\al,\bt)\colon\bigl((y_i,y_j),(z_i,z_j)\bigr)\mapsto\bigl((y_i',y_j'),(z_i',z_j')\bigr)$ as 
\begin{empheq}{equation}
\label{q1d0map}
\begin{gathered}
\myj=\ds \Upsilon_1, \qquad 
\mupi=\ds \frac{\mqa-\mpb}{\Upsilon_1-\Upsilon_2}-\frac{\mpa-\mqa}{\mxj-\Upsilon_1}, \\
\myi=\ds \Upsilon_2, \qquad 
\mupj=\ds \frac{\mpb-\mqa}{\Upsilon_1-\Upsilon_2}-\frac{\mpb-\mqb}{\mxi-\Upsilon_2}.
\end{gathered}
\end{empheq}
According to Theorem \ref{thm:maintheorem}, this is a solution of the FYBE \eqref{MYBE}.

Next, the expression for \eqref{CUMAT1} is given by
\begin{equation}
\begin{split}
U_{\bu\bv}\bigl(x_i,x_j;x_i',x_j'\bigr)=
(\rua-\rvb-\ii x_j)\Log(x_i)-(\rub-\rvb-\ii x'_j)\Log(-x_i)\phantom{.} \\
+(\rub-\rva-\ii x'_j)\Log(x'_i)-(\rua-\rva-\ii x_j)\Log(-x'_i).
\end{split}
\end{equation}
Setting $z_i=U_{\bu\bv,i}(x_j',x_j;x_i)$, $z_j=\EXP^{U_{\bu\bv,j}(x_i',x_i;x_j)}$, $z'_i=U_{\bu\bv,i'}(x_j,x_j';x_i')$, and $z'_j=\EXP^{U_{\bu\bv,j'}(x_i,x_i';x_j')}$ (two of the exponentials in \eqref{UmattoYBM} are not needed), and applying the change of variables \eqref{q1cov} according to \eqref{UmattoYBMcov}, then gives the desired expression for the Yang-Baxter map $U(\al,\bt)\colon \bigl((y_i,y_j),(z_i,z_j)\bigr)\mapsto\bigl((y_i',y_j'),(z_i',z_j')\bigr)$ as 
\begin{empheq}{equation}
\label{h1e0altmapa}
\begin{array}{ll}
\myj=\ds -\muj\mxi, \quad &
\mupi=\ds -\mui\muj^{-1}, \\[0.1cm]
\myi=\ds  \mxj+\mxi\mui+\mpa-\mpb ,\quad &
\mupj=\ds \muj.
\end{array}
\end{empheq}
According to Theorem \ref{thm:maintheorem}, the Yang-Baxter maps \eqref{q1d0map} and \eqref{h1e0altmapa} together are a solution of the FYBE \eqref{MAYBE}.

The expression for \eqref{CUMAT2} is given by
\begin{equation}
\begin{split}
U_{\bu\bv}\bigl(x_i,x_j;x_i',x_j'\bigr)=
(\rua-\rvb-\ii x_i)\Log(x_j)-(\rua-\rva-\ii x'_i)\Log(-x_j)\phantom{.} \\
+(\rub-\rva-\ii x'_i)\Log(x'_j)-(\rub-\rvb-\ii x_i)\Log(-x'_j).
\end{split}
\end{equation}
Setting $z_i=\EXP^{U_{\bu\bv,i}(x_j',x_j;x_i)}$, $z_j={U_{\bu\bv,j}(x_i',x_i;x_j)}$, $z'_i=\EXP^{U_{\bu\bv,i'}(x_j,x_j';x_i')}$, and $z'_j={U_{\bu\bv,j'}(x_i,x_i';x_j')}$ (two of the exponentials in \eqref{UmattoYBM} are not needed), and applying the change of variables \eqref{q1cov} according to \eqref{UmattoYBMcov}, then gives the desired expression for the Yang-Baxter map $U(\al,\bt)\colon \bigl((y_i,y_j),(z_i,z_j)\bigr)\mapsto\bigl((y_i',y_j'),(z_i',z_j')\bigr)$ as 
\begin{empheq}{equation}
\label{h1e0altmapb}
\begin{array}{ll}
\myj=\ds \mxi+\mxj\muj+\mqa-\mqb, \quad &
\mupi=\ds \mui, \\[0.1cm]
\myi=\ds  -\mui\mxj ,\quad &
\mupj=\ds -\muj\mui^{-1}.
\end{array}
\end{empheq}
According to Theorem \ref{thm:maintheorem}, the Yang-Baxter maps \eqref{q1d1map} and \eqref{h1e0altmapb} together are a solution of the FYBE \eqref{MAYBE}.

\tocless\paragraph{Case 2.}

For this case the classical R-matrices are given by \eqref{CRMAT} and \eqref{CUMAT1} with the functions of the type-II solution of the CSTRs given in Table \ref{tab:alglagcombos}.  The Yang-Baxter map for \eqref{CRMAT} was already derived in \eqref{q1d0map}, so it remains to derive the Yang-Baxter map for \eqref{CUMAT1}.  The classical R-matrix \eqref{CUMAT1} is given by (this ends up being the same as \eqref{h3d0rmat})
\begin{equation}\label{h1rmat}
\begin{split}
U_{\bu\bv}\bigl(x_i,x_j;x_i',x_j'\bigr)=&(x_i-x_i')(x_j'-x_j).
\end{split}
\end{equation}
Setting $z_i={U_{\bu\bv,i}(x_j',x_j;x_i)}$, $z_j={U_{\bu\bv,j}(x_i',x_i;x_j)}$, $z'_i={U_{\bu\bv,i'}(x_j,x_j';x_i')}$, and $z'_j={U_{\bu\bv,j'}(x_i,x_i';x_j')}$ (the exponentials in \eqref{UmattoYBM} are not needed, in contrast with \eqref{h3d0rmat}), and applying the change of variables \eqref{q1cov} according to \eqref{UmattoYBMcov}, then gives the desired expression for the Yang-Baxter map $U(\al,\bt)\colon \bigl((y_i,y_j),(z_i,z_j)\bigr)\mapsto\bigl((y_i',y_j'),(z_i',z_j')\bigr)$ as 
\begin{empheq}{equation}
\label{h1e0b3mapb}
\begin{gathered}
\myj=\mxi+\muj, \qquad 
\mupi=\mui, \\
\myi= \mxj+\mui, \qquad 
\mupj=\muj. 
\end{gathered}
\end{empheq}
According to Theorem \ref{thm:maintheorem}, the Yang-Baxter maps \eqref{q1d0map} and \eqref{h1e0b3mapb} together are a solution of the FYBE \eqref{MAYBE}.

\section{Summary}

In this paper it was shown how quadrirational Yang-Baxter maps (YBMs) that are solutions to the functional (or set-theoretical) form of the Yang-Baxter equation, are directly related to the star-triangle relations, which are a special form of the Yang-Baxter equation whose solutions are given in terms of special functions.  Using explicit solutions of the classical star-triangle relations, a new set of sixteen YBMs was derived which have both two-component variables and two-component parameters.  Six of these YBMs satisfy the usual form of Yang-Baxter equation given by \eqref{MYBE}, and these Yang-Baxter maps can also be paired with different YBMs to solve the entwining form of Yang-Baxter equation given by \eqref{MAYBE}.  Apart from the elliptic case, each of these YBMs were found to be quadrirational, while only the YBMs that solve the FYBE \eqref{MYBE} were found to be reversible.  

The method of this paper relies on using a certain vertex form of the Yang-Baxter equation that is implied by the star-triangle relation. In principle there could exist solutions of the former equation that are independent of the latter, with several well-known examples at the quantum level, such as the 8-vertex model \cite{Baxter:1972hz}. Such independent solutions would likely lead to different YBMs from the ones in this paper, which have all originated from a star-triangle relation.   In particular, it is expected that extending the YBMs of this paper to multicomponent variables (specifically, $2n$-component variables, for $n=1,2,\ldots$) will require the use of such equations that are independent of the star-triangle relations, such as those from \cite{Bazhanov:2011mz}.  It is also possible that other forms of the Yang-Baxter equation that are solved by different combinations of the classical R-matrices $R_{\bu\bv}$ and $U_{\bu\bv}$ could be constructed from the same star-triangle relations that were used in this paper, which would then lead to different forms of the FYBE.  Such a different classical Yang-Baxter equation that involves only the classical R-matrix $U_{\bu\bv}$ (in the notation of this paper) appears to have arisen in the quantum group approach introduced by Bazhanov and Sergeev \cite{Bazhanov:2015gra}. The existence of other solutions of this form of the Yang-Baxter equation and the connection to Yang-Baxter maps needs further detailed investigation.  Also motivated by the latter work \cite{Bazhanov:2015gra} (see also \cite{TsuboiYBMs,DSYBM}), the author expects that it will be important to understand what are the quantum group structures, if any, associated to the YBMs of this paper.  Finally, the star-triangle relation may be interpreted as a quantum analogue of discrete integrable systems which arise in the quasi-classical limit \cite{Bazhanov:2007mh,Bazhanov:2010kz,Bazhanov:2016ajm,Kels:2018xge}, and it would be interesting to better understand what this implies for the Yang-Baxter maps of this paper.

\section*{Acknowledgements}

The author would like to thank Yang Shi for fruitful discussions and helpful suggestions for improving the manuscript.

\begin{appendices}
\numberwithin{equation}{section}

\section{List of Yang-Baxter maps}
\label{app:YBMlist}

For a variable or parameter $x$, recall the following notations used from Section \ref{sec:examples}:
\begin{equation}
\dot{x}^2=4x^3-g_2x-g_3, \qquad \breve{x}=\frac{\dot{x}^2}{4(x-\me)^2}-x-\me, \qquad \overline{x}=x+\sqrt{x^2-1},
\end{equation}
where $\wp(x)$ is the Weierstrass elliptic function with elliptic invariants $g_2,g_3$, or associated half-periods $\omega_1,\omega_2$.

\subsection{Solutions of the FYBE \eqref{MYBE}}

Each Yang-Baxter map $R(\al,\bt)\colon \bigl((y_i,y_j),(z_i,z_j)\bigr)\mapsto\bigl((y_i',y_j'),(z'_i,z'_j)\bigr)$ given below is a solution to the FYBE \eqref{MYBE}.  Each map is quadrirational satisfying Proposition \ref{prop:quadri} (except for $R6$ which is birational), and reversible satisfying Proposition \ref{prop:revers}.

\subsubsection*{\texorpdfstring{$(R6)$}{(R6)}}

\begin{equation}\nonumber
\begin{gathered}
A(\mpc,\mqc)=\frac{(\tmp+\tmq)^2}{4(\mpc-\mqc)^2}-\mpc-\mqc, \quad
B(x,\mpc,\mqc)=\tmp\mqc+\tmq\mpc-(\tmp+\tmq)x, \\
E(x,\mpc,\mqc,\gamma)=\frac{1}{E(x,\mqc,\mpc,\gamma)}=\frac{(B(x,\mpc,\gamma)+(\mpc-\gamma)\dot{x})(B(x,\mqc,\gamma)-(\mqc-\gamma)\dot{x})}{(B(x,\mpc,\gamma)-(\mpc-\gamma)\dot{x})(B(x,\mqc,\gamma)+(\mqc-\gamma)\dot{x})}, \\[0.2cm]
F(x,\mpc,\mqc)=2\dot{x}\frac{B\bigl(A(\mpc,\mqc),\mpc,\mqc\bigr)}{\mpc-\mqc}+2g_3+(A(\mpc,\mqc)+x)(g_2-4xA(\mpc,\mqc)), \\
G(x,y,\mpc,\mqc)=F(x,\mpc,\mqc)+4y(A(\mpc,\mqc)-x)^2.
\end{gathered}
\end{equation}


\begin{equation}\nonumber
\begin{split}
\Upsilon_1 &=  
\frac{\muj G(\mxj,\mxi,\mqb,\mpa)F\bigl(\mxj,\mpa,\mqa) E(\mxj,\mqa,\mqb,\mpa)^2
     -     G(\mxj,\mxi,\mpa,\mqb)F\bigl(\mxj,\mqa,\mpa)}
     {4\bigl(\mxj-A(\mpa,\mqa)\bigr)^2\bigl(G(\mxj,\mxi,\mpa,\mpb)-\muj G(\mxj,\mxi,\mqb,\mpa) E(\mxj,\mqa,\mqb,\mpa)^2\bigr)}, \\
\Upsilon_2 &=  
\frac{\mui G(\mxi,\mxj,\mqb,\mpa)F\bigl(\mxi,\mpb,\mqb) E(\mxi,\mpa,\mpb,\mqb)^2
     -     G(\mxi,\mxj,\mpa,\mqb)F\bigl(\mxi,\mqb,\mpb)}
     {4\bigl(\mxi-A(\mpb,\mqb)\bigr)^2\bigl(G(\mxi,\mxj,\mpa,\mpb)-\mui G(\mxi,\mxj,\mqb,\mpa) E(\mxi,\mpa,\mpb,\mqb)^2\bigr)}.
\end{split}
\end{equation}

\begin{empheq}[box=\fbox]{equation}\nonumber
\begin{split}
\myj=\ds \Upsilon_1, \qquad
\mupi&=\ds E(\Upsilon_1,\mpb,\mpa,\mqa)^2 
\frac{G(\Upsilon_1,\mxj,\mpa,\mqa)G(\Upsilon_1,\breve{\Upsilon}_2,\mqa,\mpb)}
     {G(\Upsilon_1,\mxj,\mqa,\mpa)G(\Upsilon_1,\breve{\Upsilon}_2,\mpb,\mqa)}, \\
\myi=\ds \Upsilon_2, \qquad
\mupj&=\ds E(\Upsilon_2,\mqb,\mqa,\mpb)^2 
\frac{G(\Upsilon_2,\mxi,\mpb,\mqb)G(\Upsilon_2,\breve{\Upsilon}_1,\mqa,\mpb)}
     {G(\Upsilon_2,\mxi,\mqb,\mpb)G(\Upsilon_2,\breve{\Upsilon}_1,\mpb,\mqa)}.
\end{split}
\end{empheq}

\subsubsection*{\texorpdfstring{$(R5)$}{(R5)}}

\begin{equation}\nonumber
\begin{split}
\Upsilon_1 = \frac{\omxj^2(1-\muj)\Bigl(\mpa^2 + \frac{\mqa^2\mqb^2}{\mpa^2}\Bigr) + (\omxj^4-\muj)\mqb^2 + (1-\muj\omxj^4)\mqa^2 - 2\mpa\mqb\mxi\omxj\Bigl(\muj - \omxj^2+ (\muj\omxj^2-1)\frac{\mqa^2}{\mpa^2} \Bigr)}{2\frac{\mqa}{\mpa}\omxj\bigl(\mqb^2(\omxj^2-\muj) + \mpa^2(1-\muj\omxj^2) - 2\mpa\mqb\mxi\omxj(\muj-1) \bigr)}, \\[0.1cm]
\Upsilon_2 = \frac{\omxi^2(1-\mui)\Bigl(\mqb^2 + \frac{\mpa^2\mpb^2}{\mqb^2}\Bigr)+ (\omxi^4-\mui)\mpb^2  + (1 - \mui\omxi^4)\mpa^2 -             2\mpa\mqb\omxi\mxj\Bigl((\mui - \omxi^2)\frac{\mpb^2}{\mqb^2}  + \mui\omxi^2-1\Bigr) }{  2\frac{\mpb}{\mqb}\omxi\bigl(\mqb^2(\omxi^2-\mui) + \mpa^2(1 - \mui\omxi^2) - 2\mpa\mqb\omxi\mxj( \mui-1)\bigr)}.
\end{split}
\end{equation}

\begin{empheq}[box=\fbox]{equation}\nonumber
\begin{gathered}
\myj=\ds \Upsilon_1, \qquad 
\mupi=\ds \frac{(\mpa^2 + \mqa^2\overline{\Upsilon}_1^2 - 2\mpa\mqa\overline{\Upsilon}_1\mxj)(\mqa^2 + \mpb^2\overline{\Upsilon}_1^2 + 2\mpb\mqa\overline{\Upsilon}_1\Upsilon_2 )} 
  {(\mqa^2 + \mpa^2\overline{\Upsilon}_1^2 - 2\mpa\mqa\overline{\Upsilon}_1\mxj)(\mpb^2 + \mqa^2\overline{\Upsilon}_1^2 + 2\mpb\mqa\overline{\Upsilon}_1\Upsilon_2)}, \\
\myi=\ds \Upsilon_2, \qquad 
\mupj=\ds \frac{(\mpb^2 + \mqb^2\overline{\Upsilon}_2^2 - 2\mpb\mqb\overline{\Upsilon}_2\mxi)(\mqa^2 + \mpb^2\overline{\Upsilon}_2^2 + 2\mpb\mqa\overline{\Upsilon}_2 \Upsilon_1)}
    {(\mqb^2 + \mpb^2\overline{\Upsilon}_2^2 - 2\mpb\mqb\overline{\Upsilon}_2\mxi)(\mpb^2 + \mqa^2\overline{\Upsilon}_2^2 + 2\mpb\mqa\overline{\Upsilon}_2 \Upsilon_1)}.
\end{gathered}
\end{empheq}

\subsubsection*{\texorpdfstring{$(R4)$}{(R4)}}


\begin{equation}\nonumber
\begin{gathered}
\Upsilon_1 =  \mxj\frac{\mpa(\mpa\mxi + \mqb\mxj) + \mqa\muj(\mpa\mxj + \mqb\mxi) }{ \mqa(\mpa\mxi+\mqb\mxj) + \mpa\muj(\mpa\mxj + \mqb\mxi)}, \\
\Upsilon_2 = \mxi\frac{\mpb(\mpa\mxj + \mqb\mxi) + \mqb\mui(\mpa\mxi + \mqb\mxj) }{\mqb(\mpa\mxj + \mqb\mxi) + \mpb\mui(\mpa\mxi+\mqb\mxj)}.
\end{gathered}
\end{equation}

\begin{empheq}[box=\fbox]{equation}\nonumber
\begin{gathered}
\myj=\ds \Upsilon_1, \qquad 
\mupi=\ds \frac{(\mpa\mxj-\mqa \Upsilon_1)( \mpb \Upsilon_1 + \mqa \Upsilon_2)}{(\mpa \Upsilon_1-\mqa\mxj)( \mpb \Upsilon_2 + \mqa \Upsilon_1)}, \\
\myi=\ds \Upsilon_2, \qquad 
\mupj=\ds \frac{(\mpb \mxi - \mqb \Upsilon_2)(\mpb \Upsilon_2+\mqa \Upsilon_1  )}{(\mpb \Upsilon_2 - \mqb\mxi )(\mpb \Upsilon_1 + \mqa \Upsilon_2)}.
\end{gathered}
\end{empheq}

\subsubsection*{\texorpdfstring{$(R3)$}{(R3)}}


\begin{equation}\nonumber
\begin{split}
\Upsilon_1 =& \mxj + (\mpa-\mqa)^2\Bigl(1 - \frac{2\sqrt{\mxj}}{\mpa-\mqa}
\frac{(\mxj-\mxi+(\mpa-\mqb)^2)\frac{\muj+1}{2} + (\mpa-\mqb)(\muj-1)\sqrt{\mxj}}
     {(\mxj-\mxi+(\mpa-\mqb)^2)\frac{\muj-1}{2} + (\mpa-\mqb)(\muj+1)\sqrt{\mxj}}\Bigr)  , \\
\Upsilon_2 =& \mxi + (\mpb-\mqb)^2\Bigl(1 - \frac{2\sqrt{\mxi}}{\mpb-\mqb}
\frac{(\mxi-\mxj+(\mpa-\mqb)^2)\frac{\mui+1}{2} + (\mpa-\mqb)(\mui-1)\sqrt{\mxi}}
     {(\mxi-\mxj+(\mpa-\mqb)^2)\frac{\mui-1}{2} + (\mpa-\mqb)(\mui+1)\sqrt{\mxi}}\Bigr).
\end{split}
\end{equation}

\begin{empheq}[box=\fbox]{equation}\nonumber
\begin{gathered}
\myj=\ds \Upsilon_1, \qquad
\mupi=\ds \frac{(\mxj - ( \mpa-\mqa - \sqrt{\Upsilon_1})^2)(\Upsilon_2 - (\mpb - \mqa + \sqrt{\Upsilon_1})^2)}
               {(\mxj - (\mpa - \mqa + \sqrt{\Upsilon_1})^2)(\Upsilon_2 - ( \mpb-\mqa - \sqrt{\Upsilon_1})^2)}, \\
\myi=\ds \Upsilon_2, \qquad
\mupj=\ds \frac{(\mxi - ( \mpb-\mqb - \sqrt{\Upsilon_2})^2)(\Upsilon_1- (\mpb - \mqa + \sqrt{\Upsilon_2})^2)}
               {(\mxi - ( \mpb - \mqb + \sqrt{\Upsilon_2})^2)(\Upsilon_1 - (\mpb-\mqa - \sqrt{\Upsilon_2})^2 )}.
\end{gathered}
\end{empheq}

\subsubsection*{\texorpdfstring{$(R2)$}{(R2)}}


\begin{equation}\nonumber
\begin{split}
\Upsilon_1 &=  \mxj + (\mpa - \mqa)\frac{(\muj+1)(\mxi-\mxj) + (1-\muj)(\mpa - \mqb)}{(1-\muj)(\mxi - \mxj) + (\muj+1)(\mpa-\mqb)}, \\
\Upsilon_2 &= \mxi + (\mpb - \mqb)\frac{(\mui+1)(\mxi- \mxj) + (\mui-1)(\mpa - \mqb) }{(1-\mui)(\mxi - \mxj) - (\mui+1)(\mpa-\mqb)}.
\end{split}
\end{equation}

\begin{empheq}[box=\fbox]{equation}\nonumber
\begin{gathered}
\myj=\ds \Upsilon_1, \qquad
\mupi=\ds \frac{(\mpa - \mqa + (\mxj - \Upsilon_1))( \mpb-\mqa + (\Upsilon_1 - \Upsilon_2))}
               {(\mpa - \mqa - (\mxj - \Upsilon_1))( \mpb-\mqa - (\Upsilon_1 - \Upsilon_2))}, \\
\myi=\ds \Upsilon_2, \qquad
\mupj=\ds \frac{(\mpb - \mqb + (\mxi - \Upsilon_2))(\mpb - \mqa - (\Upsilon_1 - \Upsilon_2))}
               {(\mpb - \mqb - (\mxi - \Upsilon_2))(\mpb - \mqa + (\Upsilon_1 - \Upsilon_2))}.
\end{gathered}
\end{empheq}

\subsubsection*{\texorpdfstring{$(R1)$}{(R1)}}

\begin{equation}\nonumber
\begin{split}
\Upsilon_1= \mxj + \frac{(\mxi-\mxj)(\mpa-\mqa)}{\muj(\mxi-\mxj)+(\mpa-\mqb)}, \qquad
\Upsilon_2= \mxi + \frac{(\mxi-\mxj)(\mpb-\mqb)}{\mui(\mxi-\mxj)-(\mpa-\mqb)}.
\end{split}
\end{equation}

\begin{empheq}[box=\fbox]{equation}\nonumber
\begin{gathered}
\myj=\ds \Upsilon_1, \qquad 
\mupi=\ds \frac{\mqa-\mpb}{\Upsilon_1-\Upsilon_2}-\frac{\mpa-\mqa}{\mxj-\Upsilon_1}, \\
\myi=\ds \Upsilon_2, \qquad 
\mupj=\ds \frac{\mpb-\mqa}{\Upsilon_1-\Upsilon_2}-\frac{\mpb-\mqb}{\mxi-\Upsilon_2}.
\end{gathered}
\end{empheq}


\subsection{Solutions of the FYBE \eqref{MAYBE}}

Each Yang-Baxter map $U(\al,\bt)\colon \bigl((y_i,y_j),(z_i,z_j)\bigr)\mapsto\bigl((y_i',y_j'),(z'_i,z'_j)\bigr)$ given below is a solution to the FYBE \eqref{MAYBE}, together with one of the Yang-Baxter maps $R(\al,\bt)$ given above.  Each map is quadrirational satisfying Proposition \ref{prop:quadri}, but these maps are not reversible.

\subsubsection*{\texorpdfstring{$(U5)$}{$(U5)$}}

\begin{equation}\nonumber
\begin{gathered}
\Upsilon_1 = \mxi\muj + \frac{\mqa + \mqb\muj}{2\mpa\mxj} + \mpa\mxj\frac{\mqb+\mqa\muj}{2\mqa\mqb}, \quad
\Upsilon_2 = \frac{\mqb}{\mpb}\frac{\mqb\omxi(\mui-1) + \mpa\mxj(\mui\omxi^2-1)}{\mpa\omxi\mxj(\mui-1) + \mqb(\mui - \omxi^2)}.
\end{gathered}
\end{equation}

\begin{empheq}[box=\fbox]{equation}\nonumber
\begin{split}
\myj=\ds  \Upsilon_1, \qquad 
\mupi=&\ds \frac{(\mpa\mxj - \mqa\overline{\Upsilon}_1)(\mqa + \mpb \Upsilon_2\overline{\Upsilon}_1)}{(\mpb \Upsilon_2 + \mqa\overline{\Upsilon}_1)(\mpa\mxj\overline{\Upsilon}_1-\mqa)}, \\
\myi=\ds \Upsilon_2, \qquad 
\mupj=&\ds -\frac{\mqb}{\mqa}\frac{\mqa^2 + \mpb^2\Upsilon_2^2 + 2\mpb\mqa\Upsilon_1\Upsilon_2}
                                  {\mqb^2 + \mpb^2\Upsilon_2^2 - 2\mpb\mqb\mxi \Upsilon_2}.
\end{split}
\end{empheq}

$R(\al,\bt)$ is given by $R5$.

\subsubsection*{\texorpdfstring{$(U4a)$}{(U4a)}}

\begin{equation}\nonumber
\begin{gathered}
\Upsilon_1 =  \frac{\mqa}{\mpa}\frac{\mqb\omxj(\muj-1) + \mpa\mxi(\muj\omxj^2-1)}{\mpa\mxi\omxj(\muj-1) + \mqb(\muj - \omxj^2)}, \quad
\Upsilon_2 = {\mxj}{\mui} + \mxi\frac{\mpb+\mpa\mui}{2\mqb} + \mqb\frac{\mpa+\mpb\mui}{2\mpa\mpb\mxi}, 
\end{gathered}
\end{equation}

\begin{empheq}[box=\fbox]{equation}\nonumber
\begin{gathered}
\myj=\ds  \Upsilon_1, \qquad 
\mupi=\ds -\frac{\mpa}{\mpb}\frac{\mqa^2  + \mpb^2 \Upsilon_1^2 + 2\mpb\mqa \Upsilon_1\Upsilon_2}
                                 {\mqa^2 + \mpa^2\Upsilon_1^2 - 2\mpa\mqa\mxj \Upsilon_1} \\
\myi=\ds \Upsilon_2, \qquad 
\mupj=\ds \frac{(\mpb\mxi-\mqb\overline{\Upsilon}_2)(\mqa + \mpb \Upsilon_1\overline{\Upsilon}_2)}
              {(\mpb \Upsilon_1+\mqa\overline{\Upsilon}_2)(\mpb\mxi\overline{\Upsilon}_2-\mqb)},
\end{gathered}
\end{empheq}

$R(\al,\bt)$ is given by $R4$.

\subsubsection*{\texorpdfstring{$(U4b)$}{(U4b)}}

\begin{equation}\nonumber
\Upsilon_1 = \mxi\muj + \frac{\mqa + \mqb\muj}{\mpa\mxj}, \qquad
\Upsilon_2 = \mxj\mui + \mqb\frac{\mpa + \mpb\mui}{\mpa\mpb\mxi}.
\end{equation}

\begin{empheq}[box=\fbox]{equation}\nonumber
\begin{split}
\myj=\ds  \Upsilon_1, \qquad 
\mupi=&\ds \frac{\Upsilon_1\Upsilon_2+\mqa\mpb^{-1}}{\mxj\Upsilon_1-\mqa\mpa^{-1}}, \\
\myi=\ds \Upsilon_2, \qquad 
\mupj=&\ds \frac{\Upsilon_1\Upsilon_2+\mqa\mpb^{-1}}{\mxi\Upsilon_2-\mqb\mpb^{-1}}.
\end{split}
\end{empheq}

$R(\al,\bt)$ is given by $R4$.

\subsubsection*{\texorpdfstring{$(U4c)$}{(U4c)}}

\begin{equation}\nonumber
\Upsilon_1 = \muj\mxi, \qquad
\Upsilon_2 = \mui\mxj.
\end{equation}

\begin{empheq}[box=\fbox]{equation}\nonumber
\begin{split}
\myj=&\ds  \Upsilon_1, \qquad 
\mupi=\ds \Upsilon_2\mxj^{-1}, \\
\myi=&\ds \Upsilon_2, \qquad 
\mupj=\ds \Upsilon_1\mxi^{-1}.
\end{split}
\end{empheq}

$R(\al,\bt)$ is given by $R4$.

\subsubsection*{\texorpdfstring{$(U3)$}{(U3)}}

\begin{equation}\nonumber
\begin{split}
\Upsilon_1 =& (\mpa - \mqa + \mxj)^2 + \muj\bigl(\mxi - (\mpa - \mqb + \mxj)^2\bigr), \\
\Upsilon_2 =& (\mqb-\mpb) + \frac{(\mui-1)\mxi+(\mui+1)(\mxj+\mpa-\mqb)\sqrt{\mxi}}{(\mui-1)(\mxj+\mpa-\mqb)+(\mui+1)\sqrt{\mxi}}.
\end{split}
\end{equation}

\begin{empheq}[box=\fbox]{equation}\nonumber
\begin{split}
\myj=\ds  \Upsilon_1, \qquad 
\mupi=&\ds \frac{(\mpa - \mqa + \mxj - \sqrt{\Upsilon_1})(\mpb-\mqa + \Upsilon_2 + \sqrt{\Upsilon_1})}
{(\mpa - \mqa + \mxj + \sqrt{\Upsilon_1})(\mpb-\mqa + \Upsilon_2 - \sqrt{\Upsilon_1})}, \\
\myi=\ds \Upsilon_2, \qquad 
\mupj=&\ds \frac{\Upsilon_1-(\mpb - \mqa + \Upsilon_2)^2}{\mxi - (\mpb-\mqb + \Upsilon_2)^2}.
\end{split}
\end{empheq}

$R(\al,\bt)$ given by $R3$.

\subsubsection*{\texorpdfstring{$(U2a)$}{(U2a)}}

\begin{equation}\nonumber
\begin{split}
\Upsilon_1 =& (\mqa-\mpa) + \frac{(\muj-1)\mxj + (\muj+1)(\mxi + \mpa - \mqb)\sqrt{\mxj}}{(\muj-1)(\mxi +\mpa - \mqb ) + (\muj+1)\sqrt{\mxj}}, \\
\Upsilon_2 =& ( \mpb-\mqb + \mxi)^2 + \mui\bigl(\mxj - (\mpa-\mqb + \mxi)^2\bigr).
\end{split}
\end{equation}

\begin{empheq}[box=\fbox]{equation}\nonumber
\begin{split}
\myj=\ds  \Upsilon_1, \qquad 
\mupi=&\ds \frac{\Upsilon_2 - (\mpb-\mqa + \Upsilon_1)^2}{\mxj - (\mpa-\mqa + \Upsilon_1)^2}, \\
\myi=\ds \Upsilon_2, \qquad 
\mupj=&\ds \frac{(\mpb - \mqb + \mxi - \sqrt{\Upsilon_2})(\mpb-\mqa + \Upsilon_1 + \sqrt{\Upsilon_2})}
                {(\mpb - \mqb + \mxi + \sqrt{\Upsilon_2})(\mpb-\mqa + \Upsilon_1 - \sqrt{\Upsilon_2})}
                 .
\end{split}
\end{empheq}

$R(\al,\bt)$ is given by $R2$.

\subsubsection*{\texorpdfstring{$(U2b)$}{(U2b)}}

\begin{equation}\nonumber
\begin{split}
\Upsilon_1=& -\mxj-(\mpa-\mqa) + \muj\bigl( \mpa - \mqb + \mxi + \mxj\bigr), \\
\Upsilon_2=& -\mxi-(\mpb-\mqb) + \mui\bigl( \mpa - \mqb + \mxi + \mxj\bigr).
\end{split}
\end{equation}

\begin{empheq}[box=\fbox]{equation}\nonumber
\begin{gathered}
\myj=\ds  \Upsilon_1, \qquad 
\mupi=\ds \frac{\mpb - \mqa + \Upsilon_1 + \Upsilon_2}{\mpa - \mqa + \mxj + \Upsilon_1}, \\
\myi=\ds \Upsilon_2, \qquad 
\mupj=\ds \frac{\mpb - \mqa + \Upsilon_1 + \Upsilon_2}{\mpb - \mqb + \mxi + \Upsilon_2}.
\end{gathered}
\end{empheq}

$R(\al,\bt)$ is given by $R2$.

\subsubsection*{\texorpdfstring{$(U2c)$}{(U2c)}}

\begin{equation}\nonumber
\begin{split}
\Upsilon_1= \mxi+\muj\mxj+\mqa-\mqb, \qquad
\Upsilon_2= -\mxj\mui.
\end{split}
\end{equation}

\begin{empheq}[box=\fbox]{equation}\nonumber
\begin{split}
\myj=\ds \Upsilon_1, \qquad &
\mupi=\ds -\Upsilon_2\mxj^{-1}, \\[0.1cm]
\myi=\ds \Upsilon_2,\qquad &
\mupj=\ds (\Upsilon_1-\mxi-\mqa+\mqb)\Upsilon_2^{-1},
\end{split}
\end{empheq}

$R(\al,\bt)$ is given by $R2$.

\subsubsection*{\texorpdfstring{$(U1a)$}{(U1a)}}

\begin{equation}\nonumber
\begin{split}
\Upsilon_1= -\mxi\muj, \qquad
\Upsilon_2= \mxj + \mui\mxi + \mpa - \mpb.
\end{split}
\end{equation}

\begin{empheq}[box=\fbox]{equation}\nonumber
\begin{split}
\myj=\ds \Upsilon_1, \qquad &
\mupi=\ds (\Upsilon_2-\mxj-\mpa+\mpb)\Upsilon_1^{-1}, \\[0.1cm]
\myi=\ds \Upsilon_2,\qquad &
\mupj=\ds \Upsilon_1 \mxi^{-1},
\end{split}
\end{empheq}

$R(\al,\bt)$ is given by $R1$.

\subsubsection*{\texorpdfstring{$(U1b)$}{(U1b)}}

\begin{equation}\nonumber
\begin{split}
\Upsilon_1= \muj+\mxi, \qquad
\Upsilon_2= \mui+\mxj.
\end{split}
\end{equation}

\begin{empheq}[box=\fbox]{equation}\nonumber
\begin{gathered}
\myj=\Upsilon_1, \qquad 
\mupi=\Upsilon_2-\mxj, \\
\myi= \Upsilon_2, \qquad 
\mupj=\Upsilon_1-\mxi. 
\end{gathered}
\end{empheq}

$R(\al,\bt)$ is given by $R1$.

\subsection{Summary}

\begin{table}[htb!]
\centering
\begin{tabular}{|c|c|c|c|c|c|c|c|c|c|c|c|}
\hline 
$R(\al,\bt)$ & 
$R6$ & 
$R5$ & 
\multicolumn{3}{c|}{$R4$} & 
$R3$ & 
\multicolumn{3}{c|}{$R2$} & 
\multicolumn{2}{c|}{$R1$} 
\\[0.0cm]
\hline
$U(\al,\bt)$ & 
 - & 
$U5$ & 
$U4a$ & 
$U4b$ & 
$U4c$ & 
$U3$ & 
$U2a$ & 
$U2b$ & 
$U2c$ & 
$U1a$ & 
$U1b$ \\[0.0cm]
\hline 
\end{tabular}
\caption{
Pairs of Yang-Baxter maps $R(\al,\bt)$ and $U(\al,\bt)$ that satisfy the FYBE \eqref{MAYBE}.  Each Yang-Baxter map $R(\al,\bt)$ also satisfies the FYBE \eqref{MYBE}.}
\label{table-ABSYBM3}
\end{table}

\end{appendices}


{\small
\bibliography{MComp}
 \bibliographystyle{utphys}
}

\end{document}